\documentclass[longauth]{aa}

\newcommand{\starPublished}{33\,812\,183}

\newcommand\hip{\textit{Hipparcos}}
\newcommand\gaia{\textit{Gaia}}

\newcommand\gdrtwo{\gaia~DR2}
\newcommand\gdrethree{\gaia~EDR3}
\newcommand\gdrthree{\gaia~DR3}
\newcommand\gdrfour{\gaia~DR4}
\newcommand\RVS{\textit{Radial Velocity Spectrometer}}

\newcommand{\grvs}{G_\mathrm{RVS}}
\newcommand{\onboardgrvs}{G_\mathrm{RVS}^{\rm{on-board}}}
\newcommand{\extgrvs}{G_\mathrm{RVS}^{\rm{ext}}}

\newcommand{\gmag}{G}
\newcommand{\grpmag}{G_\mathrm{RP}}

\newcommand{\Teff}{T_{\rm eff}}
\newcommand{\logg}{\log g}
\newcommand{\FeH}{[{\rm Fe/H}]}

\newcommand\kms{\ensuremath{\text{km~s}^{-1}}}
\newcommand\ms{\ensuremath{\text{m~s}^{-1}}}

\newcommand{\esec}{$\mathrm{e}^{-}\mathrm{s}^{-1}$}

\usepackage{graphicx}
\usepackage{txfonts}
\usepackage[draft]{hyperref}
\begin{document}

   \title{\gaia\ Data Release 3}

   \subtitle{Properties and validation of the radial velocities}

   \author{D.~Katz\inst{\ref{gepi}}
\and P.~Sartoretti\inst{\ref{gepi}}
\and A.~Guerrier\inst{\ref{cnes}}
\and P.~Panuzzo\inst{\ref{gepi}}
\and G. M.~Seabroke\inst{\ref{mssl}}
\and F.~Th\'{e}venin\inst{\ref{oca}}
\and M.~Cropper\inst{\ref{mssl}}
\and K.~Benson\inst{\ref{mssl}}
\and R.~Blomme\inst{\ref{brussels}}
\and R.~Haigron\inst{\ref{gepi}}
\and O.~Marchal\inst{\ref{strasbourg}}
\and M.~Smith\inst{\ref{mssl}}
\and S.~Baker\inst{\ref{mssl}}
\and L.~Chemin\inst{\ref{chile}}
\and Y.~Damerdji\inst{\ref{alger},\ref{liege}}
\and M.~David\inst{\ref{antwerpen}}
\and C.~Dolding\inst{\ref{mssl}}
\and Y.~Fr\'{e}mat\inst{\ref{brussels}}
\and E.~Gosset\inst{\ref{liege},\ref{fnrs}}
\and K.~Jan{\ss}en\inst{\ref{aip}}
\and G.~Jasniewicz\inst{\ref{montpellier}}
\and A.~Lobel\inst{\ref{brussels}}
\and G.~Plum\inst{\ref{gepi}}
\and N.~Samaras\inst{\ref{brussels},\ref{prague}}
\and O.~Snaith\inst{\ref{gepi}}
\and C.~Soubiran\inst{\ref{bordeaux}}
\and O.~Vanel\inst{\ref{gepi}}
\and T.~Zwitter\inst{\ref{ljubljana}}
\and T.~Antoja\inst{\ref{barcelona}}
\and F.~Arenou\inst{\ref{gepi}}
\and C.~Babusiaux\inst{\ref{grenoble},\ref{gepi}}
\and N.~Brouillet\inst{\ref{bordeaux}}
\and E.~Caffau\inst{\ref{gepi}}
\and P.~Di~Matteo\inst{\ref{gepi}}
\and C.~Fabre\inst{\ref{atos},\ref{cnes}}
\and C.~Fabricius\inst{\ref{barcelona}}
\and F.~Frakgoudi\inst{\ref{garching}}
\and M.~Haywood\inst{\ref{gepi}}
\and H.E.~Huckle\inst{\ref{mssl}}
\and C. Hottier\inst{\ref{gepi}}
\and Y.~Lasne\inst{\ref{thales},\ref{cnes}}
\and N.~Leclerc\inst{\ref{gepi}}
\and A.~Mastrobuono-Battisti\inst{\ref{gepi},\ref{lund}}
\and F.~Royer\inst{\ref{gepi}}
\and D.~Teyssier\inst{\ref{esac}}
\and J.~Zorec\inst{\ref{brussels}}
\and F.~Crifo\inst{\ref{gepi}}
\and A.~Jean-Antoine Piccolo\inst{\ref{cnes}}
\and C.~Turon\inst{\ref{gepi}}
\and Y.~Viala\inst{\ref{gepi}}
}

   \institute{
GEPI, Observatoire de Paris, Universit\'e PSL, CNRS, 5 Place Jules Janssen, 92190 Meudon, France\relax\label{gepi}
\and
CNES Centre Spatial de Toulouse, 18 avenue Edouard Belin, 31401 Toulouse Cedex 9, France\relax\label{cnes}
\and 
Mullard Space Science Laboratory, University College London, Holmbury St Mary, Dorking, Surrey RH5 6NT, United Kingdom\relax\label{mssl}
\and
Universit\'{e} C\^{o}te d'Azur, Observatoire de la C\^{o}te d'Azur, CNRS, Lagrange UMR 7293, CS 34229, 06304, Nice Cedex 4, France\relax\label{oca}
\and 
Royal Observatory of Belgium, Ringlaan 3, 1180 Brussels, Belgium\relax\label{brussels}
\and
Observatoire Astronomique de Strasbourg, Universit\'{e} de Strasbourg, CNRS, UMR 7550, 11 rue de l'Universit\'{e}, 67000 Strasbourg, France\relax\label{strasbourg}
\and
Unidad de Astronom\'ia, Fac. Cs. B\'asicas, Universidad de Antofagasta, Avda. U. de Antofagasta 02800, Antofagasta, Chile\relax\label{chile}
\and 
CRAAG - Centre de Recherche en Astronomie, Astrophysique et G\'{e}ophysique, Route de l'Observatoire Bp 63 Bouzareah 16340, Alger, Alg\'erie\relax\label{alger}
\and 
Institut d'Astrophysique et de G\'{e}ophysique, Universit\'{e} de Li\`{e}ge, 19c, All\'{e}e du 6 Ao\^{u}t, B-4000 Li\`{e}ge, Belgium\relax\label{liege}
\and 
Universiteit Antwerpen, Onderzoeksgroep Toegepaste Wiskunde, Middelheimlaan 1, 2020 Antwerpen, Belgium\relax\label{antwerpen}
\and
F.R.S.-FNRS, Rue d'Egmont 5, 1000 Brussels, Belgium\relax\label{fnrs}
\and
Leibniz Institute for Astrophysics Potsdam (AIP), An der Sternwarte 16, 14482 Potsdam, Germany\relax\label{aip}
\and 
Laboratoire Univers et Particules de Montpellier, Universit\'{e} Montpellier, CNRS, Place Eug\`{e}ne Bataillon, CC72, 34095 Montpellier Cedex 05, France\relax\label{montpellier}
\and
Astronomical Institute, Faculty of Mathematics and Physics, Charles University, V Hole\v{s}ovi\v{c}k\'{a}ch 2, CZ-180 00 Praha 8, Czech Republic\relax\label{prague}
\and
Laboratoire d'astrophysique de Bordeaux, Universit\'{e} de Bordeaux, CNRS, B18N, all{\'e}e Geoffroy Saint-Hilaire, 33615 Pessac, France\relax\label{bordeaux}
\and
Faculty of Mathematics and Physics, University of Ljubljana, Jadranska ulica 19, 1000 Ljubljana, Slovenia\relax\label{ljubljana}
\and 
Institut de Ci\`{e}ncies del Cosmos, Universitat  de  Barcelona  (IEEC-UB), Mart\'{i} i  Franqu\`{e}s  1, 08028 Barcelona, Spain\relax\label{barcelona}
\and
Univ. Grenoble Alpes, CNRS, IPAG, 38000 Grenoble, France\relax\label{grenoble}
\and
ATOS for CNES Centre Spatial de Toulouse, 18 avenue Edouard Belin, 31401 Toulouse Cedex 9, France\relax\label{atos}
\and
Max Planck Institute for Extraterrestrial Physics, High Energy Group, Gie{\ss}enbachstra{\ss}e, 85741 Garching, Germany 
\relax\label{garching}
\and
Thales Services for CNES Centre Spatial de Toulouse, 18 avenue Edouard Belin, 31401 Toulouse Cedex 9, France\relax\label{thales}
\and
Department of Astronomy and Theoretical Physics, Lund Observatory, Box 43, SE--221 00, Lund, Sweden\relax\label{lund}
\and
Telespazio UK S.L. for European Space Agency (ESA), Camino bajo del Castillo, s/n, Urbanizacion Villafranca del Castillo, Villanueva de la Ca\~{n}ada, 28692 Madrid, Spain\relax\label{esac}
}

   \date{Received TBD; accepted TBD}

 
  \abstract
   {Gaia Data Release 3 (\gdrthree) contains the second release of the combined radial velocities. It is based on the spectra collected during the first 34 months of the nominal mission. The longer time baseline and the improvements of the pipeline made it possible to push the processing limit, from $\grvs = 12$ in \gdrtwo, to $\grvs = 14$~mag.}
   {In this article, we describe the new functionalities implemented for \gdrthree, the quality filters applied during processing and post-processing and the properties and performance of the published velocities.}
   {For \gdrthree, several functionalities were upgraded or added to the spectroscopic pipeline. The calibrations were improved, in order to better model the temporal evolution of the straylight and of the instrumental Point Spread Function (PSF). The overlapped spectra, which were mostly discarded in \gdrtwo, are now handled by a dedicated module. The hot star template mismatch, which did prevent their publication in \gdrtwo, has been largely mitigated, down to $\grvs = 12$~mag. The combined radial velocity of the stars brighter than or equal to $\grvs = 12$~mag is calculated in the same way as in \gdrtwo, that is, as the median of the epoch radial velocity time series. The combined radial velocity of the fainter stars is measured from the average of the cross-correlation functions.}
   {\gdrthree\ contains the combined radial velocities of \starPublished\ stars. With respect to \gdrtwo, the interval of temperature has been expanded from $\Teff \in [3600, 6750]$~K to $\Teff \in [3100, 14500]$~K for the bright stars ($\grvs \leq 12$~mag) and $[3100, 6750]$~K for the fainter stars. The radial velocities sample a significant part of the Milky Way: they reach a few kilo-parsecs beyond the Galactic centre in the disc and up to about 10-15~kpc vertically into the inner halo. The median formal precision of the velocities is of 1.3~\kms\ at $\grvs = 12$ and 6.4~\kms\ at $\grvs = 14$~mag. The velocity zero point exhibits a small systematic trend with magnitude starting around $\grvs = 11$~mag and reaching about 400~\ms at $\grvs = 14$~mag. A correction formula is provided, which can be applied to the published data. The \gdrthree\ velocity scale is in satisfactory agreement with APOGEE, GALAH, GES and RAVE, with systematic differences that mostly do not exceed a few hundreds \ms. The properties of the radial velocities are also illustrated with specific objects: open clusters, globular clusters as well as the Large Magellanic Cloud (LMC). For example, the precision of the data allows to map the line-of-sight rotational velocities of the globular cluster 47~Tuc and of the LMC.}
   {}


   \maketitle


\section{Introduction\label{sect:intro}}
The pioneering space astrometry mission, \hip, was equipped with a two-band photometer, but no spectrograph. Radial velocities were collected from the ground to complement \hip\ proper motions and provide the third component of the velocity vectors. A huge observing endeavour was in particular conducted with the CORAVEL spectrographs, which resulted in the publication of new radial velocity measurements for about 13~500 F-G dwarfs \citep[the \textit{Geneva-Copenhagen Survey},][]{Nordstrom2004} and more than 6~500 K-M giants \citep{Famaey2005}, representing in total more than 15\% of the \hip\ catalogue. Due to the very large number of sources targeted by \gaia, about 1.5 billion, or 4 orders of magnitude more than \hip, observing a similar fraction of its targets from the ground would have been extremely complex and expensive, if not impossible. This is why, early in the mission design, a spectrograph, the \RVS\ (RVS) was included in \gaia's payload \citep{Perryman2001, Katz2004, Cropper2018}.

The RVS exposure time is 13.2~s per transit, combining the 3 spectra acquired at each crossing of the RVS focal plane. The signal collected allows to derive single epoch radial velocities of G-K type stars down to $\grvs \sim 12-13$~mag. For fainter sources, it is necessary to accumulate and combine observations. The second Gaia data release (\gdrtwo) was based on the first 22 months of the mission \citep{GaiaCollaboration2018}. It was also the first one containing radial velocities and for this "premiere", the measurement and publication of the line-of-sight velocities were limited to $\grvs = 12$~mag \citep{Sartoretti2018, Katz2019}. The Early Third Gaia Data Release (\gdrethree) contained no new radial velocities, but for a convenient use of the database, it included a copy of the \gdrtwo\ radial velocities. The re-publication of the data also provided the opportunity to re-examine the reliability of the measurements (in particular the possible contaminations by bright neighbours) and led to the rejection of slightly fewer than 15~000 stars \citep{Seabroke2021}.

The Third Gaia Data Release (\gdrthree) is based on 34~months of data \citep{DR3-DPACP-185}, that is, adding 12 months to \gdrtwo\ and therefore a proportional number of observations per source. The number of spectroscopic observations in \gdrthree\ is further increased by a new functionality which "de-blends" the overlapping spectra \citep{DR3-DPACP-154}, while in \gdrtwo\ most of them were discarded. As a consequence, \gdrthree\ extends two magnitudes fainter than \gdrtwo\ and  contains \starPublished\ combined radial velocities. It is planned to publish the radial velocities of all sources down to the RVS limiting magnitude of $\grvs \sim 16$~mag, in the Fourth Gaia Data Release (\gdrfour) which should process 66 months of data.
 
The present paper is devoted to the description and validation of the radial velocities published in \gdrthree. Two companion papers discuss specific aspects of the \gdrthree\ radial velocities. \citet{DR3-DPACP-151} and \citet{DR3-DPACP-161} describe the dedicated methods implemented to derive the radial velocities of respectively the hot stars and the double-line spectroscopic binaries. Additionally, three articles present the other products of the spectroscopic pipeline: $\grvs$~magnitudes \citep{DR3-DPACP-155}, rotational broadening \citep{DR3-DPACP-149} and spectra \citep{DR3-DPACP-154}. Finally, \citet{DR3-DPACP-127} present a global overview of the validation and properties of the \gdrthree\ data, including the spectroscopic products.

The paper is structured as follows. Section~\ref{sect:rvs} recalls the main characteristics of the RVS. Section~\ref{sect:pipeline} provides an overview of the spectroscopic pipeline and summarises the new functionalities implemented to produce \gdrthree. Section~\ref{sect:filters} presents the filters applied on the input data, during the processing and during the validation phase. Sections~\ref{sect:catalogue} to \ref{sect:precision} describe the properties of the radial velocities as well as their performance: accuracy, formal uncertainties and median formal precision. Section~\ref{sect:HVS} discusses the specific case of the high velocity stars. Section~\ref{sect:specobj} illustrates the properties of the radial velocities with specific objects: open clusters, globular clusters and the Large Magellanic Cloud. Section~\ref{sect:varperf} presents the performance of the variability indices. We conclude in Sect.~\ref{sect:conclusions}.


\section{The Radial Velocity Spectrometer\label{sect:rvs}}
This section recalls briefly the main characteristics of the RVS. For a full description of the instrument, the reader is referred to \citet{Cropper2018}.

The RVS is a medium resolving power $R = \lambda / \Delta \lambda \sim 11\ 500$, near infrared $\lambda \in [845, 872]$~nm spectrograph. The dispersion is oriented parallel to the scan direction (hereafter referred to as the along-scan direction). The RVS is illuminated by the two \gaia\ telescopes, which are imaged on the same block of twelve CCDs (three along-scan times four across-scan), located at the end of the focal plane. As a consequence, each time a star is observed (hereafter referred to as a transit), three spectra are recorded, that is one per CCD along-scan. The exposure time is 4.42 s per CCD, or 13.2 s per transit. \gaia\ is continuously scanning the sky and on average the RVS records about 8 transits per star per year. It should however be noted that the individual number of transits is a strong function of the location on the celestial sphere (see Sect.~\ref{sect:distrib}) and that the effective number of transits used to derive the radial velocities is reduced by about 25\% by the combination of the dead-time (see Sect.~\ref{sect:ope}) and of the processing filters (see Sect.~\ref{sect:filters}).

The satellite is not operated in pointing mode, but is spinning at a constant speed, making a full rotation every 6 hours. The CCDs are therefore not recording static images, but are working in \textit{Time Delay Integration Mode} (TDI), that is the charges are continuously transferred from CCD line to CCD line (and continuously read, when they reach the read-out register), following the motion of the sources as they cross the fields of view.

The specificity of the RVS, which is pivotal in recording a very large number of spectra, is that it is an integral field spectrograph. It disperses all the light entering its two field-of-views. Rectangular windows are selected around the sources of interest, stored in the on-board memory and transferred during the contacts with the ground-stations. The windows were initially 1260 pixels along-scan by 10 pixels across-can, but the along-scan dimension was extended to 1296 pixels in June 2015 to improve the recording and correction of the background light. For the stars brighter than $\grvs = 7$~mag, the full two-dimensional windows (called window class~0 or WC0) are transmitted to the ground segment. For the fainter stars, the across-scan dimension is collapsed at CCD-level and one-dimensional windows (window class~1 or WC1) are transferred. The pixels outside the windows are "flushed", that is, they are clocked through the readout register, but are not read. The maximum number of windows that can be read and stored at any one time is limited to 72 per CCD (and fewer when stars brighter than $\grvs = 7$~mag are observed, as these stars use 10 resources each out of the 72 available).

Because the RVS is an integral field spectrograph, spectra of very close sources will overlap. For the stars fainter than $\grvs = 7$~mag, this generally results in the truncation of the windows containing the spectra. Over the area where the windows are in conflict, fewer than 10~pixels are assigned to each window. In many cases, this produces non-rectangular windows. A detailed description of the on-board window conflict resolution strategy is provided in \citet{DR3-DPACP-154}.


\section{Spectroscopic pipeline\label{sect:pipeline}}
The core of the \gdrthree\ spectroscopic pipeline is similar to the pipeline operated for \gdrtwo\ and described in \citet{Sartoretti2018}. However, many new functionalities were added to improve the quality of the measurements and produce the new spectroscopic products published in \gdrthree, that is, the radial velocities in the magnitude range $\grvs \in [12, 14]$~mag, the hot star radial velocities \citep{DR3-DPACP-151}, the rotational broadening \citep{DR3-DPACP-149}, the $\grvs$ magnitudes \citep{DR3-DPACP-155} and the spectra \citep{DR3-DPACP-154}. After a rapid overview of the \gdrthree\ spectroscopic pipeline (Sect.~\ref{sect:overview}) and a presentation of the different estimates of $\grvs$ (Sect.~\ref{sect:grvs}), the other parts of this section summarise the novelties implemented for the new release (with an emphasis on those which are relevant for the derivation of the radial velocities; Sect.~\ref{sect:straylight} to \ref{sect:varindices}) and presents the main aspects of the operations (Sect.~\ref{sect:ope}). The on-line documentation provides a more thorough description of the \gdrthree\ pipeline \citep{Sartoretti2022doc}.

\subsection{Overview\label{sect:overview}}
The \gdrthree\ spectroscopic pipeline is made of two technical and four scientific workflows.

The two technical workflows, "SourceInit" and "EpochInit", are in charge of preparing the data for the downstream workflows. In particular, "SourceInit" is tasked to gather, when available, the atmospheric parameters (effective temperature, surface gravity and metallicity) of the sources that will be processed. At this stage, the atmospheric parameters can have two origins. They may come from an internal compilation of ground-based catalogues or from the processing of the \gaia\ XP\footnote{The XP spectra are the low resolution spectra collected by \gaia's blue (BP: $\lambda \in [330, 680]$~nm) and red (RP: $\lambda \in [640, 1050]$~nm) spectro-photometers \citep{GaiaCollaboration2016}.} and RVS spectra performed by the General Stellar Parametrizer from photometry \citep[GSP-Phot;][]{DR3-DPACP-156} and from spectrocopy \citep[GSP-Spec;][]{DR3-DPACP-186}. Each set of atmospheric parameters is later used in the spectroscopic pipeline to select the synthetic spectrum used as template to derive the radial velocity (see Sect.~\ref{sect:stamta}).\\

The four scientific workflows perform the following tasks:
\begin{enumerate}
\item Straylight workflow (Sect.~\ref{sect:straylight}): It measures the background light level.
\item Calibration workflow (Sect.~\ref{sect:calibration}): It selects, reduces and cleans the sources suitable for the self-calibration of the RVS and calibrates the following characteristics of the instrument: the wavelength scale, the along-scan Line Spread Function (LSF) profile, the across-scan LSF profile, the across-scan location of the spectra and the $\grvs$ zero-point.
\item FullExtraction workflow (Sect.~\ref{sect:fullext}): It reduces and cleans the raw spectra to produce the calibrated spectra. This includes for each raw spectrum: the subtraction of the bias and bias non-uniformity \citep{Hambly2018}, the flagging of the saturated samples, the multiplication by the CCD gain, the subtraction of the dark current, the subtraction of the background light, the estimation of the flux lost outside of the window, the flagging of the spectra overlapping with cosmetic defects, the flagging of the spectra with a bright neighbour, the detection and removal of the cosmic rays, the collapse of the two-dimensional windows (assigned on-board to stars brighter than $\grvs = 7$~mag), the de-blending of the spectra contained in overlapping windows, the transformation from pixel to wavelength, the measurement of the $\grvs$ magnitude from RVS spectra, the division by the filter response, the normalisation of the fluxes, the detection and flagging of the emission lines and specific features (discontinuities or steep slope), and the determination of the atmospheric parameters (if unavailable in "SourceInit").
\item STAMTA workflow (Sect~\ref{sect:stamta}): It processes the calibrated spectra to extract the astrophysical information such as the radial velocity or the broadening velocity. It works source per source and is made of two parts: Single Transit Analysis (STA) which analyses the data transit per transit and Multiple Transit Analysis (MTA) which combines and uses together the transits recorded for each source.
\end{enumerate}

\subsection{$\grvs$ magnitudes\label{sect:grvs}}
Several estimates of the $\grvs$ magnitude are used in the spectroscopic pipelines and are referred to in the present paper: the on-board $\grvs$ ($\onboardgrvs$), the external $\grvs$ ($\extgrvs$) and the internal $\grvs$ (\verb|grvs_mag|).

The on-board $\grvs$ is estimated by Gaia's on-board software, prior to the observation by the RVS. It is either, based  on the signal in two specific samples in the red photometer, or derived from the $G$ magnitude. This magnitude estimate is used on board for the windowing and sampling decisions. It is transmitted to the ground together with the other data collected by the satellite.

The external $\grvs$ is derived from \gdrtwo\ $\gmag$ and $\grpmag$ magnitudes \citep{Evans2018, Riello2018}, when available, following equations (2) and (3) from \citet{GaiaCollaboration2018}. About 2.5\% of the stars processed by the spectroscopic pipeline did not possess a $\gmag$ and/or $\grpmag$ measurements. In that case, the on-board $\onboardgrvs$ is used instead. It should be noted that the definition of ($\extgrvs$) in \gdrthree\ differs from the one adopted in \gdrtwo\ \citep[see][]{Sartoretti2018, Katz2019}. The latter used the $\grvs$ from the \textit{Initial Gaia Source List} \citep[IGSL,][]{SmartNicastro2014}.

The internal $\grvs$ is the median of the $\grvs$ magnitudes evaluated at each transit, by measuring the flux contained in the RVS spectra in the wavelength range $[846, 870]$~nm. The internal magnitudes are published in \gdrthree\ in the field \verb|grvs_mag|, contained in the \verb|gaia_source| table \citep{DR3-DPACP-155}. 

The selection of the stars entering the different workflows is based on the $\extgrvs$ magnitudes. In particular, only stars with $\extgrvs \leq 14$~mag are processed by the "STAMTA" workflow, thus setting the limit for the derivation and publication of the radial velocities in \gdrthree. The internal $\grvs$ magnitudes are calculated within the "STAMTA" workflow, therefore too late to play the role assigned to the external magnitudes. However, the internal $\grvs$ is used to define the threshold between the two methods calculating the combined radial velocity (see Sect.~\ref{sect:stamta}). In Sect.~\ref{sect:accuracy} to \ref{sect:precision}, the accuracy, formal uncertainties and median formal precisions are presented as a function of \verb|grvs_mag|.

\subsection{Straylight workflow\label{sect:straylight}}
The \gaia\ commissioning revealed that the level of background light was higher than expected prior to launch. This is due to stray-light from the Sun and the brightest stars diffracted at the edge of the Sun-shield, entering the satellite through the apertures of the telescopes and reflected via different optical paths up to the focal plane.

In \gdrtwo, a single calibration of the background was performed, based on the data from the first 28~days of the mission. It produced a set of three maps (one for each group of four RVS CCDs aligned across-scan) of the background light as a function of the satellite rotation phase and of the across-scan location in the focal plane. These maps were used to correct the full 22 month dataset. This calibration alone could not account for the temporal variations of the background and led to under- and over-corrections of the fluxes. In \gdrthree, the background maps are produced every 30 hours of mission time, using 72 hours of data (that is, 21 hours before and 21 hours after).

The first beneficiary of the temporal sampling is the measurement of the internal $\grvs$ magnitudes, thus based on better corrected estimates of the stellar fluxes. The combined radial velocities also benefit from the improved background calibration. Indeed, in both \gdrtwo\ and \gdrthree\ pipelines, a spectrum from which an over-estimated sky background is subtracted and whose total flux becomes negative (which could occur for the faintest stars) is rejected from the processing. Therefore, the improvement of the precision of the calibration of the background light increases the number of spectra combined to derive the \gdrthree\ radial velocities.

\subsection{Calibration workflow\label{sect:calibration}}
In the \gdrtwo\ pipeline, two static models of the along-scan LSF were used. The first one was calibrated using reference spectra collected during the first 28 days of the mission and was applicable up until the first decontamination on 23 September 2014 (see~Sect.~\ref{sect:ope}). The second one was based on pre-launch measurements and was used for the remaining 20~months of data. In the \gdrthree\ pipeline, the along-scan LSF calibration has been upgraded to better monitor the temporal variations of the instrumental profile. As described in \citet{Sartoretti2018, Sartoretti2022doc} an along-scan LSF is represented mathematically as a weighted sum of fixed profiles, also called basis functions \citep{Lindegren2009}. The purpose of the along-scan LSF calibration is to constrain the weights. In \gdrthree, the 34 months are split in 10 periods, whose boundaries correspond to discontinuities in the RVS instrument calibrations. In each period, the weights are modelled with either constant or linear functions, depending on what is the most appropriate. The templates used in the "STAMTA" workflow, to measure the transit radial velocities (see Sect.~\ref{sect:stamta}), are produced by convolving synthetic spectra with the along-scan LSF. The upgrade of this calibration improves the match between the templates and the observed spectra and consequently improves the radial velocities.

In the \gdrthree\ pipeline, two new characteristics of the RVS instrument are also calibrated: the across-scan LSF profile and the across-scan location of the spectra in the focal plane. Both pieces of information are required to model and disentangle the fluxes of the spectra contained in overlapped windows (see Sect.~\ref{sect:fullext}).

\subsection{FullExtraction workflow\label{sect:fullext}}
It happens that RVS spectra overlap, either because the sources are very close on the sky or because two sources are observed each by a different telescope\footnote{We recall that Gaia observes simultaneously with two telescopes which are imaged on the same focal plane.}, but, by misfortune, their images end up close in the focal plane. If the sources are brighter than the RVS limiting magnitude $\grvs = 16.2$~mag and if the limit of window resources is not exceeded (see Sect.~\ref{sect:rvs}), a window is allocated to each. The windows then contain a mix of the fluxes of the sources. Moreover, if the sources are fainter than $\grvs = 7$~mag, the windows are usually truncated in the across-scan dimension. The truncated windows can be "thinner" but still rectangular if they share the same along-scan boundaries, but most of the time they present an L-shape (in the case of a conflict between two windows) or a more complex geometry (if more than two sources are involved). In the \gdrtwo\ pipeline, the only windows processed were the rectangular ones. All the others were discarded. In the \gdrthree\ pipeline, a new functionality, \textit{Deblending}, has been implemented to handle truncated windows. It uses the mix of fluxes contained in the windows together with the calibrations of the across-scan LSF profile and of the across-scan location of the spectra to reconstruct separately the spectrum of each source. The de-blended spectra represent a little more than 25\% (about 540 million out of 2 billion) of the spectra processed by the "STAMTA" workflow. Of the \starPublished\ stars with a combined radial velocity published in \gdrthree, about 96\% have at least one de-blended transit.The on-board window handling strategy and the \textit{Deblending} method are described in details in \citet{DR3-DPACP-154}.

\subsection{STAMTA workflow\label{sect:stamta}}
The "STAMTA" workflow is the last stage of the pipeline, which, in particular, measures the combined radial velocities. This section provides a summary of the different steps of the derivation of the radial velocities and presents two new \gdrthree\ functionalities, which respectively selects the templates of "hot" stars and derives the combined radial velocities of the "faint" stars. The "STAMTA" workflow is described in more details in the on-line documentation \citep{Sartoretti2022doc}.

\subsubsection{Single Transit Analysis\label{sect:sta}}
The "STAMTA" workflow is made of two parts. The first one, the \textit{Single Transit Analysis} (STA) processes the calibrated spectra per source and per transit to measure the epoch radial velocities and the epoch rotational broadenings \citep[the latter are provided by a new functionality implemented in \gdrthree,][]{DR3-DPACP-149}. The first task performed by STA, is to detect the double-line spectra, which are analysed by a dedicated method \citep{DR3-DPACP-161}. A radial velocity value is derived for each component. They are then used within the Gaia "non-single star" processing \citep[][]{DR3-DPACP-161}, but not published in \gdrthree. The bulk of the spectra are single-line. They are analysed by three different modules, which are all based on the comparison between a reference template shifted step by step in radial velocity and the three observed spectra collected per transit. Two of the modules quantify the match between the template and the observation with cross-correlation functions (one in direct space and the other in Fourier space), while the third relies on a chi-square minimum-distance method \citep{David2014, Sartoretti2018}. The epoch radial velocity is calculated as the median of the estimates provided by the three methods. The epoch radial velocities are not published in \gdrthree, except for a selected sample of slightly less than 2000 Cepheids and RR-Lyrae (contained in the table \verb|vari_epoch_radial_velocity|). The publication of the radial velocity time series is planned for \gdrfour.

The choice of the template does influence the quality of the epoch radial velocities. Indeed, a significant mismatch between the template and the observed spectra can degrade the precision and to a greater extent the accuracy. The template mismatch prevented the publication of the radial velocities of the hot stars ($\Teff \geq 7000$~K) in \gdrtwo. The first step to select a template is to assign atmospheric parameters (effective temperature, surface gravity and metallicity) to the source processed. The atmospheric parameters can have different origins and the first available in the following list is used: (1) an internal compilation of ground-based catalogues \citep[see][for the full list of catalogues]{Sartoretti2022doc}, (2) an early run of the \gaia\ \textit{General Stellar Parametrizer from Spectroscopy} \citep[GSP-spec,][]{DR3-DPACP-186}, (3) an early run of the \gaia\ \textit{General Stellar Parametrizer from Photometry} \citep[GSP-phot,][]{DR3-DPACP-156}. If none of these is available, the module \textit{determineAP} is called (as part of the "FullExtraction" workflow). It handles bright and faint stars differently. For stars with $\extgrvs \leq 12$~mag, it cross-correlates a limited set of 28 synthetic spectra (convolved by the along-scan LSF profile) with the observed spectra and returns the parameters of the synthetic spectrum producing the tallest peak. One set of parameters is derived per transit and the set with the most frequent occurrence is assigned to the star, in order to use the same template for all transits from the same source. Stars with $\extgrvs > 12$~mag are too noisy and they are assigned default parameters ($\Teff = 5500$~K, $\logg = 4.5$ and $\FeH = 0.0$~dex) by \textit{determineAP}. Once the atmospheric parameters are set, the synthetic spectrum with the closest parameters (following equation~11 of \citet{Sartoretti2018}), from a library of 6772 spectra, is chosen as template. Although, this selection is appropriate for "cool" stars, it suffers from the same template mismatch issue as in \gdrtwo\ for "hot" stars. For \gdrthree, a new module, \textit{reDetermineApHotStars} was implemented to refine the selection of their templates. It is called for stars with a first determination of the template $\Teff \geq 6500$~K (if the atmospheric parameters are from the compilation of catalogues) or 7000~K otherwise. It compares, source per source, the observed spectra to the full subset of the library of synthetic spectra with $\Teff \geq 6500$~K. The parameters of the best-matched synthetic spectrum (in the chi-square sense) supersede those of the first determination. The module \textit{reDetermineApHotStars} is called for stars brighter than $\extgrvs = 12$~mag, but not for fainter stars whose spectra are too noisy. The design and performance of \textit{reDetermineApHotStars} is the subject of a companion paper \citep{DR3-DPACP-151}. The parameters of the selected templates are stored in the fields: \verb|rv_template_teff|, \verb|rv_template_logg| and \verb|rv_template_fe_h| in the \verb|gaia_source| table. The origin of the atmospheric parameters used to choose the template are encoded in the field \verb|rv_atm_param_origin|.

\subsubsection{Multiple Transit Analysis\label{sect:mta}}
The second part of the "STAMTA" workflow, the \textit{Multiple Transit Analysis} (MTA), works source per source, using the time series to compute: the combined radial velocities, the \verb|grvs_mag| magnitudes \citep{DR3-DPACP-155}, the combined spectra \citep{DR3-DPACP-154} and variability indices. The combined radial velocity of the stars brighter than \verb|grvs_mag|~$= 12$~mag is calculated in the same way as in \gdrtwo\ \citep{Sartoretti2018, Katz2019}, that is as the median of the epoch radial velocities (expressed in the solar system barycentre reference frame). The calculation of the formal uncertainty on the measurement of the combined radial velocity is very similar too, with only a minor change in the constant term:
\begin{equation}
\epsilon_{V_{\rm R}} = [(\sqrt{{\pi}\over{2 N}} \sigma_{V_{\rm R}^{\rm t}})^2 + (0.113)^2]^{0.5}
\label{eq:uncertainty}
\end{equation}
where $N$ is the number of transits used to derive the median radial velocity, $\sigma_{V_{\rm R}^{\rm t
}} = \sqrt{\frac{1}{N - 1} \sum_{i=1}^{N} (V_{\rm R}^{\rm t}(i) - \overline{V_{\rm R}^{\rm t}}\ )^2}$ the standard deviation of the epoch radial velocity time series, $V_{\rm R}^{\rm t}(i)$ the $i^{th}$ transit radial velocity within the time series, and $\overline{V_{\rm R}^{\rm t}}$ the mean of the time series. The constant term, 0.113 \kms, is meant to take into account the wavelength calibration errors and similar sources of uncertainties. It has been estimated using a subset of the catalogue of radial velocity stable stars of \citet{Soubiran2018a}.

Beyond \verb|grvs_mag| $= 12$~mag, the epoch radial velocities are not considered reliable enough to derive the combined radial velocities. A new method has therefore been implemented in the \gdrthree\ pipeline to process the stars in the magnitude range \verb|grvs_mag| $\in [12, 14]$~mag. Source by source: (i) it loads the Fourier space cross-correlation functions derived in STA for all the transits of that source, (ii) it shifts them by the barycentric correction, (iii) it averages them and (iv) it measures the combined radial velocity as the location of the maximum of the averaged cross-correlation function. The formal uncertainty is derived from the sharpness of the summit of the cross-correlation function and implements the formula proposed by \citet{Zucker2003} in his Sect.~2.3, quadratically summed with the same constant term as in Eq.~\ref{eq:uncertainty}:
\begin{equation}
\epsilon_{V_\mathrm{R}} = \left[ \left( \frac{C^{2}-1}{\mathrm{rv\_nb\_transits} \times N \times C^{\prime\prime} \times C} \right) + (0.113)^2 \right]^{0.5}
\end{equation}
where $N$ is the number of pixels in the spectrum, $C^{\prime\prime}$ is the value of the second derivative of the cross-correlation function estimated at its summit and $C$ is the maximum value of the cross-correlation function.

In the magnitude range \verb|grvs_mag| $\in [11, 12]$~mag, just before the transition between the two methods, the radial velocity formal uncertainties show an extended tail \citep[see Fig.~4 in][]{DR3-DPACP-127}. It is sparsely populated and do not produce any significant discontinuity on the radial velocity precision, which is estimated using the median of the formal uncertainties (see Sect.~\ref{sect:precision}). However, the threshold at \verb|grvs_mag| $= 12$~mag might be slightly lowered in \gdrfour.

The combined radial velocities and their formal uncertainties are respectively stored in the fields \verb|radial_velocity| and \verb|radial_velocity_error| in the \verb|gaia_source| table. The method used is provided by the field \verb|rv_method_used|: \verb|1| for the median of the epoch radial velocities (hereafter referred to as the "Modelling" method) and \verb|2| for the combination of the cross-correlation functions (hereafter referred to as the "Robust" method).

\subsection{Variability indices\label{sect:varindices}}
Except for a very small number of variable stars, the epoch radial velocities are not published in \gdrthree, but will be in \gdrfour. However, \gdrthree\ contains two variability indices, based on the properties of the time series.

\verb|rv_chisq_pvalue| is the P-value for the constancy of the radial velocity time series. It ranges from zero for the low probability of constancy to one for the high probability (that is, the scatter of the epoch radial velocities is entirely due to measurement errors and not to intrinsic properties of the source).

\verb|rv_renormalised_gof| is calculated as the F2-value \citep[see vol.~1 sect.~2.1 of][]{1997ESASP1200.....E} of the renormalised unit-weight error of the radial velocity time series. For constant stars, it should follow a normal distribution of standard deviation equal to one. Variable stars will exhibit "large" values (compared to unity).

These two variability indices require precise measurements of the epoch velocities. For this reason, they are only calculated down to magnitude \verb|grvs_mag| $= 12$~mag. \verb|rv_chisq_pvalue| is also restricted to stars with \verb|rv_nb_transits|~$\geq 3$ and \verb|rv_renormalised_gof| to stars with \verb|grvs_mag|~$\geq 5.5$~mag and \verb|rv_template_teff| $< 14500$~K. Moreover, the reliability of the two indices increases with the number of measurements. It is therefore recommended to use them for stars that have been observed at least ten times (\verb|rv_nb_transits|~$\geq 10$). Finally, both indices rely on the comparison of the scatter of the epoch radial velocities to the epoch radial velocity formal uncertainties. The latter are less precise for \verb|rv_template_teff|~$< 3900$ and $> 8000$~K. The variability indices will be more reliable in between these two values.

The two indices are in good agreement and can be combined in a single variability criterion. The following conservative criterion can be used to identify variable stars: \verb|rv_nb_transits|~$\geq 10$ \& \verb|rv_template_teff|~$\in [3900, 8000]$ \& \verb|rv_chisq_pvalue|~$\leq 0.01$ \& \verb|rv_renormalised_gof|~$> 4$. The performances of this criterion are assessed in Sect.~\ref{sect:varperf}.

In addition to the two variability indices, \gdrthree\ also provides the measurement of the peak to peak amplitude of the radial velocity time series (after filtering of the outliers): \verb|rv_amplitude_robust|.

The three fields are stored in the table \verb|gaia_source|.

\subsection{Operations\label{sect:ope}}
\gdrthree\ is based on the data collected during the 34 first months of the nominal mission, that is from 25 July 2014 to 28 May 2017. During this period, several on-board operations were performed. The main ones were the three decontaminations, during which specific parts of the payload were reheated to sublimate the remnants of water ice (23 September 2014, 3 June 2015 and 22 August 2016) and the two refocus (24 October 2014 and 3 August 2015). These operations, as well as some other minor events, either prevented the acquisition or degraded the data (\url{https://www.cosmos.esa.int/web/gaia/dr3-data-gaps}). In total, 92.2\% of the 34 months were ingested in the spectroscopic pipeline, while the remaining 7.8\% were either unavailable or considered unfit for processing.

The data were processed by the 2500 cores of the Hadoop cluster of the \textit{Centre National d'\'Etudes Spatiales} (CNES) in Toulouse, France. The 2.8 billion spectra of the stars brighter than $\extgrvs = 14$~mag were ingested in the spectroscopic pipeline (plus approximately 12~billion spectra of fainter stars, potentially needed by the \textit{Deblending} module and used for that purpose only). Approximately 855 million were filtered-out within the "FullExt" workflow (see Sect.~\ref{sect:filtproc}) and the remaining two billion were processed by the "STAMTA" workflow. The full processing required about 3 million CPU hours or 120 days in real time. It occupied approximately 300~TB of disc space for the input, intermediate and output data. It produced 37~499~608 combined radial velocities as well as broadening velocities \citep{DR3-DPACP-149}, \verb|grvs_mag| magnitudes \citep{DR3-DPACP-155} and combined spectra \citep{DR3-DPACP-154}.


\section{Filtering non-reliable data\label{sect:filters}}
Quality checks were carried out throughout the production of the radial velocity catalogue: first on the input data, then during processing and finally on the catalogue itself during the validation phase. The spectra, transits or sources that failed the tests were discarded. The filters are described in the sections below.

\subsection{Input data}
The first series of filters was applied to the input data, before their transmission to the scientific processing:

\textit{Bad data intervals.} The transits recorded during the 7.8\% of the time considered as unfit for processing (see Sect.~\ref{sect:ope} and \url{https://www.cosmos.esa.int/web/gaia/dr3-data-gaps}) were removed.

\textit{Low quality astrometry.} The astrometry is used together with the calibration of the satellite attitude to derive, for each pixel of a spectrum, the coordinates of the source in the field of view (the along and across field angles $\eta$ and $\zeta$), when the pixel crosses the CCD central line (also called fiducial line). The field angles are used within the "FullExtraction" workflow to apply the wavelength calibration. Accurate radial velocities require an accurate wavelength calibration and therefore accurate astrometry. The sources were therefore removed in the following cases: (i) missing or not converged \textit{Astrometric Global Iterative Solution} \citep[AGIS; ][]{Lindegren2021a} coordinates, (ii) AGIS has flagged the source as duplicated, (iii) AGIS has used less than 5 observations, (iv) the source \verb|astrometric_excess_noise| is larger than 20~mas or, (v) the source \verb|astrometric_sigma5d_max| > 100~mas.

\textit{Low quality field angles.} The RVS dispersion is oriented along-scan. As a consequence, an error on the along-scan field angle $\eta$ propagates linearly to the wavelength zero-point. The transits with uncertainties on $\eta$ larger than 200~mas (corresponding to about 29~\kms) were removed.

\subsection{Processing\label{sect:filtproc}}
The following filters were applied in the course of the processing:

\textit{Large number of saturated pixels.} The spectra containing more than 40 saturated pixels were removed.

\textit{Negative total flux.} The spectra with a negative total flux (after subtraction of the bias, bias non-uniformity and background) were removed.

\textit{High background.} The background is subtracted from the spectra in the "FullExtraction" workflow. Unfortunately, there is no way to correct for the extra photon-noise it adds. The spectra with a background (i) higher than 100~$e^{-}~\text{pixel}^{-1}~\text{s}^{-1}$ or (ii) higher than 40~$e^{-}~\text{pixel}^{-1}~\text{s}^{-1}$ and an uncertainty on the background calibration larger than 0.4~$e^{-}~\text{pixel}^{-1}~\text{s}^{-1}$, were considered as significantly contaminated and were removed.

\textit{CCD cosmetic defects.} The spectra containing a CCD column affected by a cosmetic defect were removed.

\textit{Neighbour with no window.} It may happen that a star has a bright neighbour that does not have a window. In this case, the new \textit{Deblending} functionality \citep[Sect.~\ref{sect:fullext},][]{DR3-DPACP-154} is not triggered, because it requires the (truncated) windows of both sources. To mitigate the risk of contamination, an area corresponding to about 2500~pixels along-scan by 20~pixels across-scan was monitored around each spectrum. If at least a star with no RVS window and a magnitude smaller than that of the observation plus three (that is: $\extgrvs(neighbour) < \extgrvs(observation) + 3$) was found in this area, the spectrum was considered as significantly contaminated and was removed.

\textit{Large number of cosmic rays.} The spectra containing 100 or more pixels hit by cosmic rays were removed.

\textit{Non-deblended truncated windows.} There are two main reasons that can prevent the \textit{Deblending} module from working \citep[see][]{DR3-DPACP-154}: either one of the truncated windows is missing or the information contained in the windows is insufficient to reconstruct the spectra reliably (this can in particular happen when the separation between the sources is very small and the spectra are almost co-located). The spectra contained in truncated windows and which could not be successfully processed by the \textit{Deblending} module were removed.

\textit{Emission-line stars.} The spectroscopic pipeline includes a module to detect the emission-line stars. However, it does not have templates reproducing these stars. The resulting template mismatch can produce systematic errors on the radial velocity of several hundreds kilometre per second. As a consequence, the combined radial velocities of about 25~000 sources, with 40\% or more transits flagged as emission-lines, were not computed.

\textit{Corrupted spectra.} The module detecting the emission-line stars also performs sanity checks on the spectra. Those presenting instrumental or numerical artefacts (such as a sudden discontinuity) were removed.

Over the 2.8 billion spectra processed in \gdrthree, about 855 million were removed within the "FullExtraction" workflow and 2~billion were transmitted to the "STAMTA" workflow. The two main causes of rejection of the spectra were the inability to de-blend them ($\sim$540 million spectra) and the presence of a bright neighbour without window ($\sim$135 million spectra).

\subsection{Validation\label{sect:filtval}}
The "STAMTA" workflow produced 37~499~608 combined radial velocities. These were examined during the validation phase and a last set of filters was applied. About 3.7 million measurements were removed, while \starPublished\ combined radial velocities passed successfully the quality checks and are published in $\gdrthree$. The filters are presented in the sections below.

\subsubsection{Stellar and galactic properties\label{sect:stell}}
The \gdrthree\ spectroscopic pipeline processed all the spectra recorded during the first 34 months of the mission down to $\extgrvs \leq 14$~mag, without any selection on the stellar type or on colour indices. However, some objects require specific analyses which are not or not yet implemented in the spectroscopic pipeline or are implemented in other \gaia\ processing pipelines. Some selections were therefore performed a-posteriori.

\textit{Double-line spectroscopic binaries.} The combined radial velocity does not provide a good estimate of the systemic velocity of double-line spectroscopic binaries, and of course, it is not suitable for describing the variations in the velocities of the two components. Approximately 40~000 sources with 10\% or more transits flagged as double-line were considered SB2 candidates and their combined radial velocities were discarded. The epoch radial velocities of these sources were transmitted to the \gaia\ "non-single star" group, who processed them and publishes their orbital elements in \gdrthree\ \citep[][]{DR3-DPACP-161} 

\textit{Emission-line stars.} The pipeline did filter stars with 40\% or more transits flagged as emission-lines. In validation, this threshold was lowered to 30\%. A few more emission-line stars identified visually were also discarded. In total, another 3000 combined radial velocities were removed. It is planned to upgrade the library of templates with emission-line spectra in \gdrfour.

\textit{Cool stars.} The validations showed that the accuracy and precision of the cool stars down to \verb|rv_template_teff|~=~3100~K were of good enough quality to be published in \gdrthree. The combined radial velocities of about 243~000 stars cooler than 3100~K were removed.

\textit{Hot stars.} The module \textit{reDetermineApHotStars} allows to refine the selection of the template of the hot stars \citep[see~Sect.~\ref{sect:sta} and][]{DR3-DPACP-151}. It was used to process sources down $\extgrvs \leq 12$~mag, but not the fainter ones, because the transit spectra then become too noisy. As a consequence, the faint hot stars suffer from the same template mismatch issue as in \gdrtwo\ \citep{Katz2019}, which prevents their publication. The combined radial velocities of approximately 1.7 million stars with \verb|grvs_mag| $> 12$ and \verb|rv_template_teff|~$\geq 7000$~K were removed.

The module \textit{reDetermineApHotStars} allows to mitigate the template mismatch of the bright stars for \verb|rv_template_teff| in the range [6500, 14500]~K \citep{DR3-DPACP-151}. Beyond that temperature, significant radial velocity systematics persist, which will require further improvements in \gdrfour. For \gdrthree, the combined radial velocities of approximately 66~000 stars with \verb|grvs_mag| $\leq 12$ and \verb|rv_template_teff|~$> 14~500$~K were removed.

Several \gaia\ pipelines are interdependent, the output of one being the input of the other. As a result, they often cannot operate in parallel, but are time-shifted. Toward the end of the validation phase, the \gaia\ \textit{Extended Stellar Parameterizer –
Hot Stars} \citep[ESP-HS;][]{DR3-DPACP-157} completed its final run and own validation. The effective temperatures derived by ESP-HS were compared to the template temperatures. It revealed a group of 20~470 stars having cool templates (\verb|rv_template_teff|~$\leq 7000$~K), but effective temperatures estimated by ESP-HS larger than or equal to 7500~K. This is the range of temperature where a template mismatch can produce very significant radial velocity systematics. For safety, their combined radial velocities were removed.

\textit{Extra-galactic sources.} The spectroscopic pipeline processes all spectra assuming the source is a star. Although the limiting magnitude of $\extgrvs \leq 14$~mag is bright for galaxies and quasars, a few may have made their ways into the processing. No suitable treatment is provided for these objects and their combined radial velocities would be erroneous. Two groups of (potential) extra-galactic objects were identified and removed. The first is made of 112 sources with \verb|phot_bp_rp_excess_factor|~$> 13$ and \verb|bp_rp|~$\in ]1.2, 1.8[$~mag, which was interpreted as an extra-galactic signature. The second one is made of nine sources confirmed individually as galaxies or quasars. It should be noted that 4027 sources contained in the \verb|qso_candidates| table and 160 contained in the \verb|galaxy_candidates| table, also possess a combined radial velocity in the \gdrthree\ catalogue, because it has not been possible to confirm or deny their extra-galactic nature during the validation phase. These sources are briefly discussed in \citet{DR3-DPACP-101}. The classification of the sources, including the extra-galactic sources, is presented in \citet{DR3-DPACP-158} and in \citet{DR3-DPACP-165} for the variable sources.

\subsubsection{Noisy data\label{sect:noise}}
To reach $\extgrvs = 14$~mag, the spectroscopic pipeline did process very low signal to noise ratio (SNR) spectra, sometimes below $SNR = 1$ per pixel (after combining the information from all the transits). In the very low SNR regime, spurious secondary cross-correlation peaks can exceed the (true) main peak, leading to an erroneous radial velocity measurement. Several filters were applied to mitigate this problem.

\textit{Faint grvs\_mag magnitudes.} The stars processed in \gdrthree\ were selected on the basis of their external $\extgrvs$ magnitudes, because the internal \verb|grvs_mag| magnitudes, computed by the "STAMTA" workflow, were not yet available. During the validation phase, the two measurements were compared and showed an overall good agreement \citep{Sartoretti2018}. However, some stars exhibited an internal magnitude significantly fainter than 14. In order to achieve a relatively sharp cut-off in magnitudes (both external and internal) for \gdrthree, the stars fainter than \verb|grvs_mag| $= 14.5$~mag were excluded from the release. The stars whose transits have all been de-blended constitute a special case. Their mean flux (recorded in the wavelength range $[846, 870]$~nm) is measured, but their internal magnitude is not derived. An alternative, but equivalent, filter was applied to these stars. Those with a mean flux lower than 525~\esec were discarded. In total, approximately 21~000 combined radial velocities were removed.

\textit{Suspicious cross-correlation functions.} The noise in the spectrum propagates to the cross-correlation function and modifies its morphology. This information can be used to detect spurious radial velocities. The \gdrthree\ spectroscopic pipeline measures several characteristics of the combined cross-correlation function, such as: maximum value, full-width at half maximum, kurtosis, skewness, ratio and distance between the two highest peaks and a few more. These quantities are not published in the release, but were available for the validation phase. These parameters (complemented by the combined radial velocity formal uncertainty) were used, together with a set of APOGEE~DR14 data \citep{Abolfathi2018}, to "train" a model to identify the spurious combined radial velocities. First, the training set was divided between "valid" and "invalid" radial velocities, based on the residuals: \gdrthree\ minus APOGEE. Then, thresholds were defined for the different parameters, that allowed to separate the valid and invalid radial velocities. For \gdrthree, the model was trained in "the old-fashioned way" by visualising the data in various data spaces. For \gdrfour, machine learning approaches are explored. The combined radial velocities of the stars with \verb|grvs_mag| $\geq 12$~mag are measured from the combined cross-correlation function (see Sect.~\ref{sect:mta}). The model was therefore applied over the same magnitude range. About 2.9 million stars were flagged as invalid and their combined radial velocities removed. In a second step, still during the validation phase, the model was improved with two more quantities to characterise the cross-correlation function. The application of the upgraded model required the regeneration and characterisation off line of the combined cross-correlation functions. The off-line processing capabilities being more limited than those of the nominal processing, the application of the model was restricted to the stars with an absolute value of the combined radial velocity larger than or equal to 300~\kms\ (but with no selection on \verb|grvs_mag|). This off-line run led to the removal of about 34~500 additional combined radial velocities.

\textit{Low signal-to-noise ratios.} The filter on the morphology of the cross-correlation function removed 2.9 million stars and was effective in removing spurious measurements. However, at very low signal to noise ratio, the estimated rate of invalid radial velocities (estimated from the shape of the radial velocity distribution, see Sect.~\ref{sect:HVS}) was still high to very high: between 5 and 10\% for SNR~$\in [1.5, 2.0[$ and several tens of percent for SNR~$\in [0.0, 0.5]$. As a consequence, the combined radial velocities of the $\sim$270~000 sources with a SNR \verb|rv_expected_sig_to_noise|~$< 2$, were removed.

\textit{Large radial velocity formal uncertainties.} As described above, the combined radial velocity formal uncertainties were used together with the characteristics of the cross-correlation functions to identify spurious velocities in two data sets, that is respectively, the stars with \verb|grvs_mag| $\geq 12$~mag and with $|V_R| \geq 300$~\kms. The stars with a combined radial velocity formal uncertainty larger than 40~\kms\ were flagged as invalid. For consistency, the same filter was applied to the stars whose cross-correlation function had not been examined. This filter removed approximately 65~800 additional combined radial velocities. This criterion not only discards spurious measurements, but also large amplitude binaries and variables. Therefore, the epoch radial velocities of these stars were transmitted to the \gaia\ "non-single star" and "variability analysis" groups, for analysis and publication of the genuine large amplitude sources \citep{DR3-DPACP-178, DR3-DPACP-168}.

\textit{Large scatter of the epoch radial velocities.} The combined radial velocities of about 2~500 stars, brighter than \verb|grvs_mag| $\leq 12$~mag, and  showing a scatter of the epoch radial velocities larger than 577~\kms, were removed. This conservative threshold corresponds to the standard deviation of a uniform distribution over the range $[-1000, +1000]$~\kms.

\textit{Inconsistent results from different methods.} In the nominal run, the faint stars (\verb|grvs_mag| $\geq 12$~mag) combined radial velocities are derived using a single method, based on the analysis of the average Fourier space cross-correlation functions (see Sect.~\ref{sect:mta}). However, test and validation runs were conducted with complementary methods using respectively: (i) the average direct space cross-correlation function, (ii) the average chi-square minimum-distance function and (iii) the combined epoch spectra. The bulk of the stars showed a satisfactory agreement between the different estimators. However, the combined radial velocities of about 48~500 stars with a large discrepancy between the nominal run's method and the other methods, were removed.

\subsubsection{Contaminated data\label{sect:cont}}
Because the RVS is an integral field spectrograph, neighbouring stars can contaminate each others. If the overlapping spectra are close enough, they will be either handled by the \textit{Deblending} module (if all windows are available; see Sect.~\ref{sect:fullext}) or filtered-out (if some windows are missing; see Sect.~\ref{sect:filtproc}). However, if a source is bright enough, it can contaminate its neighbours at larger distances than the area considered by the nominal processing. Several filters were used to identify and remove the invalid radial velocities resulting from the contamination by these stars.

\begin{figure}[h!]
\centering
\includegraphics[width=0.99 \hsize]{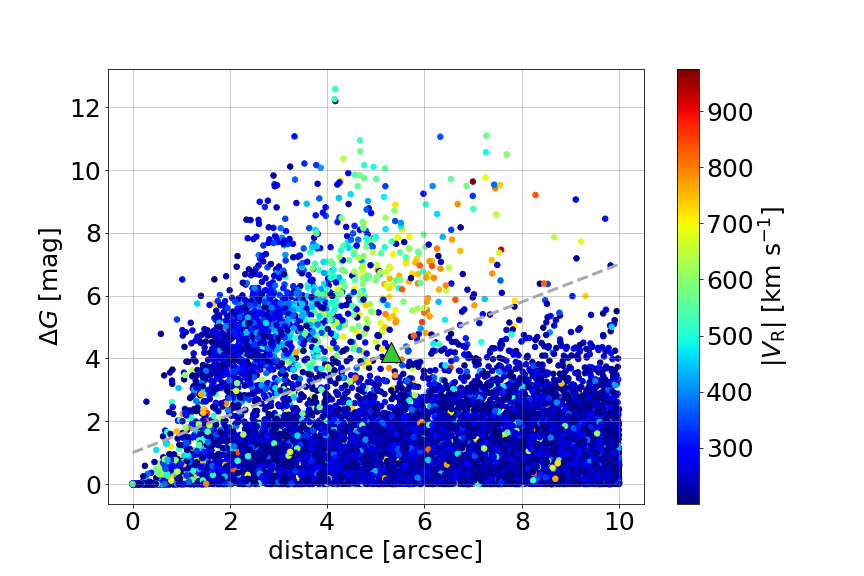}
\caption{Differences in G-band magnitude between the stars and their neighbours as a function of their separations on the sky, for the 17~697 sources with $|V_R| > 200$~\kms\ and one or more neighbours with a brighter G-band magnitude, within a radius of 10 arcsec. The colour code provides the absolute value of the star \gdrthree\ combined radial velocities. The dashed grey line materialises the separation between the spurious and valid velocity measurements. The green triangle shows the location of the star \gdrthree\ 4657994321463473792, which was kept in the release despite the fact that it is located slightly above the dividing line (see text).\label{fig:neigh}}
\end{figure}

\textit{Stars with bright neighbours.} When, in a given window, the flux of a neighbour exceeds the flux of the object targetted, the epoch radial velocity measured is usually the one of the contaminant. Moreover, this radial velocity is shifted by the offset between the wavelength reference frames of the two spectra (contaminated and contaminant), which, to first order, is proportional to the along-scan distance between the sources: $\Delta \mathrm{frames} \sim 145 \times d_{AL}$ (where the offset $\Delta \mathrm {frames}$ is expressed in \kms\ and the along scan distance $d_{AL}$ is in arcsec). This can produce very large spurious values, as in the case of the star \gdrtwo\ \verb|5932173855446728064| \citep{Boubert2019}. For along scan separations larger than about 6.9~arcsec, the offset between the wavelength reference frames goes out of the radial velocity measurement interval of $[-1000, +1000]$~\kms\ and the epoch radial velocity becomes pseudo-random. The contamination happens at transit level. Yet, if the sources are close enough, or if they are observed repeatedly in the same configuration (that is, similar along scan separations) or if the contaminating flux largely exceeds the one of the observation, the combined radial velocity can be spoiled. The stars with large radial velocities (positive or negative) are few in the Milky Way and even a small number of outliers can produce a significant contamination rate. Therefore, a specific attention was paid to the stars with $|V_\mathrm{R}| > 200$~\kms. There are 17~697 stars with a \gdrthree\ combined radial velocity in this range and one or more neighbours with a brighter G-band magnitude, within a radius of 10 arcsec. Figure~\ref{fig:neigh} shows the difference in G-band magnitude between these stars and their neighbours ($\Delta G = G_\mathrm{star} - G_\mathrm{neigh}$) as a function of their separations on the sky. The colour code provides the absolute value of the star \gdrthree\ combined radial velocity. The stars mainly fall into two groups: one with a small difference in magnitude and the other with a larger difference. In the top group, the absolute value of the combined radial velocity correlates with the distance between the star and its neighbour. As described above, this is precisely the behaviour expected in the case where the flux recorded in a window is dominated by a contaminant. The lower group does not show significant signs of correlation, which indicates that the measured velocities are not spoiled by contaminants. The separation between the two groups was modelled by a straight line ($\Delta G = 0.6 \times d + 1.0$) adjusted "manually". It is represented by the dashed grey line in Fig.~\ref{fig:neigh}. Of the 1624 stars located on or above this limit, 1623 were removed. The star \gdrthree\ \verb|4657994321463473792|, which seats almost on the dividing line, and has very consistent velocities in APOGEE DR17 \citep{Abdurrouf2022} and in \gdrthree\ (respectively 273.5~\kms\ and 273.3~\kms), was kept. It should also be noted that this filter does not handle possible contamination by bright sources contained in the conjugate field of view. Another 25 contaminated stars were identified with an alternative criterion and removed. Finally, using the binary catalogue of \citet{ElBadry2021}, the \gaia\ catalogue validation group also identified 57 stars separated by less than 1.6~arcsec from their companion and showing differences in radial velocities larger than 500~\kms \citep{DR3-DPACP-127}. Their combined radial velocities were removed.

\textit{Contaminated spectra.} The RVS bandpass filter has relatively steep wings. They are trimmed within the "FullExt" workflow, which keeps only the wavelength range $[846, 870]$~nm. However, on rare occasions, the combined spectra produced by the "STAMTA" workflow \citep{DR3-DPACP-154} still exhibit a wing. This is caused by a bright contaminant, which is shifted in the dispersion direction (that is, along-scan) with respect to the contaminated spectra. On the detector, the wing of the bright contaminant overlaps with the area corresponding to the wavelength range $[846, 870]$~nm of the contaminated spectra. The combined radial velocities of 18 stars presenting this feature, were removed.

\subsubsection{Other filters}
Two more filters were applied during the validation phase.

\textit{Inconsistent location of the calcium triplet lines.} The combined radial velocities of 172 stars that were inconsistent with the location of the calcium triplet lines in their combined spectrum, were removed. The reason for this inconsistency is not yet understood.

\textit{High Velocity Stars (HVS) with inconsistent bibliographic measurements.} The combined radial velocities of 14~HVS (defined as stars with $|V_R| >= 500$~\kms), which were strongly discrepant with the literature were removed.


\begin{figure*}[tbp]
\centering
\includegraphics[width=0.99 \hsize]{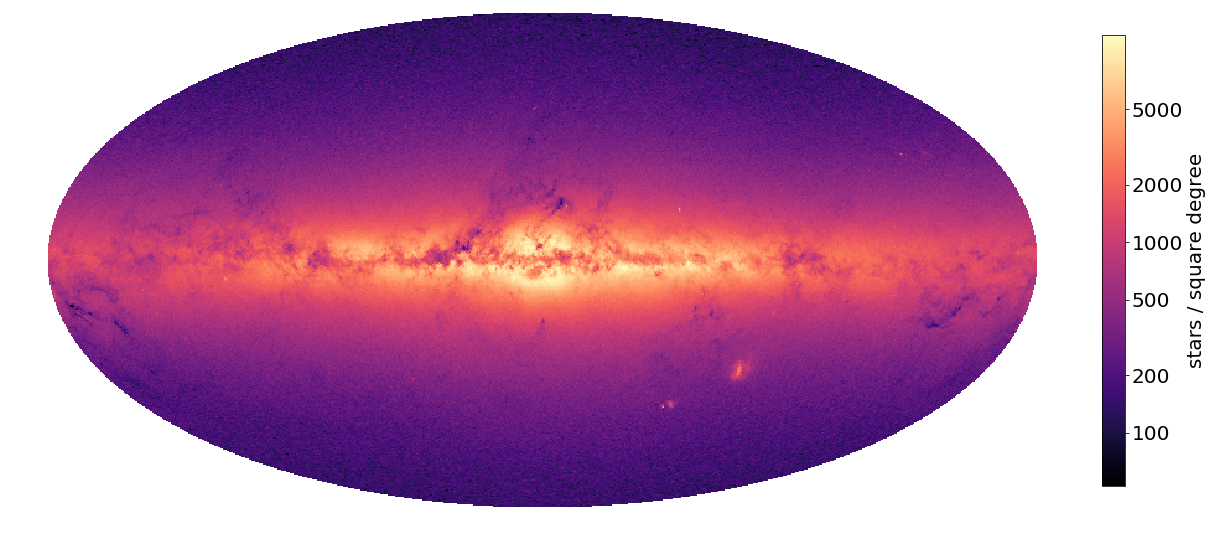}
\includegraphics[width=0.49 \hsize]{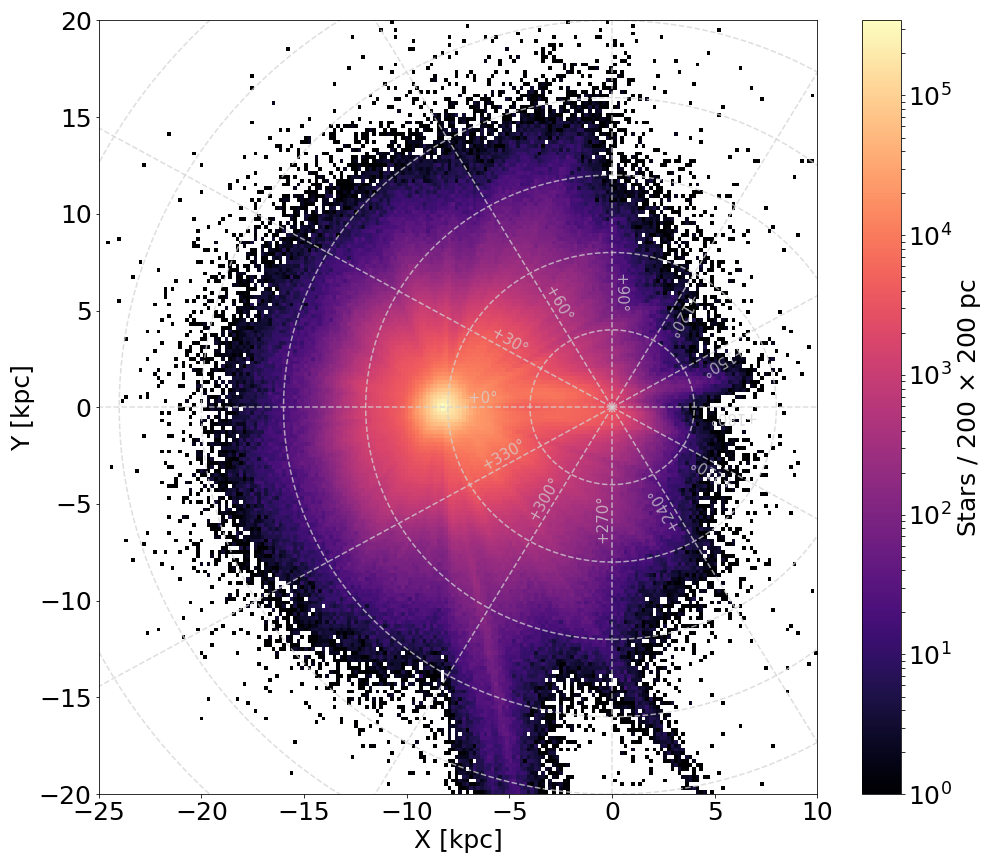}
\includegraphics[width=0.49 \hsize]{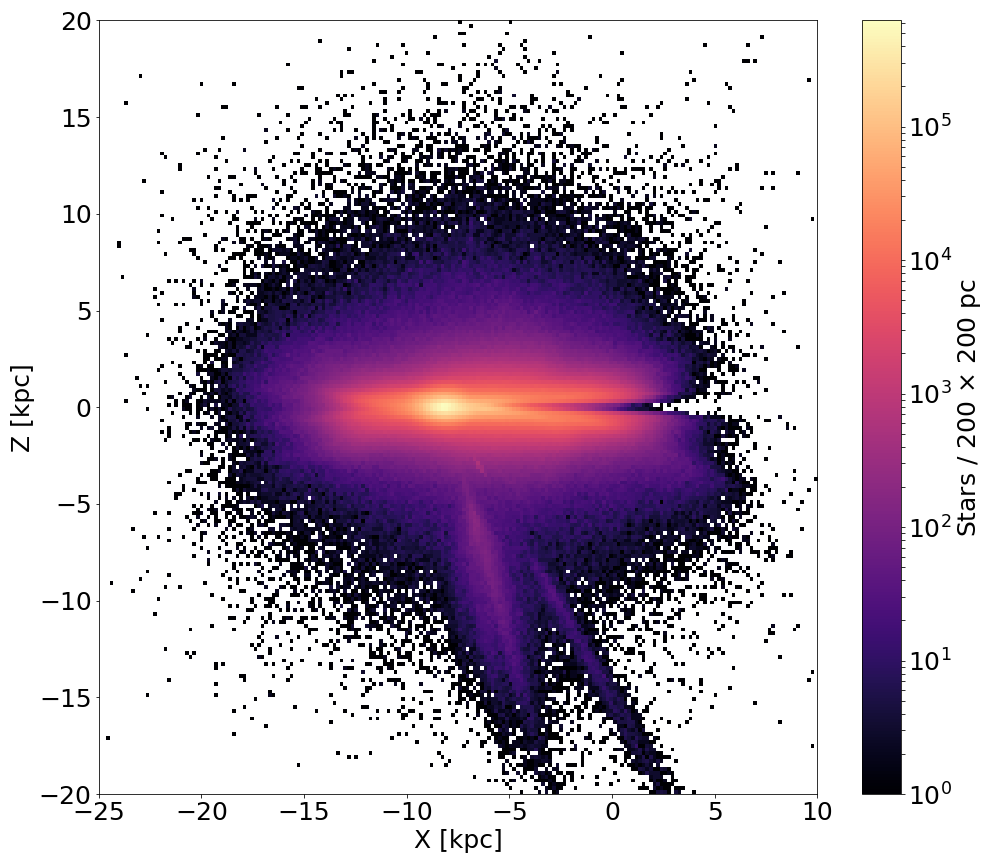}
\caption{\textit{Top:} Sky distribution of the \starPublished\ stars having a combined radial velocity published in \gdrthree. The image uses a Mollweide projection in Galactic coordinates $(l, b)$. The Galactic centre is at the centre of the figure and the Galactic longitudes increase to the left. The sampling of the map is approximately 0.2~square degree (healpix level~7). \textit{Bottom left:} "Face-on view" of a sub-sample of 29.4 million stars (see text) in Galacto-centric Cartesian coordinates (X, Y). The Galactic centre is located at $X=Y=0$~kpc and the Sun at $X=-8.277$~kpc \citep{Gravity2022} and $Y = 0$~kpc. In this image, the Milky Way is "seen" from the Galactic North Pole and rotates clockwise. The sampling is $200 \times 200$~parsec$^2$. \textit{Bottom right:} "Edge-on view" of the same sub-sample in Galacto-centric Cartesian coordinates (X, Z). The Galactic centre is located at $X=Y=0$~kpc and the Sun at $X=-8.277$~kpc and $Z=20.8$~pc \citep{BennettBovy2019}. The Z axis is oriented positive towards the Galactic North Pole. The sampling is $200 \times 200$~parsec$^2$.\label{fig:distrib}}
\end{figure*}

\begin{figure*}[tbp]
\centering
\includegraphics[width=0.49 \hsize]{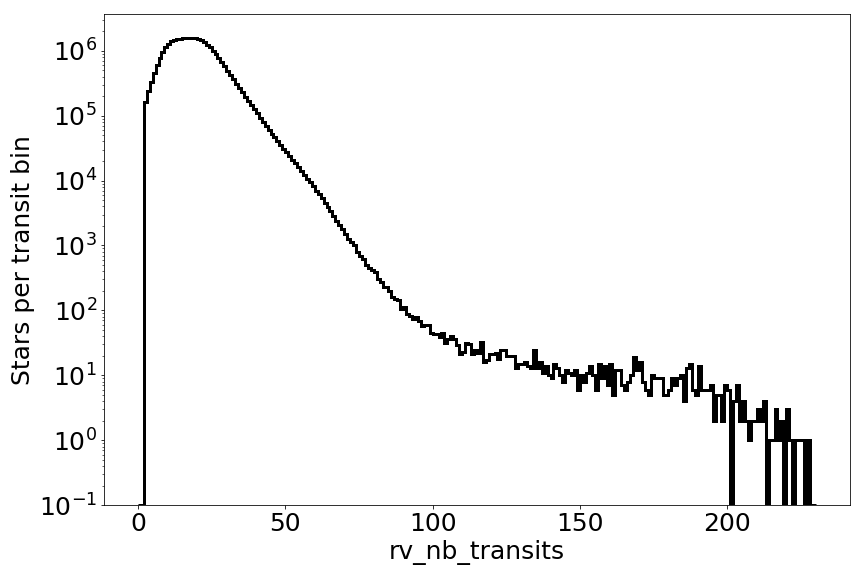}
\includegraphics[width=0.49 \hsize]{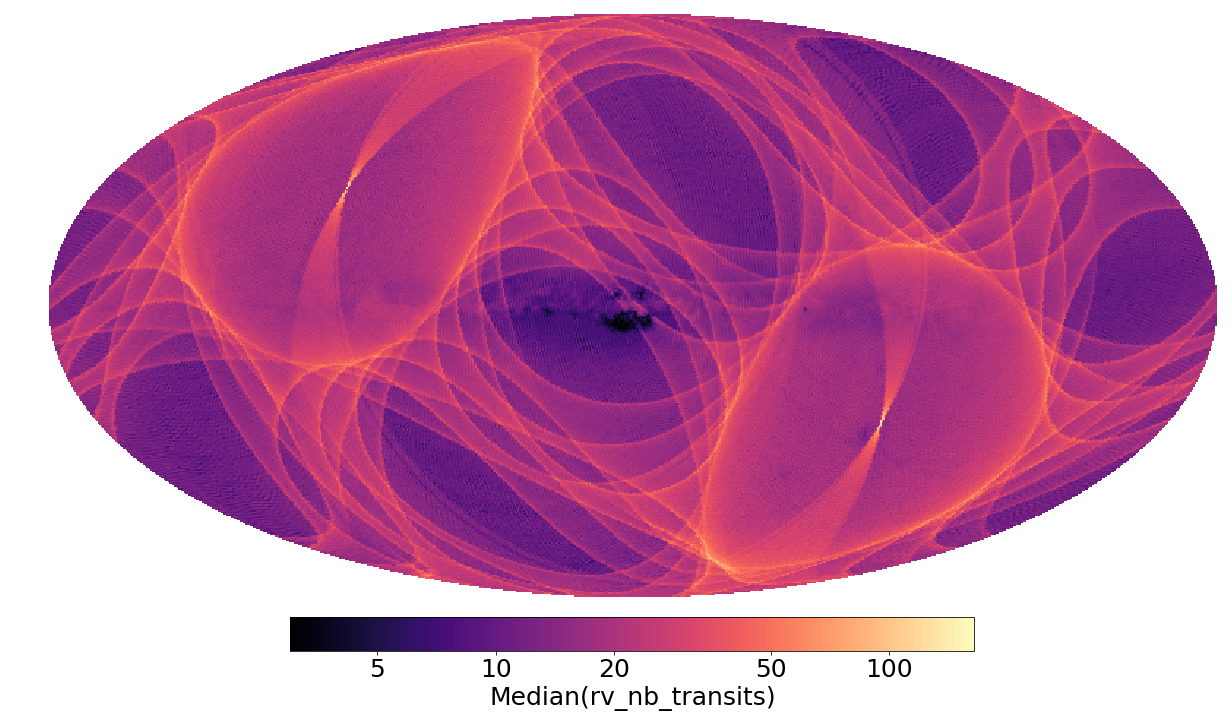}
\includegraphics[width=0.49 \hsize]{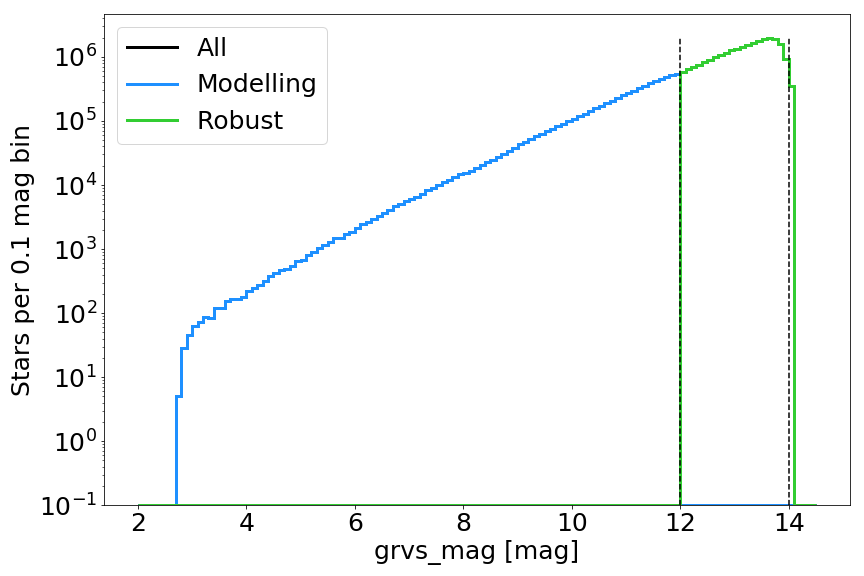}
\includegraphics[width=0.49 \hsize]{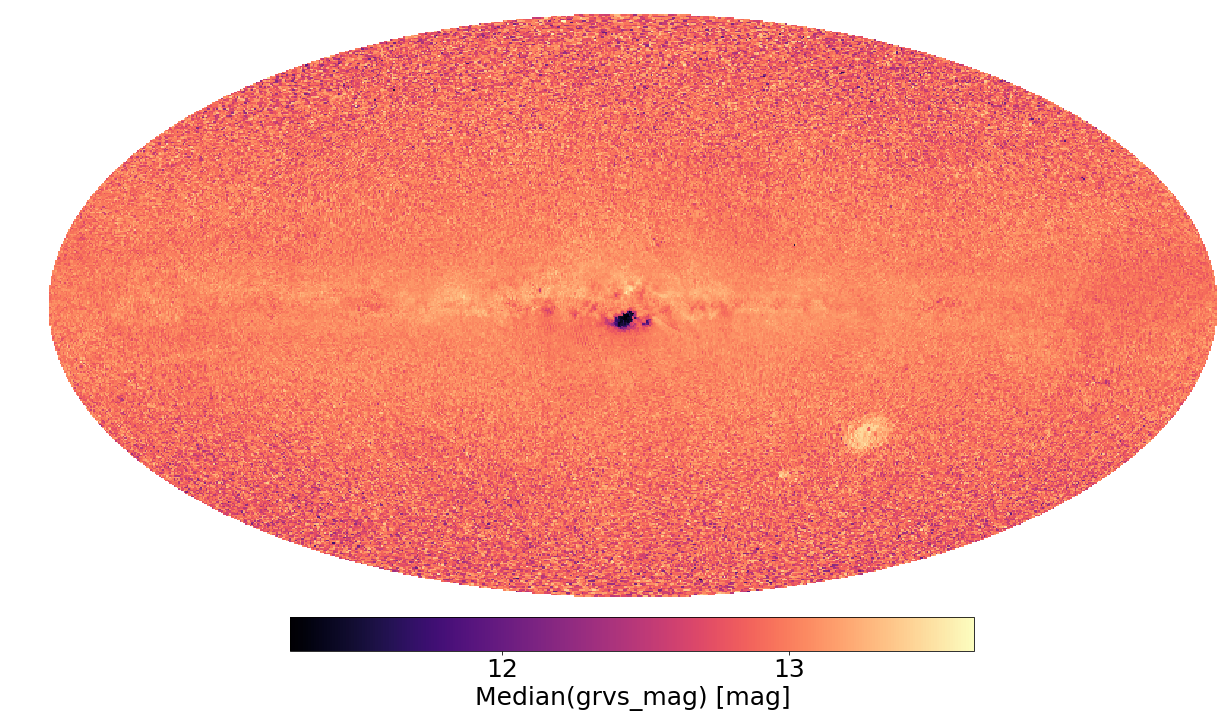}
\includegraphics[width=0.49 \hsize]{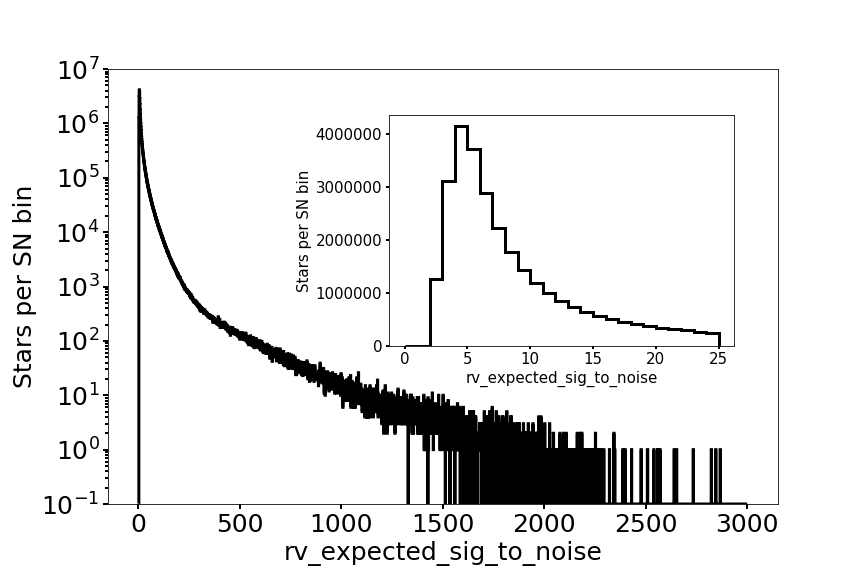}
\includegraphics[width=0.49 \hsize]{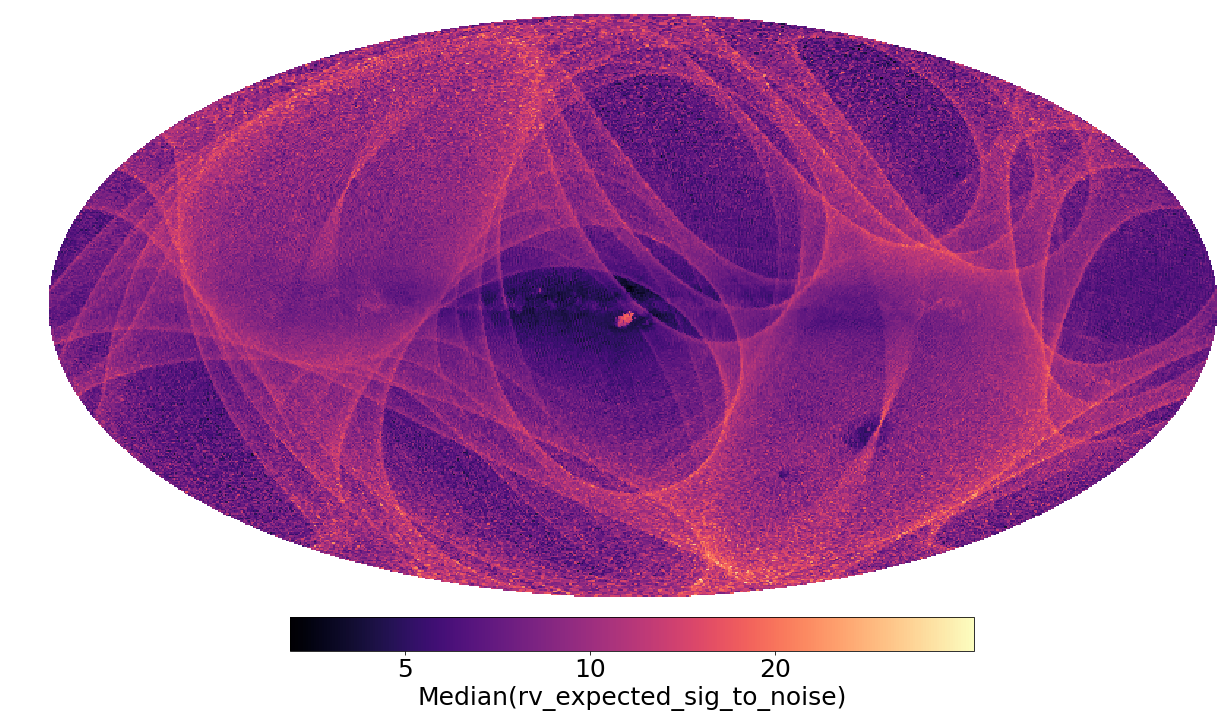}
\caption{Distributions of the numbers of transits (rv\_nb\_transits, top), grvs\_mag magnitudes (middle) and signal-to-noise ratios (rv\_expected\_sig\_to\_noise, bottom) respectively, of the stars whose combined radial velocity is published in \gdrthree. In the left column, the distributions are presented in the form of histograms. The right column presents the sky maps of the median values of these three quantities. The maps use Mollweide projections in Galactic coordinates $(l, b)$. The Galactic centre is at the centre of the images and the Galactic longitudes increase to the left. The sampling of the maps is approximately 0.2~square degree (healpix level~7).\label{fig:transits}}
\end{figure*}

\section{\gdrthree\ radial velocity catalogue\label{sect:catalogue}}
\subsection{Content overview}
\gdrthree\ contains the combined radial velocities of \starPublished\ stars. The dataset extends two magnitudes fainter than in \gdrtwo\ \citep{Katz2019} and \gdrethree\ \citep{Seabroke2021}, that is down to $\extgrvs = 14$~mag. The interval of temperature has also been expanded from \verb|rv_template_teff|~$\in [3600, 6750]$~K in \gdrtwo\ to \verb|rv_template_teff|~$\in [3100, 14500]$~K for the bright stars ($\extgrvs \leq 12$~mag) and $[3100, 6750]$~K for the fainter stars. The database also contains complementary information, such as the radial velocity formal uncertainty, the parameters of the template and variability indices. The list of products from the spectroscopic pipeline published in \gdrthree\ is provided in Tables~\ref{tab:dr3Fields} and \ref{tab:dr3Fields2}.

Figure~\ref{fig:distrib} (top) shows the distribution on the sky of the stars having a combined radial velocity in \gdrthree. As expected, the bulk of the dataset belongs to the Milky Way. However, the \gdrthree\ radial velocity catalogue also includes stars from satellite galaxies, such as the Large Magellanic Cloud (LMC, $l = 280.5^\circ$, $b = -32.9^\circ$) and the Small Magellanic Cloud (SMC, $l = 302.8^\circ$, $b = -44.3^\circ$, see Sect.~\ref{sect:LMC}) which are well visible in the lower right corner of the image, or Sagittarius (less visible in this image, but see Sect.~\ref{sect:velocities}). Moreover, a significant fraction of the known Globular Clusters (GC) are also part of \gdrthree, with a number of velocity measurements per GC ranging from a few to more than a thousand (see Sect.~\ref{sect:GC}).

In \gdrtwo, the $\extgrvs$ magnitudes used to select the stars mainly came from the \textit{Initial Gaia Source List} \citep[IGSL][]{SmartNicastro2014}. They were calculated from several catalogues, mostly the GSC2.3 \citep{Lasker2008}, Tycho-2 \citep{Hog2000} and SDSS \citep{Strauss2002}. The heterogeneity of the input photometry produced small offsets in the resulting $\extgrvs$ and some spatial fluctuations in the GSC2.3 $B_J$ and $R_F$ magnitudes (in particular in the overlapping areas of the photometric plates) were propagated to the calculated magnitudes. This produced small spatial variations in the limiting magnitude of the \gdrtwo\ spectroscopic catalogue and the footprints of the GSC2.3 and SDSS were visible in the radial velocity sky map \citep{Katz2019, Rybizki2021}. In \gdrthree, the $\extgrvs$ magnitudes were mainly calculated from $G$ and $G_\mathrm{RP}$ magnitudes (Sect.~\ref{sect:grvs}). As can be seen from the top part of Fig.~\ref{fig:distrib}, the \gdrthree\ sky map exhibits significantly fewer selection artefacts.

Figure~\ref{fig:distrib} (bottom) presents the distributions of the stars in the Galactocentric Cartesian XY (left) and XZ planes (right). The maps show a sub-sample of 29.4 million stars, after removing the sources with an astrometric renormalized unit weight error \verb|ruwe|~$>= 1.4$ or with \verb|duplicated_source| set to \verb|True|. The coordinates are calculated with the Bayesian photo-geometric distances from \citet{BailerJones2021}. As can be seen in the "face-on" view, the radial velocity catalogue now samples a significant part of the Milky Way disk. It extends a few kilo-parsecs beyond the Galactic centre and therefore allows to probe the kinematic of the bar \citep{DR3-DPACP-75}. Vertically, the catalogue encompasses the thin and thick discs and part of the inner halo, up to about 10-15~kpc. The two elongated features at negative Y and negative Z are LMC and SMC stars, whose distances are under-estimated \citep[see the discussion in][]{BailerJones2021}.

\subsection{Transits, magnitudes and signal to noise\label{sect:distrib}}
Figure~\ref{fig:transits} shows the distributions of the numbers of transits, \verb|rv_nb_transits| (top), \verb|grvs_mag| magnitudes (middle) and signal-to-noise ratios, \verb|rv_expected_sig_to_noise| (bottom), respectively, of the stars whose combined radial velocity is published in \gdrthree. In the left column, the distributions are presented in the form of histograms. The right column presents the sky maps of the median values of these three quantities.

The number of transits ranges from 2 (minimum number for the pipeline to calculate the combined radial velocity) to 227, with a median number of 18. As shown by the sky map, over most of the celestial sphere, the number of transits is driven by the satellite scan law. In particular, during the first month of the nominal mission, the satellite was operated in \textit{Ecliptic Pole Scanning Law} (EPSL) mode. During that period, the northern and southern Ecliptic poles were monitored by each telescope every 6 hours, collecting large numbers of observations of the stars in these two areas. The number of transits drops sharply in the densest regions (the less extincted ones) of the bulge and disc. Three effects combine to produce this rapid decrease. First, from approximately 36~000 stars per square degree, the maximum number of windows that can be allocated is reached (see Sect.~\ref{sect:rvs}), so that some spectra are not recorded. Then, it happens that when the satellite scans the Galactic plane for several days, the on-board memory is saturated and the lowest priority data (usually the faintest stars) must be erased before they can be transmitted to the ground. Finally, in these dense regions, a very large fraction of the spectra are blended and if the de-blending cannot be performed \citep[either because a window is missing or because the sources are too close; see][]{DR3-DPACP-154}, the radial velocity is not derived. It should be noted that these effects happen at transit level and therefore they primarily impact the number of transits, but can also affect the completeness.

The \verb|grvs_mag| magnitude is not calculated for the stars with all transits de-blended. The number of stars with a \verb|grvs_mag| magnitude published in \gdrthree\ is 32~276~087 \citep{DR3-DPACP-155}, slightly lower than the number of radial velocities. The bright limit of the velocity catalogue is \verb|grvs_mag| $= 2.76$~mag and results from the saturation of the RVS spectra. The magnitude of the faintest star is \verb|grvs_mag| $= 14.1$~mag. The transition between the two methods used to calculate the combined radial velocities ("Modelling" in blue and "Robust" in green) occurs at \verb|grvs_mag| $= 12$~mag. The sky map of the median \verb|grvs_mag| magnitudes is relatively smooth and homogeneous over much of the celestial sphere. It is fainter in the direction of the Magellanic Clouds because of the excess of intrinsically fainter sources belonging to the two galaxies. Conversely, the median magnitude is brighter in the densest regions of the bulge, because the windowing scheme and the on-board storage-deletion schemes both favour the bright stars.

The lower limit of the signal-to-noise ratios was set at 2 during the validation phase, in order to minimise the contamination by spurious radial velocity values (see Sect.~\ref{sect:noise}). The largest signal to noise ratio is 2868 (for the star \gdrthree\ \verb|413828761929696256|, \verb|grvs_mag| $= 3.82$~mag and 62 transits). The distribution is very asymmetric and the large values are very few. The median of the distribution is 7.8. The map shows that, over most of the sky, the median signal to noise ratio is driven by the scanning law (through the number of transits). However, the median signal to noise ratios are lower in the direction of the Magellanic Clouds and higher in the densest regions of the bulge, because the median \verb|grvs_mag| magnitudes are respectively fainter and brighter there.

\begin{figure}[h!]
\centering
\includegraphics[width=0.99 \hsize]{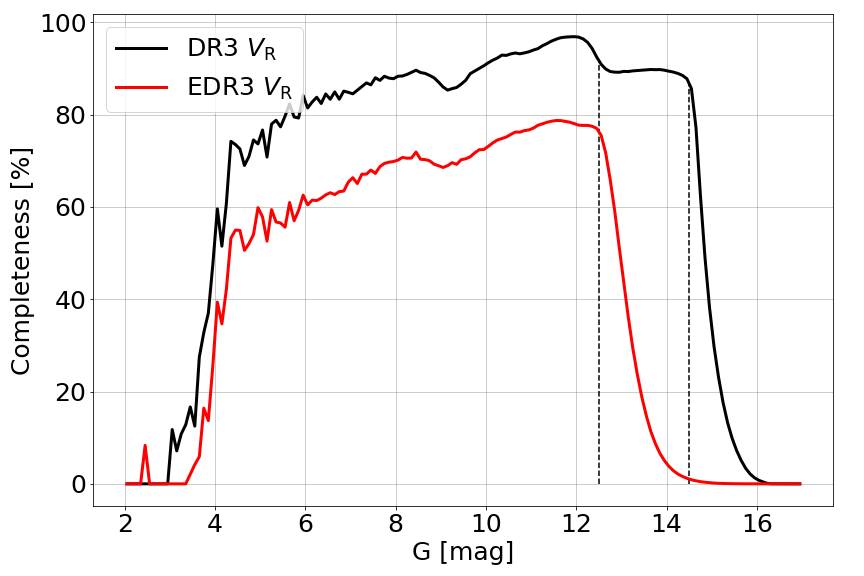}
\includegraphics[width=0.99 \hsize]{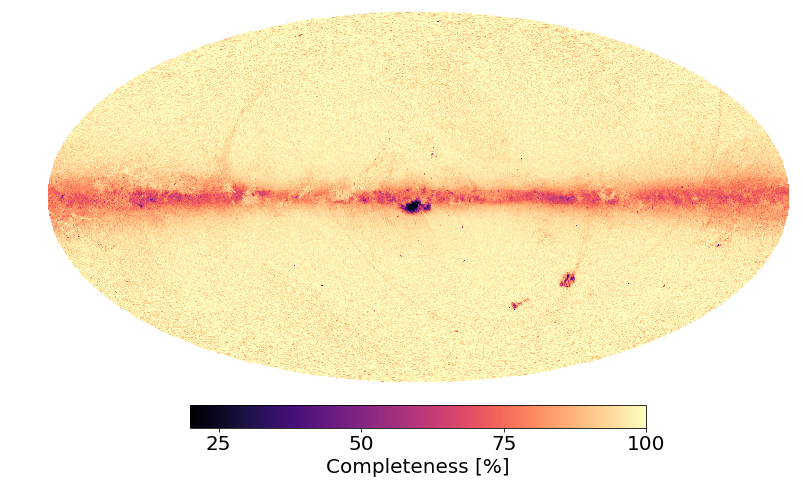}
\caption{\textit{Top:} Completeness of the \gdrthree\ (black curve) and \gdrethree\ (red curve) combined radial velocity catalogues, as a function of $G$~magnitude. Two dashed vertical lines are drawn at G = 12.5 and G = 14.5~mag, corresponding to the respective drops in completeness of each of the catalogues. \textit{Bottom:} Sky map of the completeness with respect to the \gdrthree\ catalogue (restricted to the stars with $G \leq 14.5$~mag). The image uses a Mollweide projection in Galactic coordinates $(l, b)$. The Galactic centre is at the centre of the image and the Galactic longitudes increase to the left. The sampling of the map is approximately 0.2 square degree (healpix level 7). \label{fig:completeness}}
\end{figure}

\subsection{Completeness\label{sect:completeness}}
In this section, the completeness is estimated with respect to the \gaia\ catalogue as a whole, that is, as the ratio of the number of stars having a radial velocity to the total number of stars in the catalogue, either per bin of magnitude (Fig.~\ref{fig:completeness}, top) or per healpix cell (Fig.~\ref{fig:completeness}, bottom).

Figure~\ref{fig:completeness} (top) compares the completeness of the \gdrthree\ (black curve) and \gdrethree\ (red curve) radial velocity catalogues, as a function of $G$~magnitude. The longer time baseline and the upgrade of the \gdrthree\ pipeline have improved the completeness with respect to the previous release. Besides the offset between the two curves, they exhibit similar trends with an increase in completeness over the entire range, from $G \sim 4$ to $\sim$12~mag. The sharp drop below $G \sim 4$~mag is caused by the saturation of the spectra. On the other side, the \gdrthree\ completeness drops around $G = 14.5$~mag, that is two magnitudes fainter than \gdrethree\ (consistent with the extension of the processing limit from $\extgrvs = 12$ to $\extgrvs = 14$). It should be noted that the majority of the stars have fainter $G$ than $\grvs$ magnitudes (an un-reddened G2V star has $G - \grvs \sim 0.65$~mag), which explains that the drop happens at $G \sim 14.5$~mag. The moderate decrease around $G \sim 8.5$~mag is produced by the transition from 2D windows (WC0) to 1D windows (WC1). The former are never truncated, while the latter can. The 1D windows therefore suffer from a higher rejection rate, corresponding to the cases where the spectra cannot be de-blended (see Sect.~\ref{sect:filtproc}). Between $G \sim 12$ and 12.5~mag, the completeness decreases a little and then stays roughly flat up to 14.5~mag. The decrease is the result of filters applied in validation, in particular the removal of (i) the hot stars with \verb|grvs_mag| $> 12$~mag and \verb|rv_template_teff|~$\geq 7000$~K (see Sect.~\ref{sect:stell}) and of (ii) the stars with spurious cross-correlation functions (see Sect.~\ref{sect:noise}).

Figure~\ref{fig:completeness} (bottom) shows the sky map of the completeness with respect to the \gdrthree\ catalogue (restricted to the stars with $G \leq 14.5$~mag). The completeness varies inversely with stellar density. It is maximum outside the Galactic plane, decreases in the disc and drops in the least extinct, and therefore densest, areas of the bulge.

\subsection{Radial velocities\label{sect:velocities}}
\begin{figure}[h!]
\centering
\includegraphics[width=0.99 \hsize]{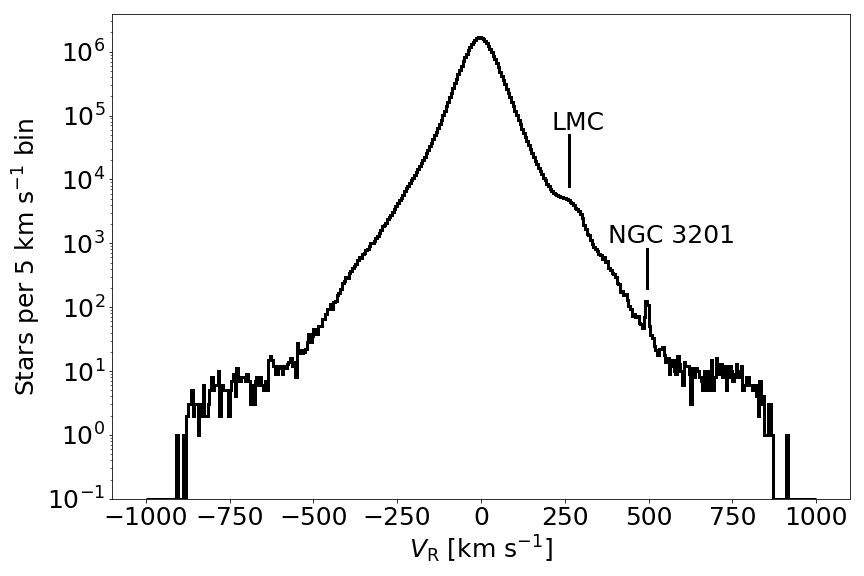}
\includegraphics[width=0.99 \hsize]{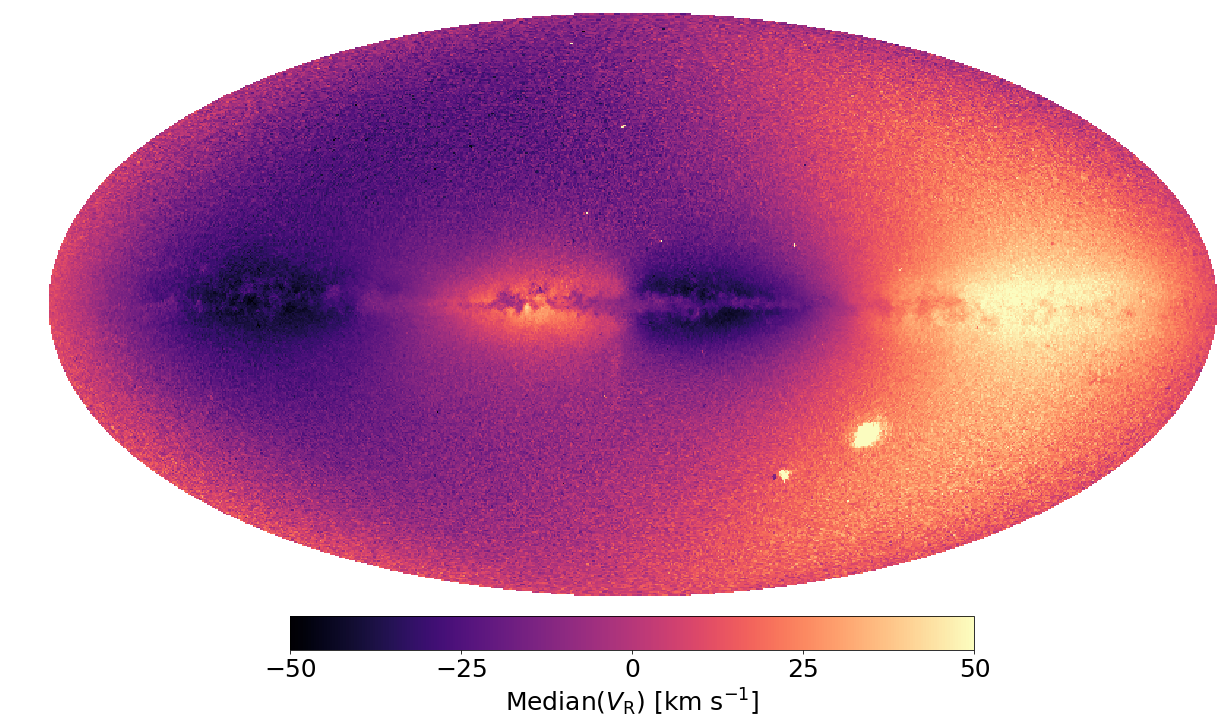}
\caption{\textit{Top:} Histogram of the \starPublished\ combined radial velocities. The sampling is 5~\kms\ per bin. \textit{Bottom:} Sky map of the median radial velocity. The image uses a Mollweide projection in Galactic coordinates $(l, b)$. The Galactic centre is at the centre of the image and the Galactic longitudes increase to the left. The sampling of the maps is approximately 0.2 square degree (healpix level 7). The limits of the colour bar have been set at $-$50 and $+$50~\kms, in order to highlight the median velocity variations in the disc.\label{fig:velocities}}
\end{figure}

Figure~\ref{fig:velocities} (top) shows the histogram of the \starPublished\ combined radial velocities. The bulk of the sources belongs to the Milky Way disc and are centred around $\sim$0~\kms. Two specific objects stand out in the right wing of the distribution. On the one hand, the LMC appears as a bump around $+$262~\kms \citep{McConnachie2012}. On the other hand, the high velocity Globular Cluster NGC~3201 is visible as a spike around $+$494~\kms \citep{Baumgardt2019}. The outer part of the wings of the distribution, beyond $|V_R| \gtrsim 500-600$~\kms, does flatten. This indicates that the proportion of spurious measurements remains significant in the very high velocity regime. The specific case of the \textit{High-Velocity Stars} (HVS) is discussed in Sect.~\ref{sect:HVS}.

Figure~\ref{fig:velocities} (bottom) shows the map of the median radial velocities as a function of Galactic longitudes and latitudes. The rotation of the Galactic disc (projected along the lines of sight) is manifested by the alternation of bright areas (with positive median velocities) and dark areas (with negative median velocities). Several objects whose radial velocities differ from those of their close environment (and which are sufficiently populated to weigh on the median value), are visible by contrast. In addition to the LMC and SMC appearing as bright spots in the lower right corner of the image, the Sagittarius dwarf galaxy is visible as a faint quasi-vertical stripe below the Galactic centre. Several Globular Clusters and compact objects appear as tiny dots in the image, such as 47~Tuc, the dark dot \citep[$V_\mathrm{R} = -17$~\kms,][]{Baumgardt2019} just left of the SMC, or Omega Cen a bright dot \citep[$V_\mathrm{R} = 234$~\kms,][]{Baumgardt2019} at $l \sim 309^\circ$ and $b \sim +15^\circ$.


\section{Accuracy\label{sect:accuracy}}
Formally, the accuracy is the systematic difference between the measured values and the true values. However, the true values, which would allow to assess the systematics in absolute terms, are not known. As a consequence, in this section, we assess the systematic differences in relative terms, by comparison to five ground-based catalogues: APOGEE~DR17 \citep{Abdurrouf2022}, GALAH~DR3 \citep{Buder2021, Zwitter2021}, GAIA-ESO Survey (GES) DR3 \citep{Gilmore2012, Randich2013}\footnote{These references describe the GES survey, but not the DR3 specifically.}, LAMOST~DR7 \citep{Zhao2012, Deng2012, Luo2015}\footnote{These references describe the LAMOST survey and pipeline, but not the DR7 specifically.} and RAVE~DR6 \citep{Steinmetz2020_1, Steinmetz2020_2}. Quality filters are applied before comparison. They are described in Appendix~\ref{app:GB}.

The "relative" accuracy is estimated as the median of the radial velocity residuals: \gdrthree\ minus ground-based catalogue. The boundaries of the 68.3\% confidence interval on this estimator are calculated as:
\begin{equation}
\epsilon_{\rm acc}^{\rm low} = \sqrt{{\pi}\over{2 N}} (\tilde{V}_{\rm R}^{\rm res} - Per(V_{\rm R}^{\rm res}, 15.85))
\end{equation}
\begin{equation}
\epsilon_{\rm acc}^{\rm upp} = \sqrt{{\pi}\over{2 N}} (Per(V_{\rm R}^{\rm res}, 84.15) -\tilde{V}_{\rm R}^{\rm res}),
\end{equation}
where $N$, $\tilde{V}_{\rm R}^{\rm res}$, $Per(V_{\rm R}^{\rm res}, 15.85)$, $Per(V_{\rm R}^{\rm res}, 84.15)$ are respectively, the number of elements, the median and the 15.85$^{\rm th}$ and 84.15$^{\rm th}$ percentiles of the distribution of radial velocity residuals.

\begin{figure*}[tbp]
\centering
\includegraphics[width=0.49 \hsize]{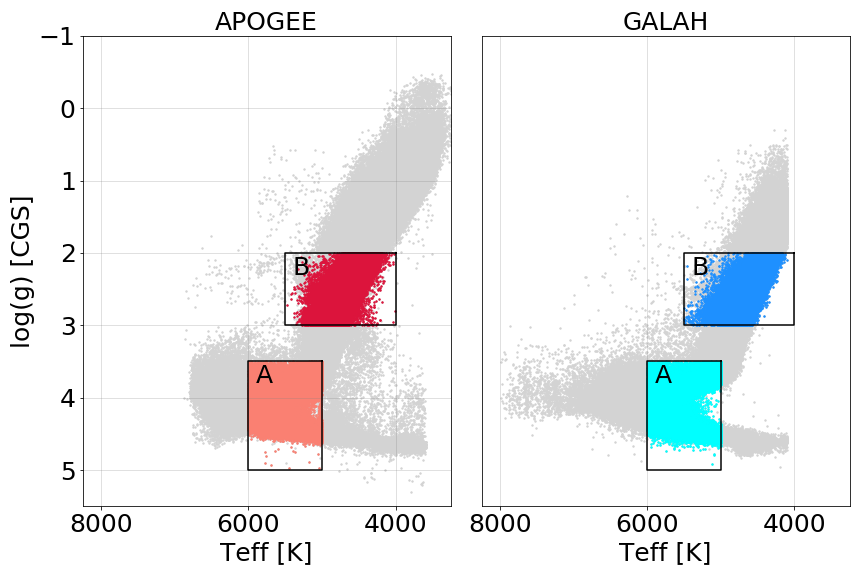}
\includegraphics[width=0.49 \hsize]{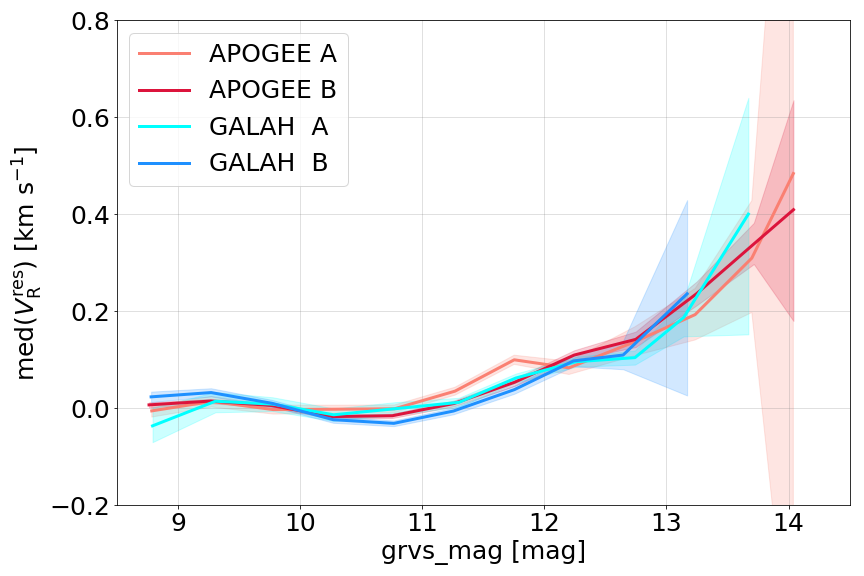}
\includegraphics[width=0.49 \hsize]{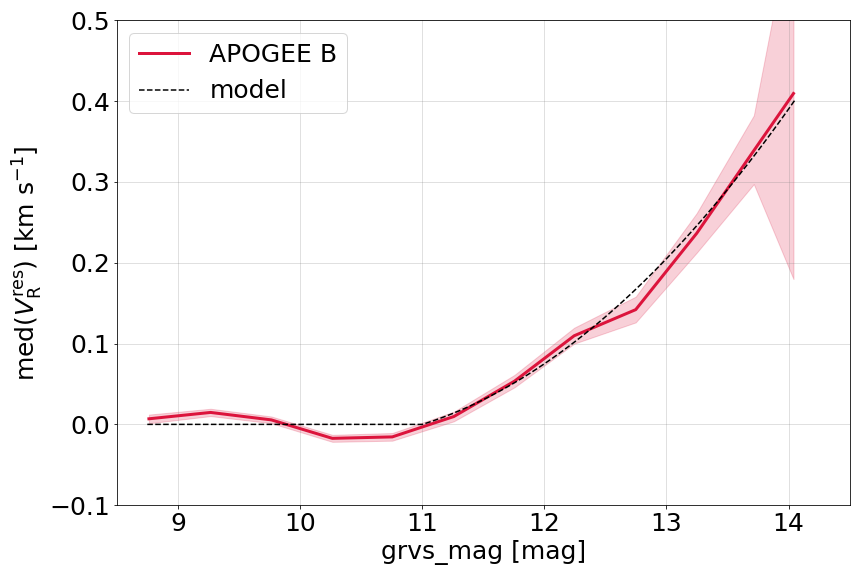}
\includegraphics[width=0.49 \hsize]{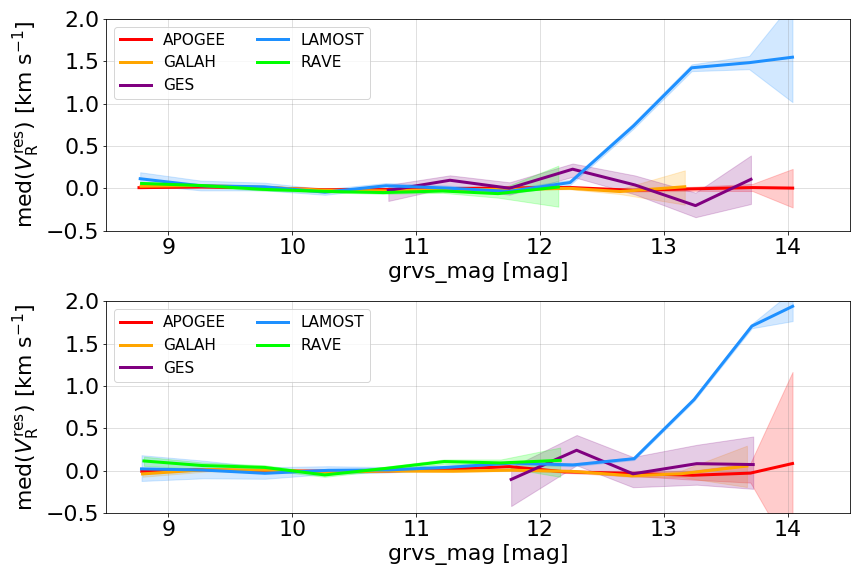}
\caption{\textit{Top left:} Kiel diagrams of the APOGEE and GALAH catalogues. The groups of stars selected to assess the magnitude trend are shown in salmon/cyan (group A: dwarfs/turn-off) and red/blue (group B: giants). \textit{Top right:} Median residuals of the radial velocity (Gaia minus catalogue) as a function of grvs\_mag magnitude, for the four samples. \textit{Bottom left:} Radial velocity correction model (dashed line) over-plotted to the median residuals of the radial velocity of the APOGEE B sample (red curve). \textit{Bottom right:} Median residuals of the radial velocity as a function of grvs\_mag magnitude, after applying the radial velocity correction, with respect to the five reference catalogues: APOGEE, GALAH, GES, LAMOST and RAVE. The upper panel shows the group B stars (giants) and the lower panel, the group A stars (dwarfs/turn-off). All curves are shifted vertically (see text) and therefore show relative trends. The 68.3\% confidence interval on the measurements of the medians is represented as shaded areas.\label{fig:accgrvs}}
\end{figure*}

\begin{figure*}[tbp]
\centering
\includegraphics[width=0.49 \hsize]{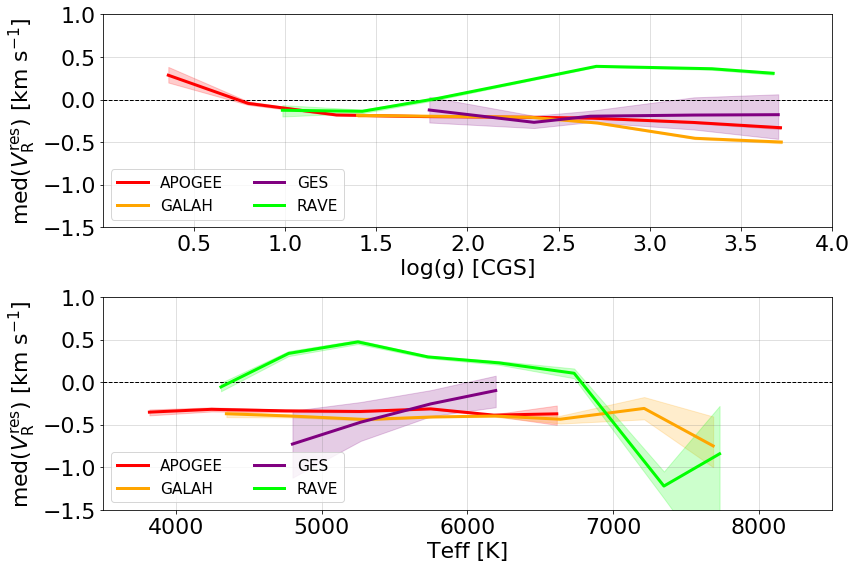}
\includegraphics[width=0.49 \hsize]{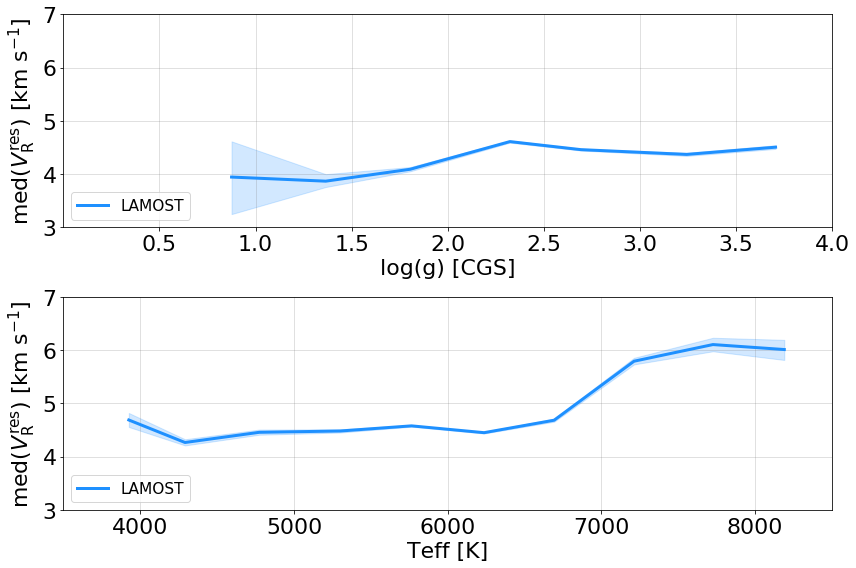}
\includegraphics[width=0.49 \hsize]{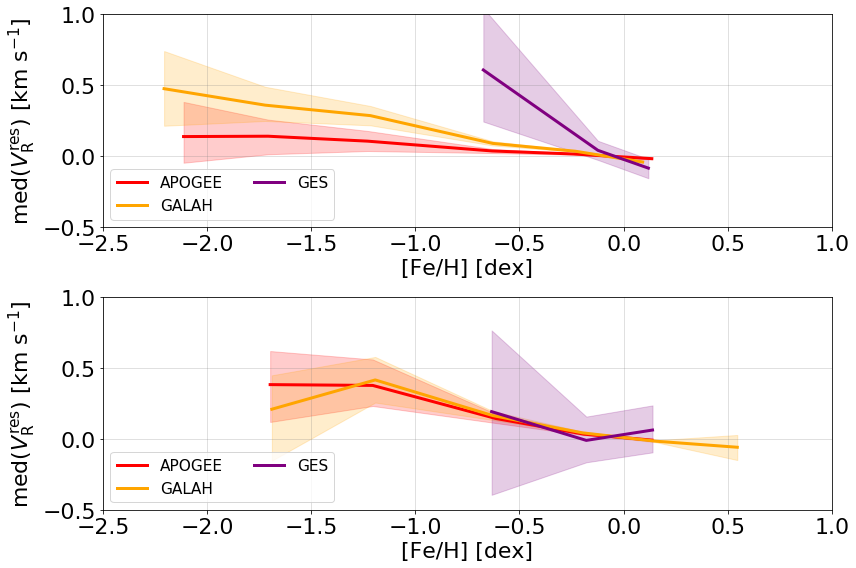}
\includegraphics[width=0.49 \hsize]{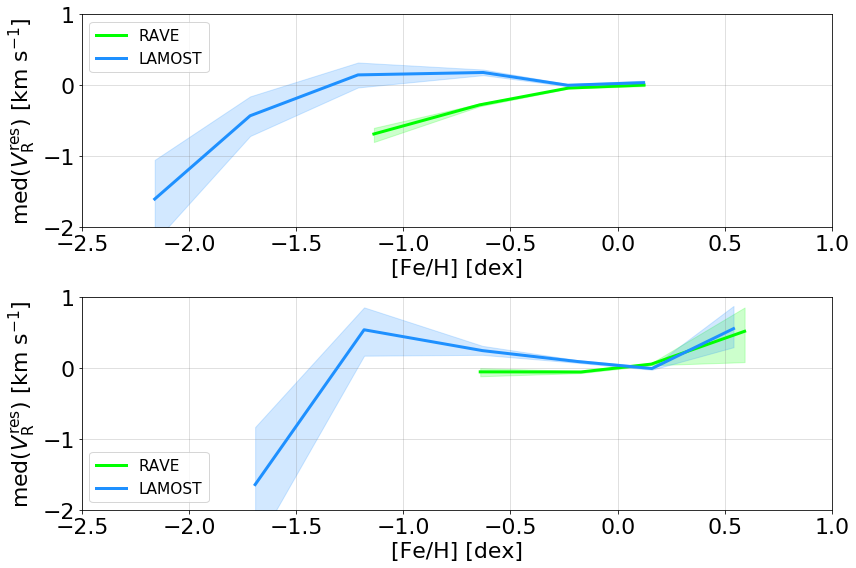}
\caption{From top to bottom, \textit{Row~1:} Median residuals of the radial velocity as a function of surface gravity, for samples of metal-rich giants. \textit{Row~2:} Median residuals of the radial velocity as a function of effective temperature, for samples of metal-rich main sequence stars. \textit{Rows~3 and 4} Median residuals of the radial velocity as a function of metallicity, for samples of giants and main sequence stars, respectively. The metallicity trends are presented relative to solar metallicity (see text). The 68.3\% confidence interval on the measurements of the medians is represented as shaded areas. \label{fig:accap}}
\end{figure*}

\subsection{Magnitude trend\label{sect:magtrend}}
In the \gdrtwo\ catalogue, the radial velocities showed a systematic magnitude trend, starting around $\grvs \sim 9$~mag and reaching about 500~$\ms$ at $\grvs \sim 11.75$~mag \citep{Katz2019, Tsantaki2022}. At the time of publication of \gdrtwo, the origin of the trend was not understood. Posterior tests showed that the amplitude of the bias decreases when the intensity of the background light increases. This strongly suggest that the trend is produced by traps in the CCD pixels. Indeed, traps can snare a part of the signal recorded, and prevent it to propagate from CCD line to CCD line. After a while, the trapped photo-electrons are eventually released. If this happens fast enough that the spectrum has moved only one or a few pixels, the released photo-electrons will skew the line profiles, in the direction opposite to the propagation of the signal. In the RVS, the shorter wavelengths lead. Traps would therefore skew the lines toward longer wavelength, mimicking a positive radial velocity shift. Pre-launch laboratory tests showed that the weaker the signal, the stronger the trapping effect. This would explain both that, the intensity of the bias increases with magnitude and that, on the contrary, the background light reduces the effect. The tests did not show a significant time dependence of the trend, at least over the first 34 months of data. It is therefore likely that the majority of the traps were present from the start of the mission and that few of them, so far, are the result of cosmic ray damage.

Early in the preparation of \gdrthree, the trend was modelled as a function of magnitude and background-light intensity. During processing, the model was used to correct the epoch radial velocities and the epoch cross-correlation functions, before computing the combined radial velocities. In order to assess if the calibration did fully remove the magnitude trend or whether there remains a residual bias, we compare the \gdrthree\ velocities to those of the APOGEE and GALAH catalogues. Figure~\ref{fig:accgrvs} (top left) shows the Kiel diagrams of the two catalogues. In each, two groups are selected, one made of dwarf-turn-off stars (group A: salmon in APOGEE and cyan in GALAH) and the other made of red giants and clump stars (group B: red in APOGEE and blue in GALAH). The two samples were restricted to the metallicity range $\FeH \in [-0.5, 0.5]$~dex, to avoid mixing potential magnitude and metallicity trends. Figure~\ref{fig:accgrvs} (top right) presents the median of the radial velocity residuals (Gaia minus catalogue) as a function of \verb|grvs_mag| magnitude, for the four samples. The curves are adjusted vertically, so that their positions coincide on the bright side, with the aim to compensate for possible offsets between dwarfs and giants or between APOGEE and GALAH. The adjustment was made by subtracting the median value of the radial velocity residuals measured in the magnitude range \verb|grvs_mag| $\in [8.5, 10.5]$~mag. The four curves exhibit a very similar trend: quasi-flat for \verb|grvs_mag| $\leq 11$~mag and then, increasing to reach about 400~\ms at \verb|grvs_mag| $= 14$~mag. There is therefore a residual trend and the consistency between the four samples, of different origins and properties, indicates that this trend is in the \gdrthree\ combined radial velocities.

The sample APOGEE B is the one containing the largest number of sources (129~598 giants) and also the one with the most precise median trend. It is therefore used to model the radial velocity bias by fitting a second order polynomial to the median trend. The model is applicable to the stars fainter than \verb|grvs_mag| $= 11$~mag. Figure~\ref{fig:accgrvs} (bottom left) shows the model (dashed line) over-plotted to the median residuals of the radial velocities of the APOGEE~B sample (red curve). The combined radial velocities published in \gdrthree\ are not corrected for the magnitude trend. To mitigate it, it is therefore recommended to subtract from \gdrthree\ \verb|radial_velocity| (for stars with \verb|rv_template_teff| $< 8500$~K):
\begin{equation}
\label{eq:vrcor}
\begin{array}{ll}
\mathrm{grvs\_mag} < 11:    V_\mathrm{R}^{corr} = & 0~\kms \\
\mathrm{grvs\_mag} \geq 11: V_\mathrm{R}^{corr} = & 0.02755 \times grvs\_mag^2 \\
                                                  & - 0.55863 \times grvs\_mag\\
                                                  & + 2.81129~\kms \\
\end{array}
\end{equation}

The hot stars also exhibit a magnitude trend, but it is quite different from the one of the cooler stars. For the stars with a template temperature \verb|rv_template_teff|~$\geq 8500$~K it is recommended to use the correction derived in \citet{DR3-DPACP-151}, instead of Eq.~\ref{eq:vrcor} above.

Figure~\ref{fig:accgrvs} (bottom right) presents the median residuals of the radial velocity as a function of \verb|grvs_mag| magnitude, after applying the radial velocity correction to the five reference catalogues: APOGEE, GALAH, GES, LAMOST and RAVE. The stars have been selected in the same two groups used to assess the magnitude trends. Group B (giants) is shown in the upper panel and group A (dwarfs/turn-off) in the lower panel. After correction, APOGEE, GALAH, GES and RAVE samples (both dwarfs and giants) do not show any statistically significant trend anymore. On the other hand, the LAMOST stars present a trend which reaches approximately 1.5 to 2 \kms\ at \verb|grvs_mag|~$= 14$~mag. Since the effect is only visible in the LAMOST DR7 data, it is likely intrinsic to this catalogue.

\subsection{Atmospheric parameter trends\label{sect:aptrends}}
In this section, we assess how the radial velocity systematic differences between \gdrthree\ and the APOGEE, GALAH, GES, RAVE and LAMOST catalogues depend on the atmospheric parameters, that is the effective temperature, surface gravity and metallicity. In order to avoid mixing the effects, specific samples are selected for each parameter.

Figure~\ref{fig:accap} presents the medians of the radial velocity residuals as a function: (row~1) of surface gravity, for samples of metal-rich giants, (row~2) of effective temperature, for samples of metal-rich main sequence stars and (rows~3 and 4) of metallicity, for samples of giants and main sequence stars, respectively. The selection of the samples is described in Appendix~\ref{app:Acc}. The surface gravity and effective temperature plots present the absolute differences between \gdrthree\ and the comparison catalogues. Conversely, the median residual curves as a function of metallicity are adjusted vertically by subtracting the median value of the radial velocity residuals measured in the metallicity range $\FeH \in [-0.25, 0.25]$~dex. Therefore, the metallicity trends are presented relative to solar metallicity. All samples are corrected for the magnitude trend (Eq.~\ref{eq:vrcor}).

APOGEE, GALAH and to some extent GES present similar trends and their median radial velocity differences with \gdrthree\ are mostly in the range $[-500, +500]$ \ms. RAVE does not show the same trends as the three first comparison catalogues, but its median differences with respect to \gdrthree\ are also mostly limited to $\pm 500$~\ms, except for $\Teff > 7000$~K where the offset between the two catalogues reaches approximately $-1$~\kms. The offsets between LAMOST and \gdrthree\ are larger, varying between $+4$ and $+6$~\kms, depending on temperature and gravity. The metallicity trend presents an amplitude of approximately 2~\kms. In this section, we compare the \gdrthree\ radial velocities to the raw LAMOST DR7 radial velocities. It should be noted that \citet{Zhang2021} provide calibrations of LAMOST DR7 radial velocity zero-points, based on \gdrtwo\ radial velocities.

The \gdrthree\ radial velocities presented here are not corrected for the gravitational redshift, which shifts the measurements by a few 10 \ms\ in giants and several hundreds \ms\ in dwarfs \citep{LindegrenDravins2003}, nor for the convective shift produced by convective motions in stellar atmospheres. We therefore compared \gdrthree\ velocities to estimates that did not take these effects into account either. It should however be noted that the GALAH catalogue provides also a measurement of the radial velocity which is corrected for the gravitational redshift and the convective shift \citep{Buder2021, Zwitter2021}.

In this section, we have considered temperature trends up to $\Teff \sim 7500$~K. \citet{DR3-DPACP-151} studied the hotter stars and showed that, after correcting for the hot star magnitude trend, for $\Teff \geq 7500$~K, the absolute value of the \gdrthree\ median radial velocity bias is mostly smaller than 3~\kms.


\section{Validation of the formal uncertainties\label{sect:uncertainties}}
The formal uncertainty on the measurement of the combined radial velocity is published in the field \verb|radial_velocity_error|. To evaluate the reliability of the formal uncertainties, we compare our data to the APOGEE DR17 catalogue \citep{Abdurrouf2022}. In the perfect case, if both \gdrthree\ and APOGEE uncertainties were the true ones and if all stars were constant (that is, included no binary, nor variable), the distribution of the radial velocity residuals divided by the combined true uncertainties should follow a normal distribution of standard deviation equal to~1. Conversely, one can define a normalised radial velocity residual as:
\begin{equation}
V_{res}^{norm} = \frac{V_\mathrm{R} - V_\mathrm{APO}}{\sqrt{(f_\sigma \times \epsilon_{V_\mathrm{R}})^2 + \epsilon_\mathrm{APO}^2}}
\label{eq:resnorm}
\end{equation}
where $V_\mathrm{R}$ and $\epsilon_{V_\mathrm{R}}$ are a \gdrthree\ combined radial velocity and its associated formal uncertainty, respectively, $V_\mathrm{APO}$ and $\epsilon_\mathrm{APO}$ the same quantities for the APOGEE catalogue and $f_\sigma$ is the multiplicative factor that should be applied to the formal uncertainty $\epsilon_{V_R}$, in order to obtain a distribution of $V_{res}^{norm}$ with a standard deviation\footnote{Here we use a robust estimator of the standard deviation: $\sigma = (Per(84.15) - Per(15.85)) / 2$, where $Per(15.85)$ and $Per(84.15)$ are the 15.85$^{th}$ and 84.15$^{th}$ percentiles of the distribution, respectively.} equal to~1. The coefficient $f_\sigma$ provides a measurement of the reliability of the formal uncertainty. As a caveat, we should note that APOGEE uncertainties too can be subject to some imprecision and that this will reflect on the value derived for $f_\sigma$. Hereafter, we use the same terminology as in \citet{DR3-DPACP-127}, and we refer to $f_\sigma$ as the standard error factor (also referred to as unit weight error in the literature).

\begin{figure}[h!]
\centering
\includegraphics[width=0.99 \hsize]{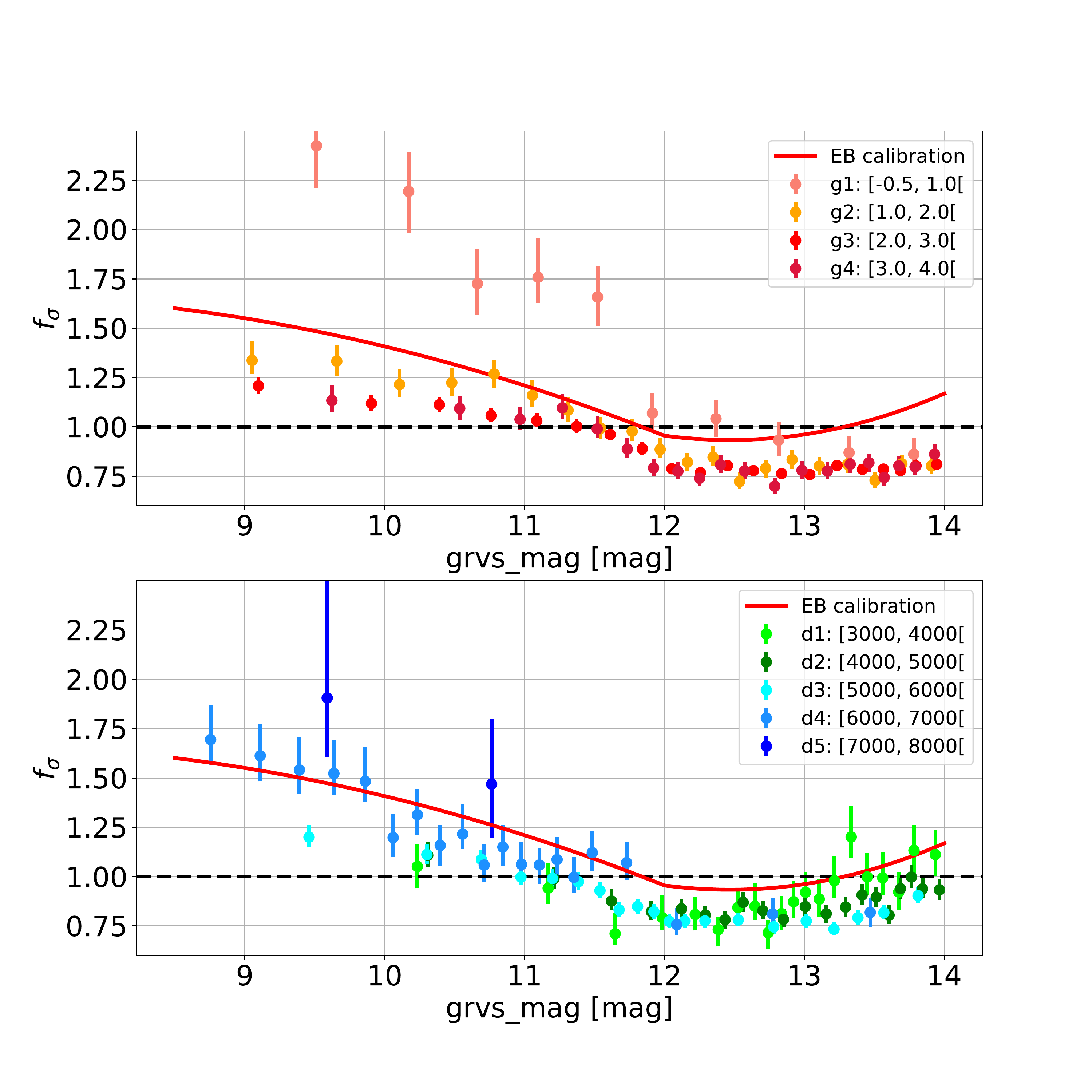}
\includegraphics[width=0.99 \hsize]{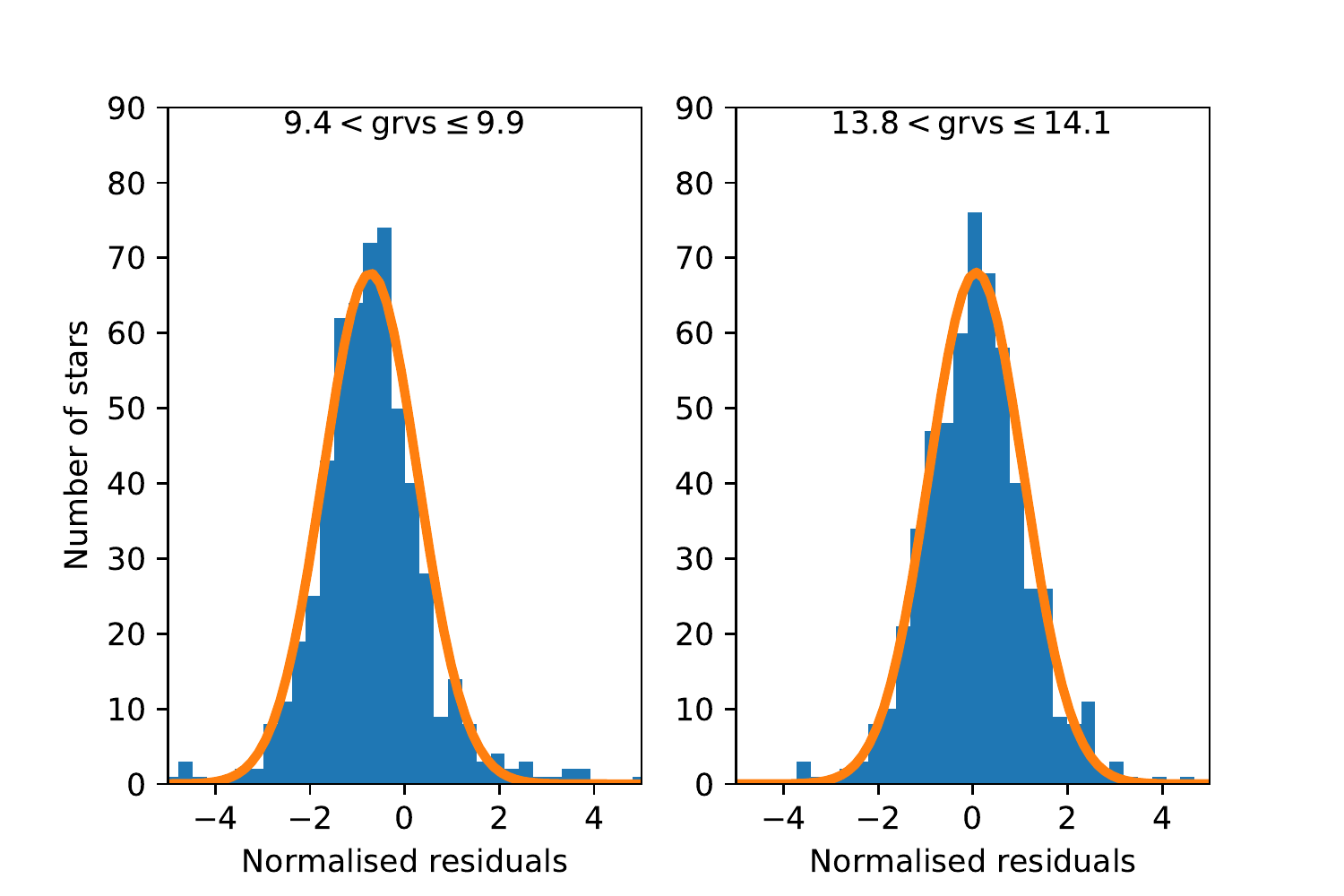}
\caption{From top to bottom, \textit{Row 1:} Standard error factor (see text) as a function of grvs\_mag magnitude, for the giant star samples. The range of surface gravity of each sample is provided in the legend. The red curve corresponds to the calibration of $f_\sigma$ provided in \citet{DR3-DPACP-127}. \textit{Row 2:} same as row 1, for the dwarf star samples. The range of effective temperature of each sample is provided in the legend. \textit{Row 3:} Distributions of the normalised residuals of two groups of stars from the g2 giant sample (see Appendix~\ref{app:Unc}), selected in the magnitude ranges grvs\_mag $\in [9.4, 9.9]$ (left) and $[13.8, 14.1]$~mag (right). Gaussian profiles with standard deviations equal to 1 are overlaid (orange curves).\label{fig:unc}}
\end{figure}

To measure the reliability of the formal uncertainties, it is important to use a sample containing as few variable stars as possible. We therefore select a sub-sample of the APOGEE catalogue, with at least four APOGEE measurements (\verb|NVISITS|~$\geq 4$) and a scatter of the individual APOGEE radial velocities \verb|VSCATTER|~$\leq 0.5$~\kms. We calculate the APOGEE uncertainties as $\epsilon_{APO} = \mathrm{VSCATTER} / \sqrt{\mathrm{NVISITS}}$. The sub-sample is then further split in several dwarf and giant star samples: from g1 at the top of the giant branch, $\logg \in [-0.5, 1.0[$, to g4 at the bottom, $\logg \in [3.0, 4.0[$, and from d1 at the cool end of the main sequence, $\Teff \in [3000, 4000[$~K, to d5 at the hot end, $\Teff \in [7000, 8000[$~K. Fig.~\ref{fig:appUnC} shows the selection of the giant and dwarf star samples in the Kiel diagram.

Figure~\ref{fig:unc} shows the standard error factor as a function of \verb|grvs_mag| magnitude, for the giant (row 1) and dwarf samples (row 2). Although in detail, the values of $f_\sigma$ vary from sample to sample, several overall trends emerge. On the one hand, the standard error factor decreases with magnitude up to \verb|grvs_mag| $= 12$~mag and then remains roughly constant in the giant star sample and even increases slightly with magnitude in the dwarf star samples. This transition corresponds to the change of method to calculate the combined radial velocities as well as their formal uncertainties (see Sect.~\ref{sect:mta}). On the other hand, the standard error factor is larger than 1 for \verb|grvs_mag| $< 11-12$~mag and mostly smaller than 1 beyond, indicating that the formal uncertainties are somewhat under-estimated on the bright side and over-estimated on the faint one. One can also note that for the bright stars, the reliability of the uncertainties improves with increasing gravities (giants) and decreasing temperature (dwarfs).

As discussed above, the potential under- or over-estimation of the APOGEE uncertainties would modify the estimation of the standard error factor. The $f_\sigma$ values provided in Fig.~\ref{fig:unc} should not be considered as calibrations, but as illustrations of the overall behaviour of the formal uncertainties. In order to by-pass the question of the reliability of the uncertainties of the comparison catalogue, the \gaia\ validation group conducted a similar study \citep{DR3-DPACP-127}, using the catalogue of wide binaries of \citet{ElBadry2021}. One of the merits of this approach is that it relies on \gdrthree\ radial velocities and formal uncertainties only. On the other hand, the selected sample is too small to split it in several groups of dwarfs and giants. \citet{DR3-DPACP-127} propose a global calibration of $f_\sigma$ as a function of magnitude. It is represented in Fig.~\ref{fig:unc} by the red curve. It exhibits a trend broadly similar to the ones of the giant and dwarf star samples.  

Figure~\ref{fig:unc} (row 3) shows two examples of distributions of normalised residuals of groups of stars from the g2 giant sample, selected in the magnitude ranges \verb|grvs_mag| $\in [9.4, 9.9]$ (left) and $[13.8, 14.1]$~mag (right), respectively. The normalised residuals are calculated following Eq.~\ref{eq:resnorm}, using the values of $f_\sigma$ of the g2 sample shown in Fig.~\ref{fig:unc} (row~1). Gaussian profiles with standard deviations equal to 1 are overlaid (orange curves). The distributions and profiles are in satisfactory agreement, indicating that, at first order, the \gdrthree\ radial velocity errors follow a Gaussian distribution.


\section{Median formal precision\label{sect:precision}}
In this section, we assess the median formal precision of the \gdrthree\ combined radial velocities as a function, on the one hand, of the magnitude and atmospheric parameters and, on the other hand, of the galactic coordinates (l, b). For a given sample of stars, the median formal precision is calculated as the median of their radial velocity formal uncertainties: that is, the median of their \verb|radial_velocity_error|. The lower and upper boundaries of the 68.3\% confidence interval on the estimation of the median formal precision are calculated as:
\begin{equation}
\epsilon_{\rm prec}^{\rm low} = \sqrt{{\pi}\over{2 N}} (\tilde{\epsilon}_{V_{\rm R}} - Per(\epsilon_{V_{\rm R}}, 15.85))
\end{equation}
\begin{equation}
\epsilon_{\rm prec}^{\rm upp} = \sqrt{{\pi}\over{2 N}} (Per(\epsilon_{V_{\rm R}}, 84.15) - \tilde{\epsilon}_{V_{\rm R}})
\end{equation}
where $N$, $\tilde{\epsilon}_{V_{\rm R}}$, $Per(\epsilon_{V_{\rm R}}, 15.85)$ and $Per(\epsilon_{V_{\rm R}}, 84.15)$ are respectively, the number of elements, the median and the 15.85$^{th}$ and 84.15$^{th}$ percentiles of the distribution of \verb|radial_velocity_error|, $\epsilon_{V_{\rm R}}$.

\begin{figure}[h!]
\centering
\includegraphics[width=0.99 \hsize]{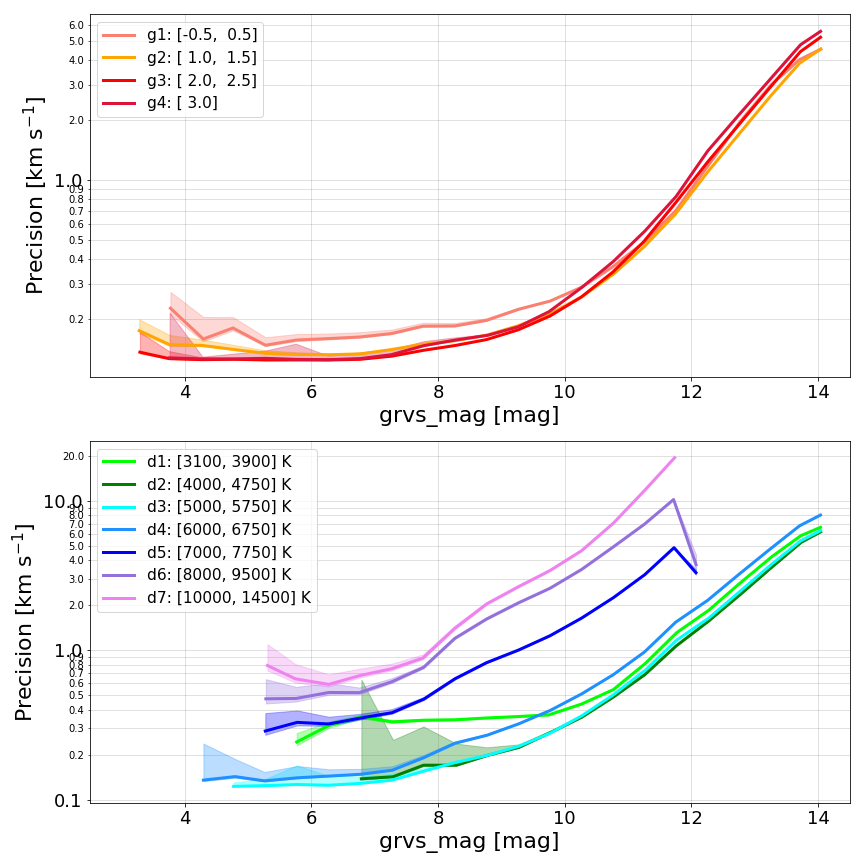}
\includegraphics[width=0.99 \hsize]{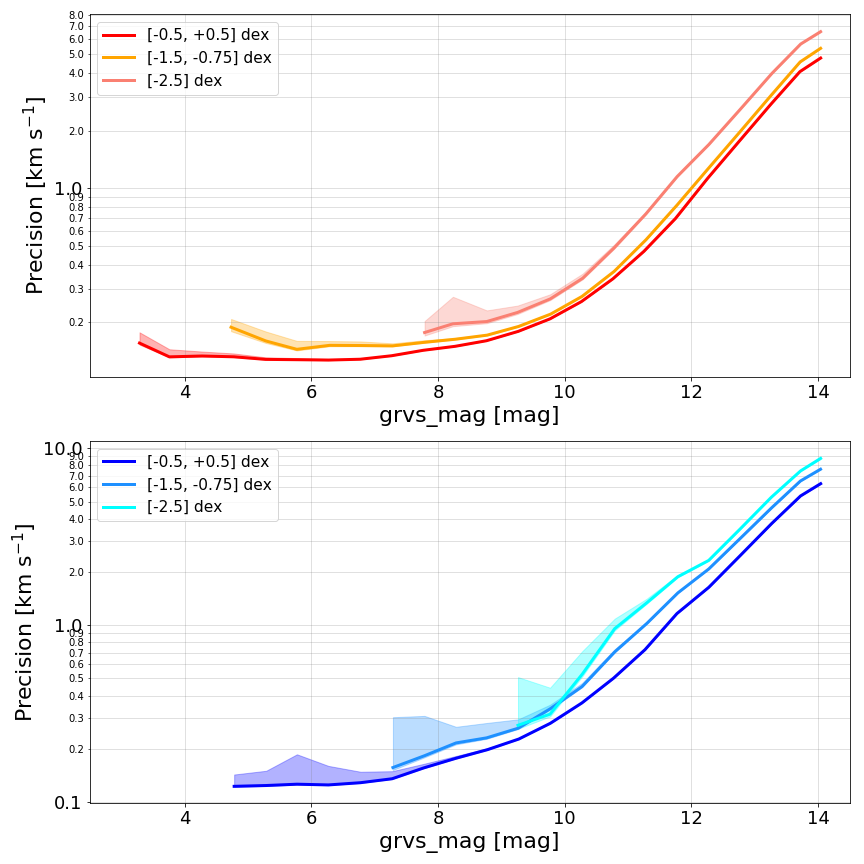}
\caption{\textit{From top to bottom, Row~1:} Median formal precision as a function of grvs\_mag magnitude, for different samples of metal-rich stars selected along the giant branch. The range of rv\_template\_logg of each sample is provided in the legend. \textit{Row~2:} Same as row~1, for samples of metal-rich stars selected along the main sequence. The range of rv\_template\_teff of each sample is provided in the legend. \textit{Row~3:} Same as row~1, for giant stars, selected in different ranges of metallicity. The range of rv\_template\_fe\_h of each sample is provided in the legend. \textit{Row~4:} Same as row~1, for dwarf stars, selected in different ranges of metallicity. The range of rv\_template\_fe\_h of each sample is provided in the legend. The 68.3\% confidence interval on the measurements of the median formal precision is represented as shaded areas.\label{fig:prec}}
\end{figure}

The median formal precision of the \gdrthree\ radial velocities considered as a whole is 1.3~\kms\ at \verb|grvs_mag| $= 12$~mag and 6.4~\kms\ at \verb|grvs_mag| $= 14$~mag.

Figure~\ref{fig:prec} presents the median formal precision as a function of magnitude, for samples: selected along the giant branch (row 1), selected along the main sequence (row 2), and of giants and dwarfs, respectively, selected in different intervals of metallicity (rows 3 and 4). The definition of the groups is based on the template parameters: \verb|rv_template_teff|, \verb|rv_template_logg| and \verb|rv_template_fe_h|. The giant star samples are ordered by increasing surface gravities, from \verb|g1| at the top of the giant branch to \verb|g4| at the bottom. The dwarf star samples are ordered by increasing temperature from \verb|d1| (\verb|rv_template_teff| $\leq 3750$~K) to \verb|d7| (\verb|rv_template_teff| $\geq 10000$~K). In row~1 and row~2, the giant and dwarf samples are restricted to the metallicity range: \verb|rv_template_fe_h| $\in [-0.5, 0.5]$~dex. In row~3 and row~4, the giant and dwarf samples combine the \verb|g2| and \verb|g3| and the \verb|d2| and \verb|d3| datasets, respectively. Fig.~\ref{fig:appPrec} shows the selection of the giant and dwarf star samples in the (\verb|rv_template_teff|, \verb|rv_template_logg|) plane.

In \gdrthree, the typical median formal precisions for a solar metallicity red clump star (\verb|g3| sample) are $\sim$125 \ms\ at \verb|grvs_mag| $= 6$~mag, $\sim$230 \ms\ at \verb|grvs_mag| $= 10$~mag and, $\sim$5.2 \kms\ at \verb|grvs_mag| $= 14$~mag. As shown in Fig.~\ref{fig:prec} (row~1), the median formal precision is weakly dependent on the surface gravity. Beyond \verb|grvs_mag| $\sim$11~mag, there is a modest improvement of the median formal precision from the bottom to the top of the giant branch. Conversely, the median formal precision improves significantly as the effective temperature decreases, in particular between 15~000 and 6000~K (Fig.~\ref{fig:prec}, row~2). This reflects the evolution of the information contained in the spectra. In hot stars, it is dominated by broad and shallow lines of the Paschen series. When the temperature decreases, these are gradually replaced by the ionized calcium triplet as well as by numerous sharp neutral lines, better suited for measuring radial velocities. Finally, the median formal precision improves with metallicity, both in giants and dwarfs (Fig.~\ref{fig:prec}, rows~3 and 4). This is the expected behaviour: the lines becoming stronger with increasing abundances.

As discussed in Sect.~\ref{sect:uncertainties}, the formal uncertainties are somewhat under-estimated for \verb|grvs_mag| $< 11-12$~mag, and over-estimated for fainter stars. However, the data used to draw the general trend do not allow an accurate definition of a correction factor as a function of both magnitude and atmospheric parameters. Therefore, in this section, the median formal precision is derived from uncorrected formal uncertainties. Applying the standard error factor ($f_\sigma$) calculated in Sect.~\ref{sect:uncertainties} would modify the typical median formal precisions of the solar metallicity red clump star considered above in the following way: at \verb|grvs_mag| $= 6$~mag, from $\sim$125 to 150~\ms\ (assuming the trend is flat for \verb|grvs_mag| $< 9$~mag), at \verb|grvs_mag| $= 10$~mag, from $\sim$230 to 260~\ms, and at \verb|grvs_mag| $= 14$~mag, from $\sim$5.2 to 4.2~\kms. The impact would be stronger on the bright hot dwarfs ($\Teff > 6000$~K) as well as on the stars at the tip of the giant branch (sample \verb|g1|).

\begin{figure}[h!]
\centering
\includegraphics[width=0.99 \hsize]{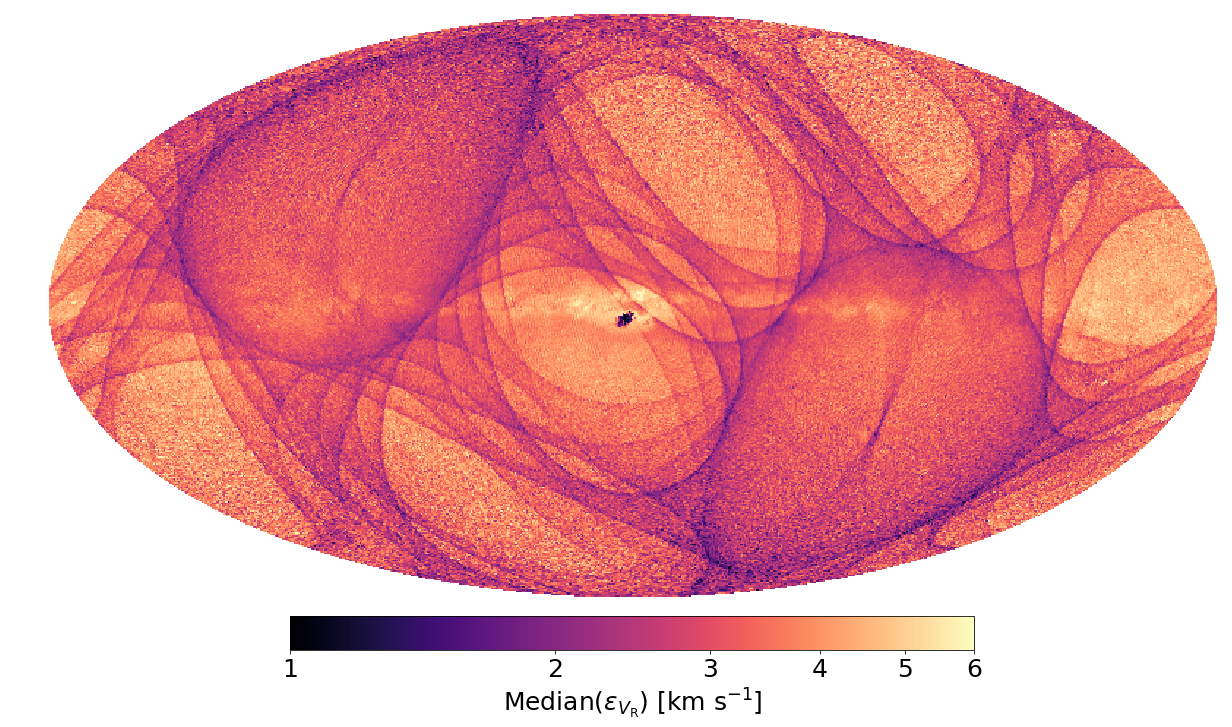}
\caption{Sky map of the median formal precision of all the combined radial velocities of the \gdrthree\ catalogue. The image uses a Mollweide projection in Galactic coordinates $(l, b)$. The Galactic centre is at the centre of the image and the Galactic longitudes increase to the left. The sampling of the map is approximately 0.2 square degree (healpix level 7).\label{fig:mapprec}}
\end{figure}

Figure~\ref{fig:mapprec} presents the map of the median formal precision as a function of Galactic longitudes and latitudes. The comparison to Fig.~\ref{fig:transits} (top) shows that, over most of the sky, the median formal precision correlates with the number of transits and is therefore driven by the scan law. Conversely, in the densest regions toward the Galactic centre, the median formal precision is of the order of 1~\kms, although the number of transits there is low. As discussed in Sect.~\ref{sect:distrib}, few faint stars are observed in these regions. This skews the magnitude distribution towards bright stars whose radial velocity measurements are more precise.

\section{High-velocity stars\label{sect:HVS}}
In order to observe the largest possible fraction of \gaia\ sources, the RVS has to collect spectra with very low signal-to-noise ratios. This situation is quite unusual. Generally, when the source is too faint, one uses a longer exposure time and/or a telescope with a larger diameter. Even after combining the transits, half of the sources with a radial velocity published in \gdrthree\ have an \verb|rv_expected_sig_to_noise| of less than 7.8 (see Sect.~\ref{sect:distrib}). As described in Sect.~\ref{sect:noise}, at very low SNR it may happen that the amplitude of the true cross-correlation peak is lower than that of the highest noise peak(s). The measured radial velocity is then random and of course false.

\begin{figure}[h!]
\centering
\includegraphics[width=0.99 \hsize]{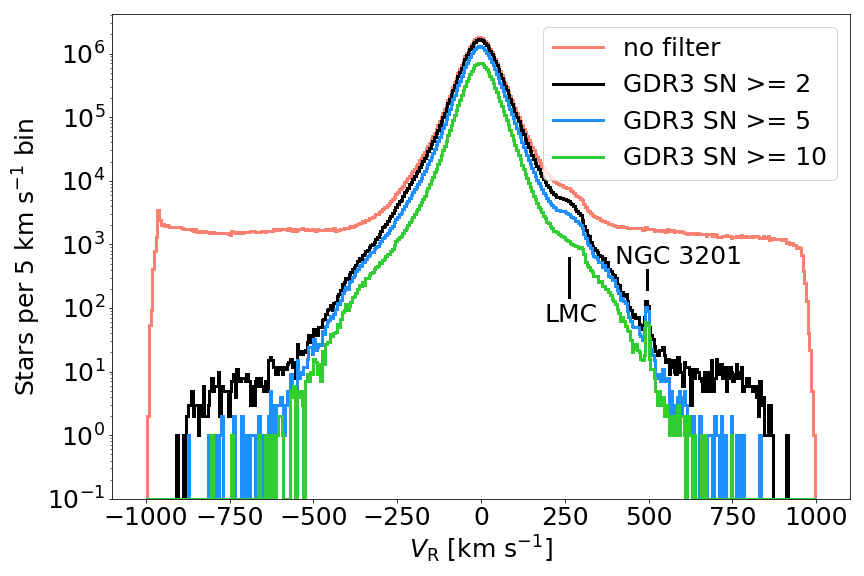}
\includegraphics[width=0.99 \hsize]{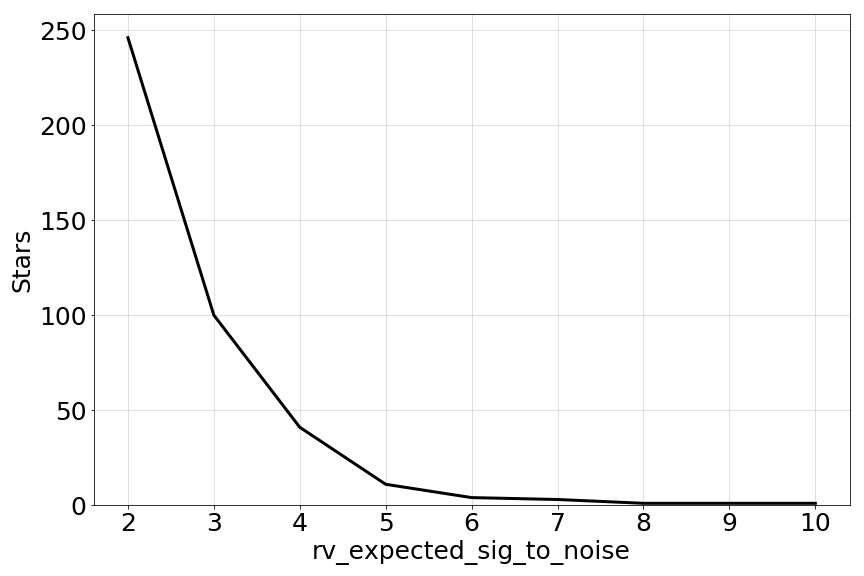}
\includegraphics[width=0.99 \hsize]{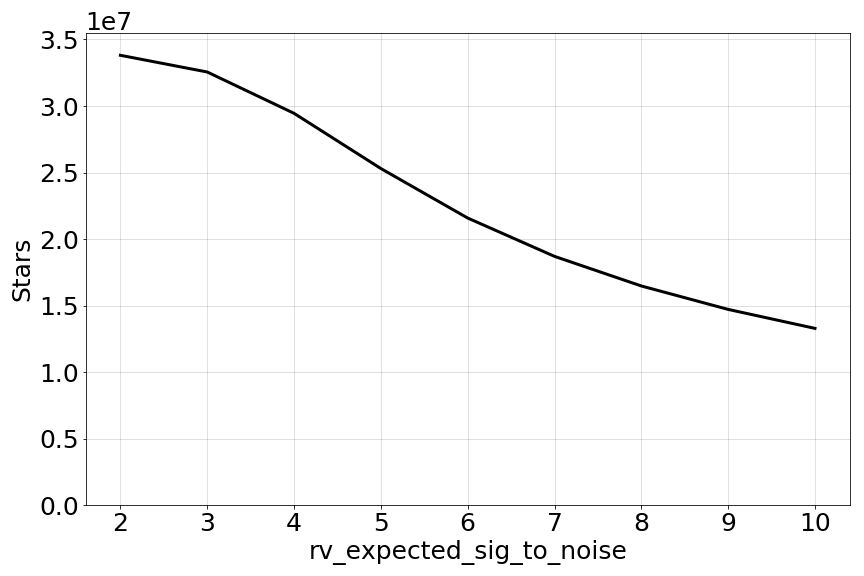}
\caption{\textit{Top:} Distributions of the radial velocities of: the 37~499~608 sources successfully processed by the pipeline (salmon), the \starPublished\ \gdrthree\ sources (black), the 25.3 million \gdrthree\ sources with rv\_expected\_sig\_to\_noise~$\geq 5$ (cyan) and the 13.3 million \gdrthree\ sources with rv\_expected\_sig\_to\_noise~$\geq 10$ (green). \textit{Middle:} Number of \gdrthree\ sources with $|V_R| \geq 750$~\kms\ and a rv\_expected\_sig\_to\_noise larger than or equal to the abscissa value. \textit{Bottom:} Number of \gdrthree\ sources with a rv\_expected\_sig\_to\_noise larger than or equal to the abscissa value.\label{fig:hvs1}}
\end{figure}

\begin{figure*}[tbp]
\centering
\includegraphics[width=0.49 \hsize]{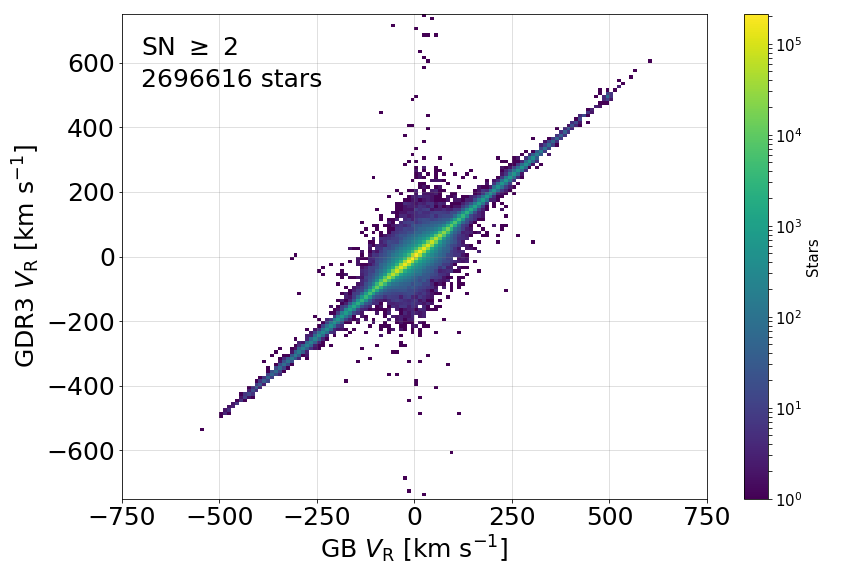}
\includegraphics[width=0.49 \hsize]{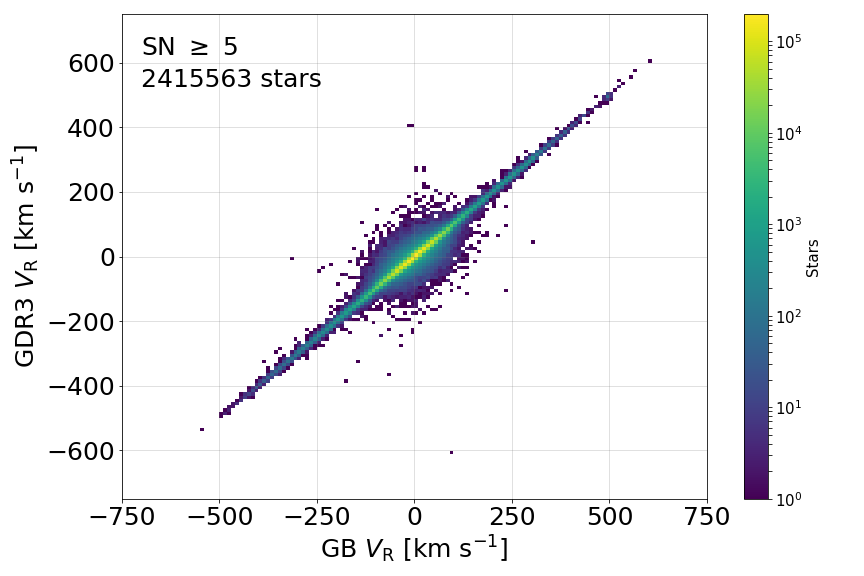}
\includegraphics[width=0.49 \hsize]{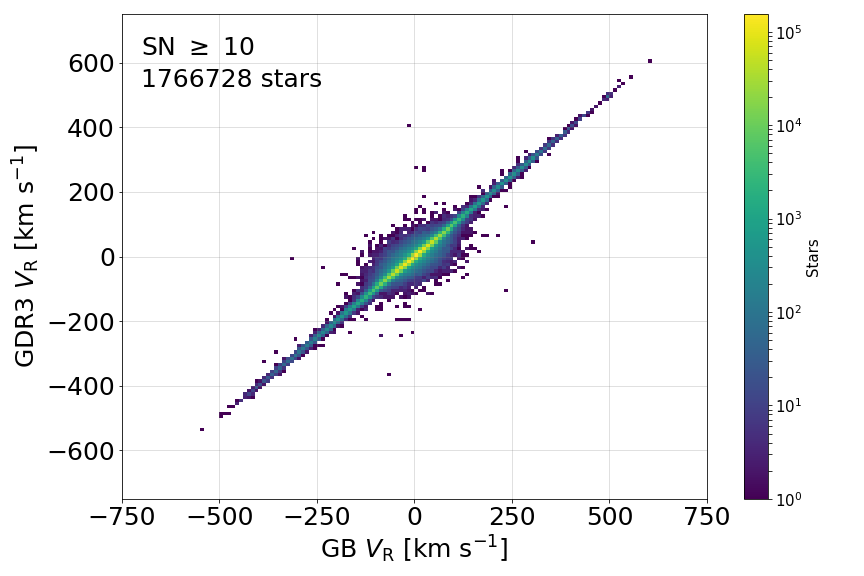}
\includegraphics[width=0.49 \hsize]{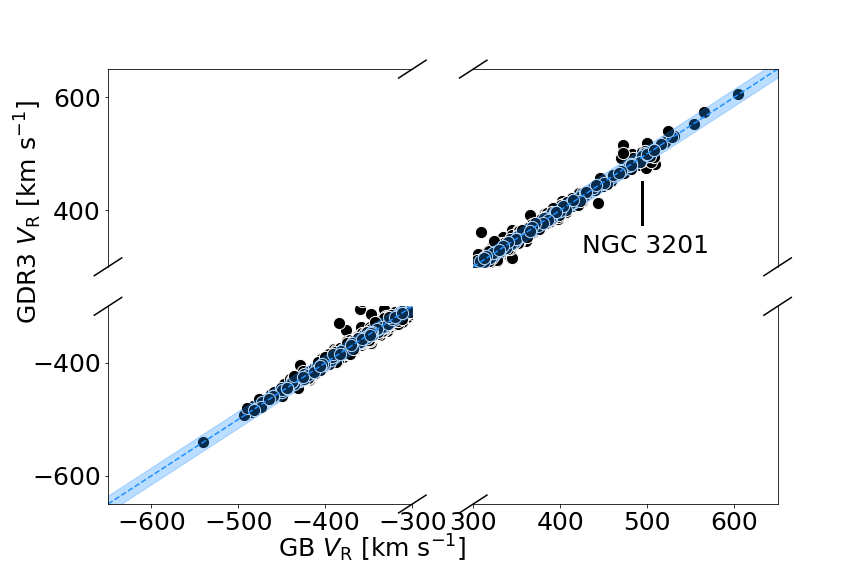}
\caption{\textit{Top left:} Comparison of the \gdrthree\ radial velocities to those of the combined dataset (see text). \textit{Top right:} Same as the top left panel, for the stars with rv\_expected\_sig\_to\_noise~$\geq 5$. \textit{Bottom left:} Same as the top left panel, for the stars with rv\_expected\_sig\_to\_noise~$\geq 10$. \textit{Bottom right:} Zoomed-view of the velocity ranges $[-600, -300]$ and $[300, 600]$~\kms\ from the top-left panel. The stars are represented individually. The blue shaded area delimits a zone of $\pm 15$~\kms around the main diagonal.\label{fig:hvs2}}
\end{figure*}

Figure~\ref{fig:hvs1} (top) presents several distributions of radial velocities. The 37~499~608 combined radial velocities produced by the pipeline are shown in salmon. The \starPublished\ sources that passed successfully the validation filters (including a cut at SNR$=2$) and have their radial velocities published in \gdrthree, are shown in black. The cyan and green histograms correspond to the 25.3 and 13.3 million sources with published velocities and \verb|rv_expected_sig_to_noise|$~\geq 5$ and 10, respectively. The salmon histogram exhibits flat extended wings. They are populated by the spurious radial velocity measurements. As expected from the random nature of false secondary cross-correlation peaks, these spurious velocities follow a pseudo-uniform distribution and they dominate in the wings of the radial velocity distribution, which are less populated than the core. The black histogram shows that the validation filters have largely removed the wings and therefore suppressed a very large part of the spurious values. It is rewarding to see the high velocity Globular Cluster NGC~3201 emerging, after applying the filters. However, as discussed in Sect~\ref{sect:velocities}, the \gdrthree\ velocities still exhibit small wings beyond $|V_R| \gtrsim 500 - 600$~\kms. 

The impact of the erroneous values depends on the science case considered. The number of outliers is rather small compared to the full sample. Therefore, it should have little influence on the studies of the overall kinematic and dynamical properties of the Milky Way. Of course, for all studies concerned directly with high velocity stars, the level of contamination in \gdrthree\ poses a problem. The main origin of the spurious measurements are the very low signal to noise ratios. Unsurprisingly, the cyan and green histograms show that further restricting the data to signal to noise ratios larger than 5 and 10, respectively, removes gradually most of the false high velocity stars. This is of course at the cost of a significant loss of completeness of the sample.

Figure~\ref{fig:hvs1} (middle) shows the number of \gdrthree\ sources with $|V_\mathrm{R}| \geq 750$~kms and \verb|rv_expected_sig_to_noise| larger than or equal to the abscissa value. Figure~\ref{fig:hvs1} (bottom) shows the total number of \gdrthree\ sources with an SNR 
larger than or equal to the abscissa value. The stars with an absolute value of the radial velocity larger than 750~\kms\ are supposedly extremely rare in the Galaxy. They are used here as a sample representative of a very strong contamination by spurious radial velocities. The \gdrthree\ catalogue contains 254 of these stars. This number drops to 13, for an SNR cut-off of 5 and 1 for a cut-off of 8. At the same time, these cuts reduce the number of sources, from initially \starPublished\ to 25~302~443 and 16~476~280, respectively. We recommend to \gdrthree\ users, working on high velocity stars, to apply a stricter filter on the signal to noise ratio. Where to set the threshold must be defined according to the science case considered, based on the curves in Fig.~\ref{fig:hvs1}.

It should also be remembered that low signal-to-noise ratios are not the only possible source of spurious high velocity stars, although it is by far the main one. Contamination by a bright neighbour can also produce large velocity outliers \citep{Boubert2019}. Specific filters have been applied to cope with this problem (see Sect.~\ref{sect:cont}), but problematic sources may have slipped through. We therefore recommend caution with high velocity stars in the vicinity of a brighter neighbour.

The ground-based surveys described in Appendix~\ref{app:GB} can provide additional insight on the spurious high velocities. For ease of comparison, the five catalogues were combined in a single dataset. A correction factor of 5~\kms\ was subtracted to the LAMOST radial velocities to adjust them approximately to the same scale as the other catalogues. Figure~\ref{fig:hvs2} (top left) compares the \gdrthree\ radial velocities to those of the combined dataset. The majority of the sources are distributed along the main diagonal, which shows the overall good agreement between the \gdrthree\ and the ground-based surveys. However, \gdrthree\ spurious high velocities are visible as a diffuse vertical sequence centred on $V_\mathrm{R}^\mathrm{GB} \sim 0$~\kms. Restricting the comparison to the sources with \verb|rv_expected_sig_to_noise| larger than or equal to 5 (Fig.~\ref{fig:hvs2}, top right), removes a significant fraction of the spurious sequence and in particular the largest values. Further restricting the dataset to SNR~$\geq 10$ removes a few more outliers (Fig.~\ref{fig:hvs2}, bottom left). 

In this section, we have so far discussed the contamination of the high velocity tails by spurious values. On the other hand, \gdrthree\ also contains numerous reliable high velocities. Figure~\ref{fig:hvs2} (bottom right) shows a zoomed-view of the velocity ranges $[-600, -300]$ and $[300, 600]$~\kms\ from the top-left panel. The blue shaded area delimits a zone of $\pm 15$~\kms\ around the main diagonal. The majority of the sources are contained within this area, illustrating the consistency of \gdrthree\ velocities with those measured from the ground, for these high velocity stars. Here again, the high velocity Globular Cluster NGC~3201 appears as an over-density.


\begin{figure}[]
\centering
\includegraphics[width=0.99 \hsize]{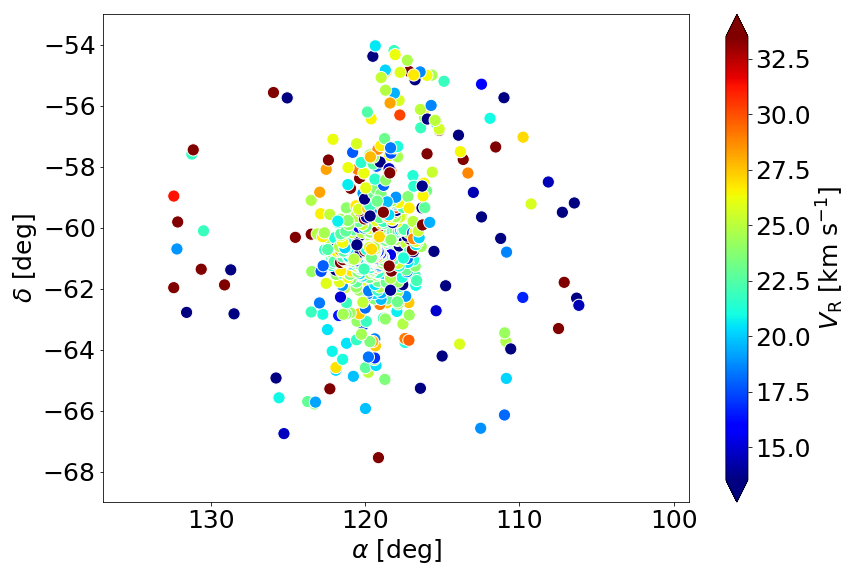}
\includegraphics[width=0.99 \hsize]{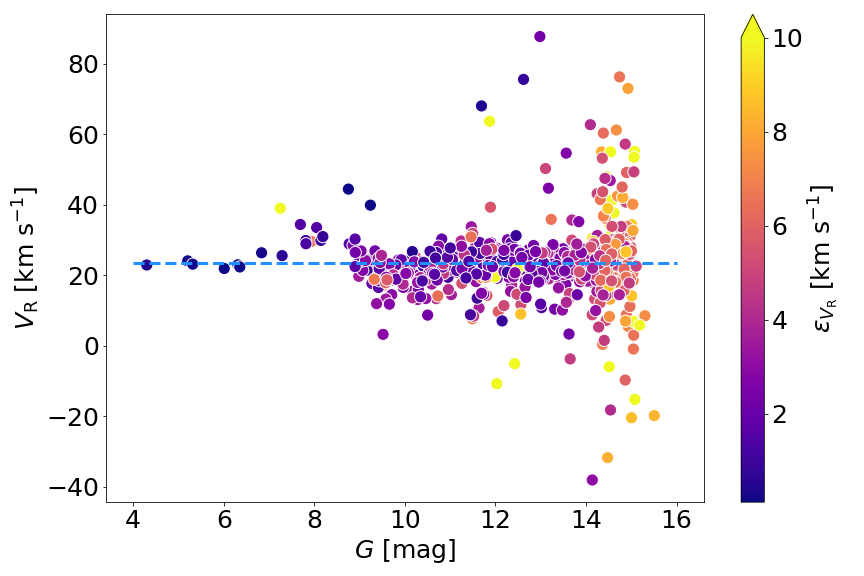}
\caption{\textit{Top:} Spatial distribution of the stars of the open cluster NGC~2516. \textit{Bottom:} \gdrthree\ combined radial velocities as a function of $G$ magnitude, for the stars of the open cluster NGC~2516. The correction of the magnitude trend of the hot stars recommended in \citet{DR3-DPACP-151} is applied.\label{fig:OC1}}
\end{figure}

\section{Some specific objects\label{sect:specobj}}
\subsection{Open clusters\label{sect:OC}}
The combined radial velocities have been tested with open clusters, as in \gdrtwo. We show here two examples, NGC~2516 and Mamajek~4. 

NGC~2516 is a well-studied and populated cluster. A mean radial velocity of 23.82$\pm$0.18 \kms\ was previously determined by \citet{Soubiran2018b}, based on the \gdrtwo\ RVS data of 132 members identified by \citet{CantatGaudin2018}. \citet{Tarricq2021} combined radial velocities from \gdrtwo\ to those from GALAH DR3 \citep{Buder2021} and the Gaia-ESO survey \citep{Gilmore2012, Randich2013} and computed a mean RV of 24.24$\pm$0.07 based on 490 members from \citet{CantatGaudin2018}. \citet{Tarricq2022} determined new memberships for this cluster on a large area of sky, based on \gdrethree, which revealed an extended tidal tail and a halo. We find that 826 of these members have a radial velocity in \gdrthree, giving a median radial velocity of 23.51~\kms. The sky distribution of the members is presented in Fig.~\ref{fig:OC1} (top). It confirms the extension of the cluster with stars in the tidal tail sharing the same motion as the cluster. Figure~\ref{fig:OC1} (bottom) shows the radial velocities of the cluster members as a function of G magnitude, after applying the correction of the magnitude trend of the hot stars recommended in \citet{DR3-DPACP-151}. It shows the consistency of the radial velocities below $G \sim 6.5$ and  above $G \sim 10.5$~mag. The range of apparent magnitude $G \in [6.5, 10.5]$~mag corresponds to the upper main sequence of NGC~2516 and is populated by hot stars with \verb|rv_template_teff| $\geq 8500$~K. They exhibit residual offsets of a few \kms, as described in \citet{DR3-DPACP-151}.

\begin{figure}[]
\centering
\includegraphics[width=0.99 \hsize]{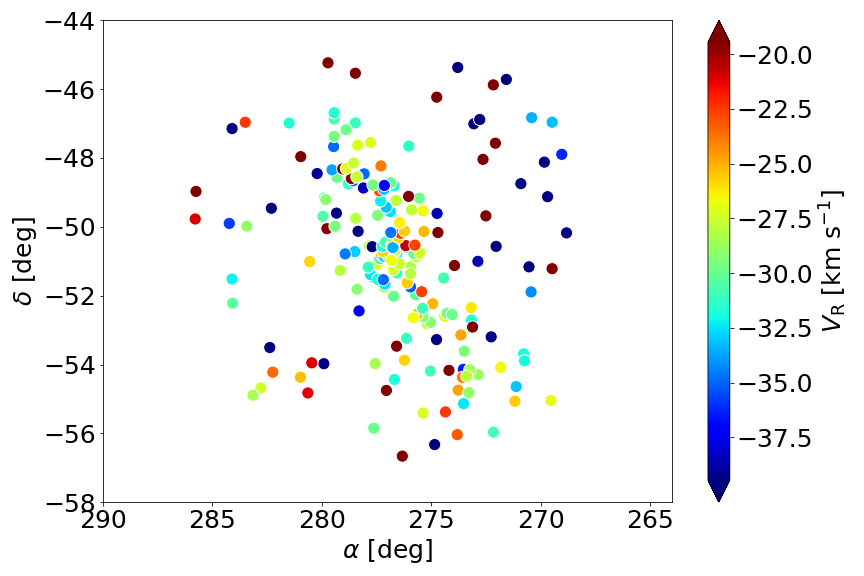}
\includegraphics[width=0.99 \hsize]{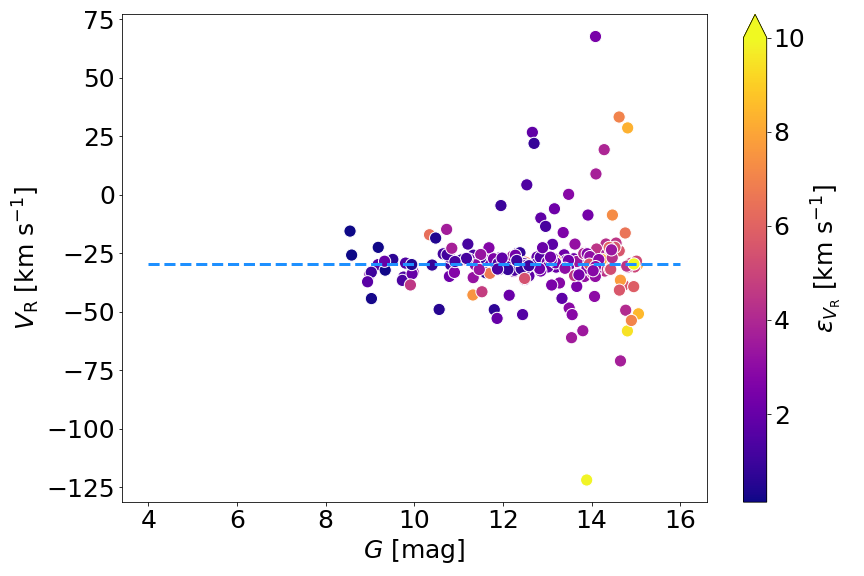}
\caption{Same as Fig.~\ref{fig:OC1} for the open cluster Mamajek~4.\label{fig:OC2}}
\end{figure}

On the contrary, Mamajek~4 is a loose and poorly studied cluster. Its mean radial velocity, $-26.32\pm$0.54~\kms, in \citet{Soubiran2018b} was based on 34 members. In \citet{Tarricq2021}, 41 members with radial velocities from \gdrtwo\ and GALAH gave a mean value of $-27.25\pm$0.34~\kms. \citet{Tarricq2022} found a tidal tail and a halo for this cluster also. On the basis of these new memberships covering a large area, we find 196 members with a combined radial velocity in \gdrthree\ giving a median value of $-29.45$~\kms. Figure~\ref{fig:OC2} presents the spatial distribution (top) and the radial velocities (also corrected for the hot star magnitude trend) as a function of $G$ magnitude (bottom) of Mamajek~4.

\subsection{Globular clusters\label{sect:GC}}
In this section, we consider as Globular Cluster members, the stars from the catalogue of \citet{VasilievBaumgardt2021} having a membership probability higher than 90\%.

\begin{figure}[]
\centering
\includegraphics[width= \hsize]{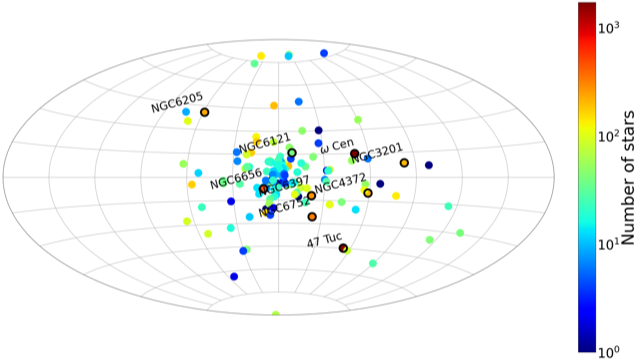}
\caption{Distribution on the sky of the 111 Globular Clusters containing stars having radial velocities published in \gdrthree. The colour-code indicates the number of stars with measured velocity in each cluster. The image uses a Aitoff projection in Galactic coordinates $(l, b)$. The Galactic centre is at the centre of the image and the Galactic longitudes increase to the left. The clusters with more than 200 members with a \gdrthree\ radial velocity are circled in black and their names are reported on the plot.\label{fig:GC1}}
\end{figure}

The increase of the limiting magnitude of the spectroscopic processing has also benefited the Globular Clusters. Out of the list of 170 Globular Clusters published in \citet{VasilievBaumgardt2021}, 111 possess radial velocities published in \gdrthree. The number of members with a \gdrthree\ radial velocity varies greatly from one cluster to another, from more than 1~000 in NGC~5139 (Omega~Cen) and NGC~104 (47~Tuc) to 5 in Terzan~5 and NGC~6522 (which is located in the Baade's window). Figure~\ref{fig:GC1} shows the distribution of the 111 Globular Clusters on the celestial sphere in Galactic coordinates $(l, b)$, as well as the number of radial velocity measurements in each of them.

\begin{figure*}[tbp]
\centering
\includegraphics[width=0.49 \hsize]{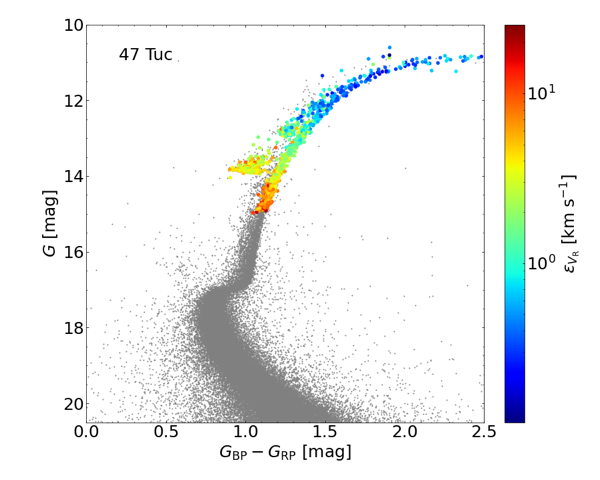}
\includegraphics[width=0.49 \hsize]{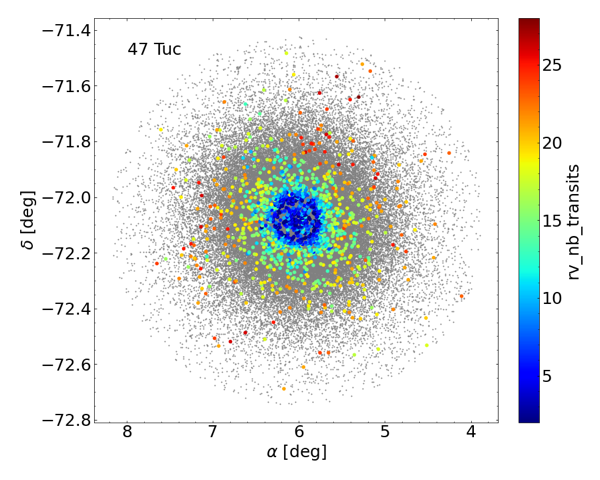}
\includegraphics[width=0.49 \hsize]{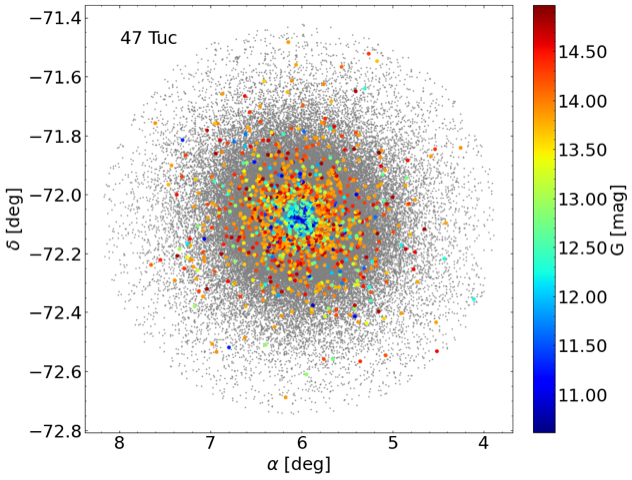}
\includegraphics[width=0.49 \hsize]{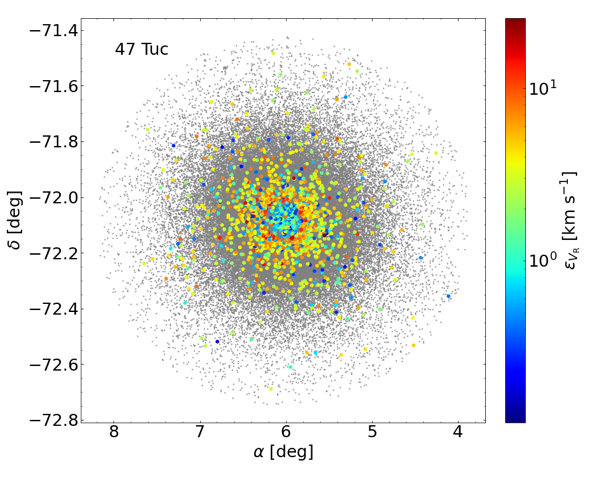}
\includegraphics[width=0.49 \hsize]{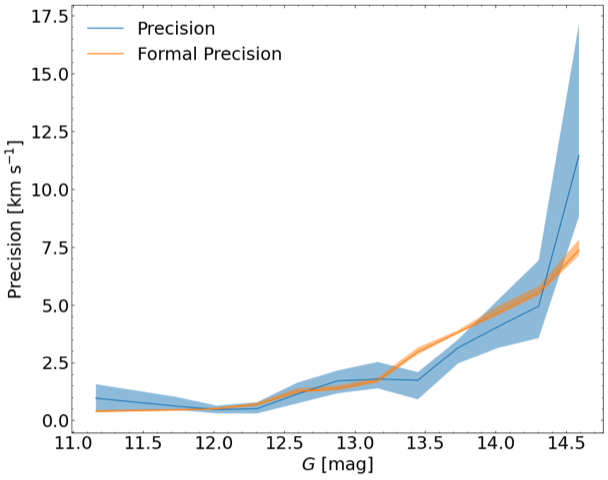}
\includegraphics[width=0.49 \hsize]{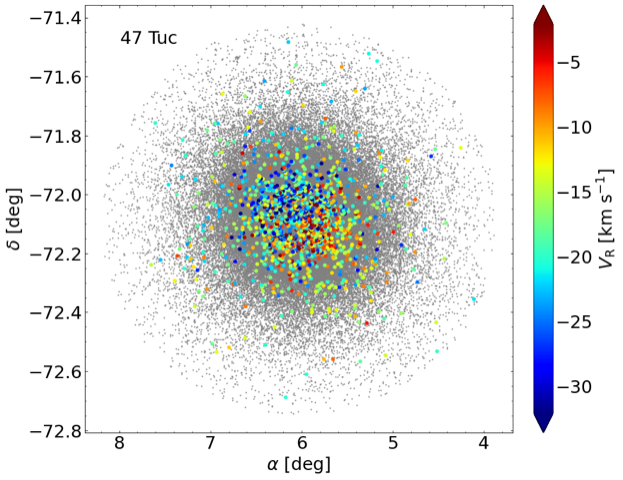}
\caption{\textit{Top left:} Colour-magnitude diagram of 47~Tuc. The stars with \gdrethree\ photometry are represented as grey dots. The stars with a \gdrthree\ radial velocity are colour-coded by radial velocity formal uncertainties. \textit{Top right:} Distribution in equatorial coordinates ($\alpha$, $\delta$) of the 47~Tuc stars. The stars with a \gdrthree\ radial velocity are colour-coded by number of transits. \textit{Middle left:} Same as top right panel, with the stars colour-coded by $G$ magnitude. \textit{Middle right:} Same as top right panel, with the stars colour-coded by radial velocity formal uncertainty. \textit{Bottom left:} Comparison of the formal and external precisions (see text) as a function of $G$ magnitude, for a sub-sample of 47~Tuc stars. The 68.3\% confidence interval on the estimations of the precisions are represented as shaded areas. \textit{Bottom right:} Same as top right panel, with the stars colour-coded by combined radial velocity.\label{fig:GC2}}
\end{figure*}

To further illustrate the properties of the radial velocities measured in Globular Clusters, we consider the case of 47~Tuc. Figure~\ref{fig:GC2} (top left) shows its colour-magnitude diagram, $G$ versus $G_\mathrm{BP} - G_\mathrm{RP}$. The stars with \gdrethree\ photometry are represented as grey dots and those with \gdrthree\ radial velocity are colour-coded according to their radial velocity formal uncertainty. The limit of the spectroscopic processing is visible as a relatively sharp cut around $G \sim 15$~mag. The radial velocity measurements cover the upper part of the red giant branch, the asymptotic giant branch and the horizontal branch. The formal uncertainties improve with magnitude and are of order a few \kms\ at the level of the horizontal branch and of a few hundreds \ms\ at the tip of the red giant branch.

The radial velocities are measured throughout the cluster, including the densest regions of the core. However, two of the issues encountered in the most crowded regions of the bulge (see Sect.~\ref{sect:distrib}) also apply to the central parts of the cluster: (i) the limit to 72 of the number of windows that can be allocated simultaneously per CCD and, (ii) the numerous conflicts between windows, part of which cannot be resolved (for example, because the sources are too close to each other). The consequences are similar to those witnessed before. The number of transits per source is very low in the central part of the cluster and increases outward (Fig.~\ref{fig:GC2}, top right). Moreover, the limiting magnitude of the sources with measured velocities drops in the core of the cluster (Fig.~\ref{fig:GC2}, middle left). Because of this selection effect, the formal uncertainties on the radial velocity measurements are globally smaller in the centre of the cluster (Fig.~\ref{fig:GC2}, middle right).

The high stellar density makes the observation conditions in the Globular Clusters peculiar. One may therefore wonder if the formal uncertainties remain reliable in these objects. To answer this question, we use a sample of stars whose velocities were measured from the ground and are compiled in \citet{BaumgardtHilker2018}. Using this sample, the radial velocity precision is assessed in two different ways. The median formal precision is calculated as the median of the radial velocity formal uncertainties (as in Sect.~\ref{sect:precision}). The external precision is computed as the robust standard deviation of the radial velocity residuals (\gaia\ minus ground-based). Figure~\ref{fig:GC2} (bottom left) compares the two precision curves. Overall they are in satisfactory agreement, although with a $\sim 1.5\sigma$ discrepancy at $G \sim 14.5$. Therefore, the radial velocity formal uncertainties appear also reliable in dense regions, such as Globular Clusters.

The large number of measured radial velocities allows to study the internal motions of the stars within the cluster. In Fig.~\ref{fig:GC2} (bottom right), the stars are colour-coded according to their \gdrthree\ combined radial velocity. Their precisions makes it possible to clearly visualise the line-of-sight rotational velocity of 47~Tuc. It is traced by the gradient of the radial velocities, which vary from yellow-red at the bottom left to blue at the top right. Combined with the proper motions, the \gdrthree\ radial velocities should provide a detailed insight on the kinematic of 47~Tuc and of similarly sampled objects.

\subsection{Large Magellanic Cloud\label{sect:LMC}}
The extension of the radial velocity catalogue down to magnitude $\extgrvs = 14$ has increased the number of stars observed in nearby galaxies. The best sampled of these objects is the Large Magellanic Cloud. In this section, we take a look at the properties of the radial velocities measured in the LMC.

The selection of the LMC stars is performed in several steps. First, the stars within a 10 degree radius from the centre of the LMC \citep[($\alpha$, $\delta$) = ($81.28^\circ$, $-69.78^\circ$),][]{vanderMarel2001} are extracted from the $\gdrthree$ database. Then the foreground stars are removed, keeping the stars with $\varpi / \epsilon_\varpi < 5$. Finally, the sample is restricted to stars with a combined radial velocity larger than 150~\kms, in order to mitigate the contamination by stars with disc velocities. The selected sample contains 29~631 stars.

\begin{figure}[h!]
\centering
\includegraphics[width=0.99\hsize]{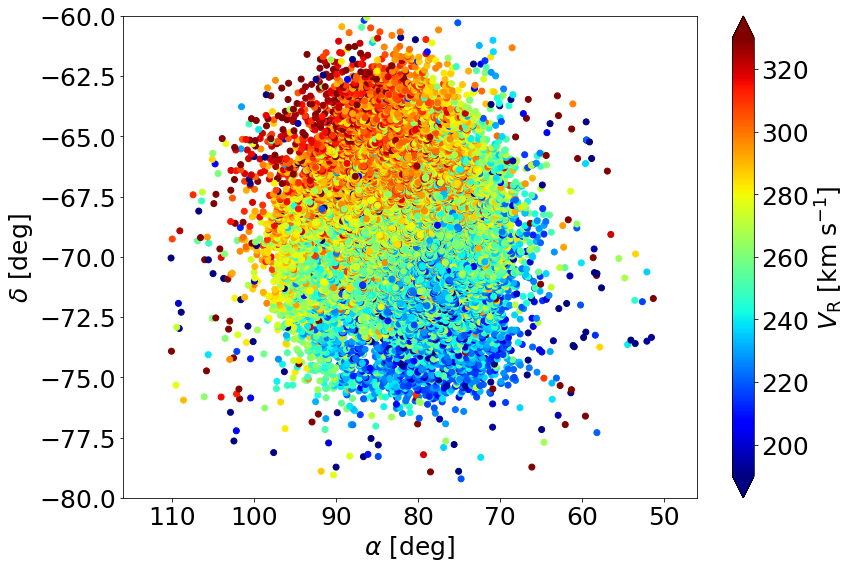}
\caption{Distribution in equatorial coordinates ($\alpha$, $\delta$) of the LMC stars, colour-coded by combined radial velocities.\label{fig:LMC}}
\end{figure}

Figure~\ref{fig:LMC} shows the distribution in equatorial coordinates of the LMC stars, colour-coded according to their combined radial velocities. They exhibit a clear gradient, from  $\sim$320~\kms\ (red) at the top to $\sim$200~\kms\ (blue) at the bottom, which maps the rotation of the LMC projected onto the line-of-sight. The combined radial velocities will complement the \gdrethree\ astrometric and photometric measurements \citep{GaiaCollaborationBrown2021, Lindegren2021a, Riello2021}, which already provided a detailed view of the structure and properties of the LMC \citep{GaiaCollaborationLuri2021}.

\section{Radial velocity variability\label{sect:varperf}}
\gdrthree\ contains two radial velocity variability indices. In Sect.~\ref{sect:varindices}, we propose to combine them into a single criterion. The stars with \verb|rv_chisq_pvalue|~$\leq 0.01$ \& \verb|rv_renormalised_gof|~$> 4$ are considered variable . This criterion is applicable to the stars with  \verb|rv_nb_transits|~$\geq 10$ \& \verb|rv_template_teff|~$\in [3900, 8000]$.

\begin{figure}[h!]
\centering
\includegraphics[width=0.99\hsize]{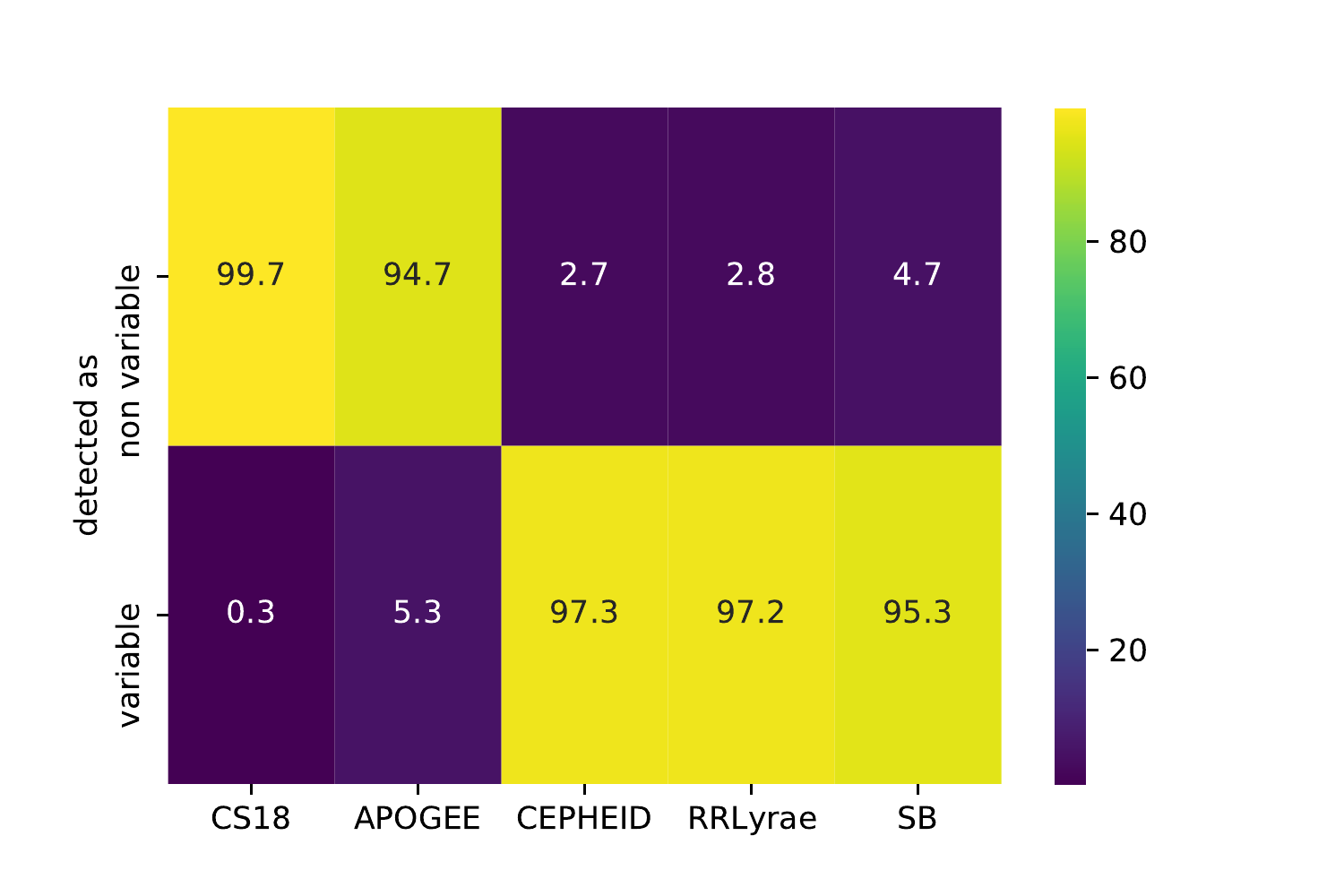}
\caption{Percentage of stars classified as constant (top row) and variable (bottom row), for the two samples of constant stars: CS18 and APOGEE, and the three samples of variable stars: Cepheids, RR-Lyrae and spectroscopic binaries.\label{fig:varperf}}
\end{figure}

To assess the performance of this criterion five datasets are used, two made of constant stars and the three others of variable stars. The first constant star sample (hereafter referred to as CS18) is extracted from the catalogue of \citet{Soubiran2018a}, selecting the stars with at least two measurements in this catalogue, with a minimum of 300 days between the first and the last measurement and with a standard deviation of the radial velocity measurements smaller than or equal to 100~\ms. The second sample of constant stars comes from the APOGEE~DR17 catalogue \citep{Abdurrouf2022}. Here also, a subsample of stars is selected based on the number of APOGEE measurements and on their scatter: \verb|NVISITS|~$\geq 4$ and \verb|VSCATTER|~$\leq 0.5$~\kms. The first two samples of variable stars are made of 744 and 492 previously known Cepheids and RR-Lyrae, respectively, provided by the \gaia-DPAC variability group (G.~Clementini private communication). The last sample of variable stars is made of spectroscopic binaries compiled by T.~Morel (private communication) from the series of papers by R.F.~Griffin \citep[see][and previous papers in the series]{Griffin2019}.

Figure~\ref{fig:varperf} presents the result of the classification in the form of a confusion matrix. For each sample (one per column), the percentages of stars classified as constant and as variable, following our proposed criterion, are provided on the top and bottom rows, respectively. The percentage of CS18 and APOGEE stars correctly classified as constant are 99.7\% and 94.7\%, respectively. The success rates are similar for the variables, 97.3\% of the Cepheids, 97.2\% of the RR-Lyrae and 95.3\% of the spectroscopic binaries are classified as variable by the criterion above.


\section{Conclusions\label{sect:conclusions}}
\gdrthree\ contains the combined radial velocities of \starPublished\ stars, down to $\extgrvs = 14$~mag. With respect to the first radial velocity catalogue published in \gdrtwo, the interval of temperature has been expanded from \verb|rv_template_teff| $\in [3600, 6750]$~K to \verb|rv_template_teff| $\in [3100, 14500]$~K for the bright stars (\verb|grvs_mag| $\leq 12$~mag) and $[3100, 6750]$~K for the fainter stars. The radial velocities sample a significant part of the Milky Way disc, extending a few kilo-parsecs beyond the Galactic centre as well as into the inner halo, up to about 10-15~kpc. The median formal precision of the velocities is of 1.3~\kms\ at \verb|grvs_mag|~$= 12$ and 6.4~\kms\ at \verb|grvs_mag|~$= 14$~mag. The velocities exhibit a small systematic trend with magnitude starting around \verb|grvs_mag|~$= 11$~mag and reaching about 400~\ms\ at \verb|grvs_mag|~$= 14$~mag and for which a correction formula is provided. The \gdrthree\ velocity scale is in satisfactory agreement with APOGEE, GALAH, GES and RAVE. The systematic differences do mostly not exceed a few hundreds~\ms.

For now, \gdrthree\ will provide the largest catalogue of stellar radial velocities. However, records are made to be broken. \gdrfour\ should process all RVS spectra down to the limiting magnitude $\onboardgrvs = 16.2$~mag and it should contain the radial velocities of more than 100 million stars. To achieve this objective, the pipeline will have to process a huge volume of extremely low signal to noise ratio spectra. The separation of the invalid and valid measurements is the challenge of the next radial velocity catalogue. Specific machine learning and deep learning methods are being developed and tested to best identify the spurious velocities, minimise the contamination and maximise the completeness.

\begin{acknowledgements}
We are very grateful to Jos de Bruijne for his thorough reading of the article and for his very helpful suggestions.

This work presents results from the European Space Agency (ESA) space mission \gaia. \gaia\ data are being processed by the \gaia\ Data Processing and Analysis Consortium (DPAC). Funding for the DPAC is provided by national institutions, in particular the institutions participating in the \gaia\ MultiLateral Agreement (MLA). The \gaia\ mission website is \url{https://www.cosmos.esa.int/gaia}. The \gaia\ archive website is \url{https://archives.esac.esa.int/gaia}.
Acknowledgements are given in Appendix~\ref{app:ackno}.

This paper made use of the NASA's Astrophysics Data System (ADS) bibliographic services, as well as of the open-source Python packages ASTROPY (\url{http://www.astropy.org}) \citep{Astropy2013, Astropy2018}, HEALPY (\url{http://healpix.sf.net}) \citep{Gorski2005, Zonca2019},  MATPLOTLIB \citep{Hunter2007}, NUMPY \citep{Harris2020}, and PANDAS \citep{McKinney2010}.
\end{acknowledgements}

%
%

\bibliographystyle{aa}
\bibliography{gaia_dr3_rv}


\begin{appendix}

\section{Gaia DPAC acknowledgements\label{app:ackno}}
This work presents results from the European Space Agency (ESA) space mission \gaia. \gaia\ data are being processed by the \gaia\ Data Processing and Analysis Consortium (DPAC). Funding for the DPAC is provided by national institutions, in particular the institutions participating in the \gaia\ MultiLateral Agreement (MLA). The \gaia\ mission website is \url{https://www.cosmos.esa.int/gaia}. The \gaia\ archive website is \url{https://archives.esac.esa.int/gaia}.

The \gaia\ mission and data processing have financially been supported by, in alphabetical order by country:
\begin{itemize}
\item the Algerian Centre de Recherche en Astronomie, Astrophysique et G\'{e}ophysique of Bouzareah Observatory;
\item the Austrian Fonds zur F\"{o}rderung der wissenschaftlichen Forschung (FWF) Hertha Firnberg Programme through grants T359, P20046, and P23737;
\item the BELgian federal Science Policy Office (BELSPO) through various PROgramme de D\'{e}veloppement d'Exp\'{e}riences scientifiques (PRODEX) grants and the Polish Academy of Sciences - Fonds Wetenschappelijk Onderzoek through grant VS.091.16N, and the Fonds de la Recherche Scientifique (FNRS), and the Research Council of Katholieke Universiteit (KU) Leuven through grant C16/18/005 (Pushing AsteRoseismology to the next level with TESS, GaiA, and the Sloan DIgital Sky SurvEy -- PARADISE);  
\item the Brazil-France exchange programmes Funda\c{c}\~{a}o de Amparo \`{a} Pesquisa do Estado de S\~{a}o Paulo (FAPESP) and Coordena\c{c}\~{a}o de Aperfeicoamento de Pessoal de N\'{\i}vel Superior (CAPES) - Comit\'{e} Fran\c{c}ais d'Evaluation de la Coop\'{e}ration Universitaire et Scientifique avec le Br\'{e}sil (COFECUB);
\item the Chilean Agencia Nacional de Investigaci\'{o}n y Desarrollo (ANID) through Fondo Nacional de Desarrollo Cient\'{\i}fico y Tecnol\'{o}gico (FONDECYT) Regular Project 1210992 (L.~Chemin);
\item the National Natural Science Foundation of China (NSFC) through grants 11573054, 11703065, and 12173069, the China Scholarship Council through grant 201806040200, and the Natural Science Foundation of Shanghai through grant 21ZR1474100;  
\item the Tenure Track Pilot Programme of the Croatian Science Foundation and the \'{E}cole Polytechnique F\'{e}d\'{e}rale de Lausanne and the project TTP-2018-07-1171 `Mining the Variable Sky', with the funds of the Croatian-Swiss Research Programme;
\item the Czech-Republic Ministry of Education, Youth, and Sports through grant LG 15010 and INTER-EXCELLENCE grant LTAUSA18093, and the Czech Space Office through ESA PECS contract 98058;
\item the Danish Ministry of Science;
\item the Estonian Ministry of Education and Research through grant IUT40-1;
\item the European Commission’s Sixth Framework Programme through the European Leadership in Space Astrometry (\href{https://www.cosmos.esa.int/web/gaia/elsa-rtn-programme}{ELSA}) Marie Curie Research Training Network (MRTN-CT-2006-033481), through Marie Curie project PIOF-GA-2009-255267 (Space AsteroSeismology \& RR Lyrae stars, SAS-RRL), and through a Marie Curie Transfer-of-Knowledge (ToK) fellowship (MTKD-CT-2004-014188); the European Commission's Seventh Framework Programme through grant FP7-606740 (FP7-SPACE-2013-1) for the \gaia\ European Network for Improved data User Services (\href{https://gaia.ub.edu/twiki/do/view/GENIUS/}{GENIUS}) and through grant 264895 for the \gaia\ Research for European Astronomy Training (\href{https://www.cosmos.esa.int/web/gaia/great-programme}{GREAT-ITN}) network;
\item the European Cooperation in Science and Technology (COST) through COST Action CA18104 `Revealing the Milky Way with \gaia\ (MW-Gaia)';
\item the European Research Council (ERC) through grants 320360, 647208, and 834148 and through the European Union’s Horizon 2020 research and innovation and excellent science programmes through Marie Sk{\l}odowska-Curie grant 745617 (Our Galaxy at full HD -- Gal-HD) and 895174 (The build-up and fate of self-gravitating systems in the Universe) as well as grants 687378 (Small Bodies: Near and Far), 682115 (Using the Magellanic Clouds to Understand the Interaction of Galaxies), 695099 (A sub-percent distance scale from binaries and Cepheids -- CepBin), 716155 (Structured ACCREtion Disks -- SACCRED), 951549 (Sub-percent calibration of the extragalactic distance scale in the era of big surveys -- UniverScale), and 101004214 (Innovative Scientific Data Exploration and Exploitation Applications for Space Sciences -- EXPLORE);
\item the European Science Foundation (ESF), in the framework of the \gaia\ Research for European Astronomy Training Research Network Programme (\href{https://www.cosmos.esa.int/web/gaia/great-programme}{GREAT-ESF});
\item the European Space Agency (ESA) in the framework of the \gaia\ project, through the Plan for European Cooperating States (PECS) programme through contracts C98090 and 4000106398/12/NL/KML for Hungary, through contract 4000115263/15/NL/IB for Germany, and through PROgramme de D\'{e}veloppement d'Exp\'{e}riences scientifiques (PRODEX) grant 4000127986 for Slovenia;  
\item the Academy of Finland through grants 299543, 307157, 325805, 328654, 336546, and 345115 and the Magnus Ehrnrooth Foundation;
\item the French Centre National d’\'{E}tudes Spatiales (CNES), the Agence Nationale de la Recherche (ANR) through grant ANR-10-IDEX-0001-02 for the `Investissements d'avenir' programme, through grant ANR-15-CE31-0007 for project `Modelling the Milky Way in the \gaia\ era’ (MOD4Gaia), through grant ANR-14-CE33-0014-01 for project `The Milky Way disc formation in the \gaia\ era’ (ARCHEOGAL), through grant ANR-15-CE31-0012-01 for project `Unlocking the potential of Cepheids as primary distance calibrators’ (UnlockCepheids), through grant ANR-19-CE31-0017 for project `Secular evolution of galaxies' (SEGAL), and through grant ANR-18-CE31-0006 for project `Galactic Dark Matter' (GaDaMa), the Centre National de la Recherche Scientifique (CNRS) and its SNO \gaia\ of the Institut des Sciences de l’Univers (INSU), its Programmes Nationaux: Cosmologie et Galaxies (PNCG), Gravitation R\'{e}f\'{e}rences Astronomie M\'{e}trologie (PNGRAM), Plan\'{e}tologie (PNP), Physique et Chimie du Milieu Interstellaire (PCMI), and Physique Stellaire (PNPS), the `Action F\'{e}d\'{e}ratrice \gaia' of the Observatoire de Paris, the R\'{e}gion de Franche-Comt\'{e}, the Institut National Polytechnique (INP) and the Institut National de Physique nucl\'{e}aire et de Physique des Particules (IN2P3) co-funded by CNES;
\item the German Aerospace Agency (Deutsches Zentrum f\"{u}r Luft- und Raumfahrt e.V., DLR) through grants 50QG0501, 50QG0601, 50QG0602, 50QG0701, 50QG0901, 50QG1001, 50QG1101, 50\-QG1401, 50QG1402, 50QG1403, 50QG1404, 50QG1904, 50QG2101, 50QG2102, and 50QG2202, and the Centre for Information Services and High Performance Computing (ZIH) at the Technische Universit\"{a}t Dresden for generous allocations of computer time;
\item the Hungarian Academy of Sciences through the Lend\"{u}let Programme grants LP2014-17 and LP2018-7 and the Hungarian National Research, Development, and Innovation Office (NKFIH) through grant KKP-137523 (`SeismoLab');
\item the Science Foundation Ireland (SFI) through a Royal Society - SFI University Research Fellowship (M.~Fraser);
\item the Israel Ministry of Science and Technology through grant 3-18143 and the Tel Aviv University Center for Artificial Intelligence and Data Science (TAD) through a grant;
\item the Agenzia Spaziale Italiana (ASI) through contracts I/037/08/0, I/058/10/0, 2014-025-R.0, 2014-025-R.1.2015, and 2018-24-HH.0 to the Italian Istituto Nazionale di Astrofisica (INAF), contract 2014-049-R.0/1/2 to INAF for the Space Science Data Centre (SSDC, formerly known as the ASI Science Data Center, ASDC), contracts I/008/10/0, 2013/030/I.0, 2013-030-I.0.1-2015, and 2016-17-I.0 to the Aerospace Logistics Technology Engineering Company (ALTEC S.p.A.), INAF, and the Italian Ministry of Education, University, and Research (Ministero dell'Istruzione, dell'Universit\`{a} e della Ricerca) through the Premiale project `MIning The Cosmos Big Data and Innovative Italian Technology for Frontier Astrophysics and Cosmology' (MITiC);
\item the Netherlands Organisation for Scientific Research (NWO) through grant NWO-M-614.061.414, through a VICI grant (A.~Helmi), and through a Spinoza prize (A.~Helmi), and the Netherlands Research School for Astronomy (NOVA);
\item the Polish National Science Centre through HARMONIA grant 2018/30/M/ST9/00311 and DAINA grant 2017/27/L/ST9/03221 and the Ministry of Science and Higher Education (MNiSW) through grant DIR/WK/2018/12;
\item the Portuguese Funda\c{c}\~{a}o para a Ci\^{e}ncia e a Tecnologia (FCT) through national funds, grants SFRH/\-BD/128840/2017 and PTDC/FIS-AST/30389/2017, and work contract DL 57/2016/CP1364/CT0006, the Fundo Europeu de Desenvolvimento Regional (FEDER) through grant POCI-01-0145-FEDER-030389 and its Programa Operacional Competitividade e Internacionaliza\c{c}\~{a}o (COMPETE2020) through grants UIDB/04434/2020 and UIDP/04434/2020, and the Strategic Programme UIDB/\-00099/2020 for the Centro de Astrof\'{\i}sica e Gravita\c{c}\~{a}o (CENTRA);  
\item the Slovenian Research Agency through grant P1-0188;
\item the Spanish Ministry of Economy (MINECO/FEDER, UE), the Spanish Ministry of Science and Innovation (MICIN), the Spanish Ministry of Education, Culture, and Sports, and the Spanish Government through grants BES-2016-078499, BES-2017-083126, BES-C-2017-0085, ESP2016-80079-C2-1-R, ESP2016-80079-C2-2-R, FPU16/03827, PDC2021-121059-C22, RTI2018-095076-B-C22, and TIN2015-65316-P (`Computaci\'{o}n de Altas Prestaciones VII'), the Juan de la Cierva Incorporaci\'{o}n Programme (FJCI-2015-2671 and IJC2019-04862-I for F.~Anders), the Severo Ochoa Centre of Excellence Programme (SEV2015-0493), and MICIN/AEI/10.13039/501100011033 (and the European Union through European Regional Development Fund `A way of making Europe') through grant RTI2018-095076-B-C21, the Institute of Cosmos Sciences University of Barcelona (ICCUB, Unidad de Excelencia `Mar\'{\i}a de Maeztu’) through grant CEX2019-000918-M, the University of Barcelona's official doctoral programme for the development of an R+D+i project through an Ajuts de Personal Investigador en Formaci\'{o} (APIF) grant, the Spanish Virtual Observatory through project AyA2017-84089, the Galician Regional Government, Xunta de Galicia, through grants ED431B-2021/36, ED481A-2019/155, and ED481A-2021/296, the Centro de Investigaci\'{o}n en Tecnolog\'{\i}as de la Informaci\'{o}n y las Comunicaciones (CITIC), funded by the Xunta de Galicia and the European Union (European Regional Development Fund -- Galicia 2014-2020 Programme), through grant ED431G-2019/01, the Red Espa\~{n}ola de Supercomputaci\'{o}n (RES) computer resources at MareNostrum, the Barcelona Supercomputing Centre - Centro Nacional de Supercomputaci\'{o}n (BSC-CNS) through activities AECT-2017-2-0002, AECT-2017-3-0006, AECT-2018-1-0017, AECT-2018-2-0013, AECT-2018-3-0011, AECT-2019-1-0010, AECT-2019-2-0014, AECT-2019-3-0003, AECT-2020-1-0004, and DATA-2020-1-0010, the Departament d'Innovaci\'{o}, Universitats i Empresa de la Generalitat de Catalunya through grant 2014-SGR-1051 for project `Models de Programaci\'{o} i Entorns d'Execuci\'{o} Parallels' (MPEXPAR), and Ramon y Cajal Fellowship RYC2018-025968-I funded by MICIN/AEI/10.13039/501100011033 and the European Science Foundation (`Investing in your future');
\item the Swedish National Space Agency (SNSA/Rymdstyrelsen);
\item the Swiss State Secretariat for Education, Research, and Innovation through the Swiss Activit\'{e}s Nationales Compl\'{e}mentaires and the Swiss National Science Foundation through an Eccellenza Professorial Fellowship (award PCEFP2\_194638 for R.~Anderson);
\item the United Kingdom Particle Physics and Astronomy Research Council (PPARC), the United Kingdom Science and Technology Facilities Council (STFC), and the United Kingdom Space Agency (UKSA) through the following grants to the University of Bristol, the University of Cambridge, the University of Edinburgh, the University of Leicester, the Mullard Space Sciences Laboratory of University College London, and the United Kingdom Rutherford Appleton Laboratory (RAL): PP/D006511/1, PP/D006546/1, PP/D006570/1, ST/I000852/1, ST/J005045/1, ST/K00056X/1, ST/\-K000209/1, ST/K000756/1, ST/L006561/1, ST/N000595/1, ST/N000641/1, ST/N000978/1, ST/\-N001117/1, ST/S000089/1, ST/S000976/1, ST/S000984/1, ST/S001123/1, ST/S001948/1, ST/\-S001980/1, ST/S002103/1, ST/V000969/1, ST/W002469/1, ST/W002493/1, ST/W002671/1, ST/W002809/1, and EP/V520342/1.
\end{itemize}

The GBOT programme  uses observations collected at (i) the European Organisation for Astronomical Research in the Southern Hemisphere (ESO) with the VLT Survey Telescope (VST), under ESO programmes
092.B-0165,
093.B-0236,
094.B-0181,
095.B-0046,
096.B-0162,
097.B-0304,
098.B-0030,
099.B-0034,
0100.B-0131,
0101.B-0156,
0102.B-0174, and
0103.B-0165;
%
%
and (ii) the Liverpool Telescope, which is operated on the island of La Palma by Liverpool John Moores University in the Spanish Observatorio del Roque de los Muchachos of the Instituto de Astrof\'{\i}sica de Canarias with financial support from the United Kingdom Science and Technology Facilities Council, and (iii) telescopes of the Las Cumbres Observatory Global Telescope Network.

\section{\gdrthree\ spectroscopy related fields\label{app:DB}}
Table~\ref{tab:dr3Fields} lists the parameters related to the radial velocity published in \gdrthree. They are all stored in the \verb|gaia_source| table. It should be noted that the computation of \verb|rv_chisq_pvalue|, \verb|rv_renormalised_gof| and, \verb|rv_amplitude_robust| requires reliable epoch radial velocities and is therefore restricted to \verb|grvs_mag|~$\leq 12$~mag. \verb|rv_chisq_pvalue| is further restricted to stars with \verb|rv_nb_transits|~$\geq 3$ and \verb|rv_renormalised_gof| to stars with \verb|grvs_mag|~$\geq 5.5$~mag and \verb|rv_template_teff| $< 14500$~K. In \gdrthree, the radial velocity time series are published for a limited sample of slightly less than 2000~Cepheids and RR-Lyrae. They can be identified using the field \verb|has_epoch_rv| which is set to \verb|True|. Their time series are stored in the table \verb|vari_epoch_radial_velocity|.

Table~\ref{tab:dr3Fields2} lists the parameters produced by the spectroscopic pipeline and presented in companion papers. They are all stored in the \verb|gaia_source| table, except \verb|rvs_mean_spectrum| which is a DataLink product.

\begin{table*}[]
\caption{Radial velocity related fields published in \gdrthree. They are all stored in the gaia\_source table.\label{tab:dr3Fields}}
\begin{center}
\begin{tabular}{l l l l} \hline\hline
Field                                        & Units    & DB column name                    & Sect.  \\ \hline
Combined radial velocity                     & \kms     & \verb|radial_velocity|            & \ref{sect:mta} \\
Combined radial velocity formal uncertainty  & \kms     & \verb|radial_velocity_error|      & \ref{sect:mta} \\
Combined radial velocity method              &          & \verb|rv_method_used|             & \ref{sect:mta} \\
Number of transits                           & transits & \verb|rv_nb_transits|             & \\
Number of de-blended transits                & transits & \verb|rv_nb_deblended_transits|   & \\
Number of visibility periods                 &          & \verb|rv_visibility_periods_used| & \\
Duration of the radial velocity time series  & days     & \verb|rv_time_duration|           & \\
Expected signal-to-noise ratio               &          & \verb|rv_expected_sig_to_noise|   & \\
Renormalised goodness of fit                 &          & \verb|rv_renormalised_gof|        & \ref{sect:varindices} \\
Chi-square P-value                           &          & \verb|rv_chisq_pvalue|            & \ref{sect:varindices} \\
Amplitude of the radial velocity time series & \kms     & \verb|rv_amplitude_robust|        & \ref{sect:varindices} \\
Template effective temperature               & K        & \verb|rv_template_teff|           & \ref{sect:sta} \\
Template surface gravity                     & dex      & \verb|rv_template_logg|           & \ref{sect:sta} \\
Template metallicity                         & dex      & \verb|rv_template_fe_h|           & \ref{sect:sta} \\
Origin of the atmospheric parameters         &          & \verb|rv_atm_param_origin|        & \ref{sect:sta} \\
Availability of radial velocity time series  &          & \verb|has_epoch_rv|               & \ref{sect:sta} \\ \hline
\end{tabular}
\end{center}
\end{table*}

\begin{table*}[]
\caption{Quantities produced by the spectroscopic pipeline and presented in companion papers. They are all stored in the gaia\_source table, except rvs\_mean\_spectrum which is a DataLink product.\label{tab:dr3Fields2}}
\begin{center}
\begin{tabular}{l l l l} \hline\hline
Field                                        & Units    & DB column name                    & Reference              \\ \hline
Median broadening velocity                   & \kms     & \verb|vbroad|                     & \citet{DR3-DPACP-149}     \\
Broadening velocity uncertainty              & \kms     & \verb|vbroad_error|               & \citet{DR3-DPACP-149}     \\
Number of vbroad transits                    & transits & \verb|vbroad_nb_transits|         & \citet{DR3-DPACP-149}     \\
Median $\grvs$                               & mag      & \verb|grvs_mag|                   & \citet{DR3-DPACP-155} \\
$\grvs$ uncertainty                          & mag      & \verb|grvs_mag_error|             & \citet{DR3-DPACP-155} \\
Number of $\grvs$ transits                   & transits & \verb|grvs_mag_nb_transits|       & \citet{DR3-DPACP-155} \\
Availability of the mean spectrum            &          & \verb|has_rvs|                    & \citet{DR3-DPACP-154}   \\
Signal-to-noise ratio per pixel of the mean spectrum & & \verb|rv_spec_sig_to_noise| & \citet{DR3-DPACP-154}   \\
Mean spectrum                                &          & \verb|rvs_mean_spectrum|          & \citet{DR3-DPACP-154}   \\
\end{tabular}
\end{center}
\end{table*}

\section{Comparison catalogues\label{app:GB}}
In Sect.~\ref{sect:accuracy} (accuracy), Sect.~\ref{sect:uncertainties} (formal uncertainties) and Sect.~\ref{sect:HVS} (high-velocity stars), the \gdrthree\ combined radial velocities are compared to the velocities of one or several of the following catalogues: APOGEE~DR17 \citep{Abdurrouf2022}, GALAH~DR3 \citep{Buder2021, Zwitter2021}, GAIA-ESO Survey (GES) DR3 \citep{Gilmore2012, Randich2013}, LAMOST~DR7 \citep{Zhao2012, Deng2012, Luo2015} and RAVE~DR6 \citep{Steinmetz2020_1, Steinmetz2020_2}. The selection of the comparison samples are described below.

\textit{APOGEE~DR17.} We use the data from the \verb|allStar-dr17-synspec_rev1.fits| file, including the cross-match with the \gdrethree\ catalogue (which is based on the same list of \verb|source_id| as \gdrthree). For each APOGEE star observed multiple times, the occurrence with the highest signal to noise ratio was kept (\verb|EXTRATARG| 4$^{th}$ binary digit set to 0) and the others were discarded. The stars with no match with a star with a published radial velocity in \gdrthree\ were removed. The stars that failed to meet the following quality criteria were also removed: (i) \verb|STARFLAG| 0$^{th}$ (\verb|BAD_PIXELS|) and 3$^{rd}$ (\verb|VERY_BRIGHT_NEIGHBOR|) binary digits set to 0, (ii) \verb|ASPCAPFLAG| 10$^{th}$ (\verb|ROTATION_WARN|) and 23$^{rd}$ (\verb|STAR_BAD|) binary digits set to 0, (iii) valid \verb|VHELIO_AVG|, \verb|TEFF|, \verb|LOGG| and \verb|FE_H| and (iv) \verb|N_COMPONENTS| equal to 1. The APOGEE~DR17 comparison sample contains 459~998 stars.

\textit{GALAH~DR3.} We use the data from the \verb|GALAH_DR3_main_allstar_v2.fits| file and from the \verb|GALAH_DR3_VAC_rv_v2.fits| file, including the cross-match with the \gdrethree\ catalogue. The duplicated \gdrethree\ \verb|source_id| and the stars with no match with a star with a published radial velocity in \gdrthree\ were removed. The stars that failed to meet the following quality criteria were also removed: (i) \verb|snr_c3_iraf| $\geq 30$, (ii) \verb|use_rv_flag|, \verb|flag_sp| and \verb|flag_fe_h| all three equal to 0 and (iii) valid \verb|rv_nogr_obst|, \verb|teff|, \verb|logg| and \verb|fe_h|. The 12~760 stars identified as double-line spectroscopic binaries by \citet{Traven2020} were also discarded. GALAH DR3 is providing several measurements of the radial velocity. We use the field \verb|rv_nogr_obst|, which is not corrected for the gravitational redshift, nor for the convective shift. The GALAH~DR3 comparison sample contains 294~976 stars.

\textit{GES~DR3.} The data were downloaded from the ESO archive: \url{https://archive.eso.org/wdb/wdb/adp/phase3_main/form?collection_name=GAIAESO&release_name=DR3}. The cross-match was provided by the \gaia\ catalogue validation group \citep{DR3-DPACP-127}. The duplicated stars and the stars with no match with a star with a published radial velocity in \gdrthree\ were removed. The following stars were also removed: (i) \verb|TECH| flag set to 9020, 9030, 9050, 15100, 15110 or 15130, (ii) \verb|PECULI| flag set to 2005, 2010, 2020, 2030, 2040 or 2070 and (iii) invalid \verb|VRAD|, \verb|TEFF|, \verb|LOGG| or \verb|FEH|. The GES~DR3 comparison sample contains 5~010 stars.

\textit{LAMOST~DR7.} We use the data from the \verb|dr7_v2.0_LRS_stellar.csv.gz| file, including the cross-match with the \gdrtwo\ catalogue. The \gdrtwo\ \verb|source_id| were converted into \gdrethree\ \verb|source_id| using the table \verb|gaiaedr3.dr2_neighbourhood|. For each LAMOST star observed multiple times (with all observations matching the same \gdrethree\ star), the occurrence with the highest \verb|snrg| was kept. The stars with multiple LAMOST observations matching different \gdrethree\ stars were removed. The duplicated \gdrethree\ \verb|source_id| and the stars with no match with a star with a published radial velocity in \gdrthree\ were removed. The stars that failed to meet the following quality criteria were also removed: (i) \verb|rv_error|~$\geq 0$, \verb|snr|~$\geq 0$, \verb|snrg|~$> 25$ and \verb|snri|~$> 25$, and (ii) valid \verb|rv|, \verb|teff|, \verb|logg| and \verb|feh|. The LAMOST~DR7 comparison sample contains 1~791~438 stars.

\textit{RAVE~DR6.} The data (including the cross-match with the \gdrethree\ catalogue) were downloaded from the tap server \url{https://www.rave-survey.org/tap}, from the tables: \verb|dr6_sparv|, \verb|dr6_x_gaiaedr3|, \verb|dr6_classification| and \verb|dr6_madera|. The RAVE stars observed multiple times (with all observations matching the same \gdrethree\ star and having the same value of the \verb|flag_1| flag), the occurrence with the highest \verb|snr_med_sparv| was kept. The stars with multiple RAVE observations matching different \gdrethree\ stars or having different values of the \verb|flag_1| flag were removed. The duplicated \gdrethree\ \verb|source_id| and the stars with no match with a star with a published radial velocity in \gdrthree\ were removed. The stars with \verb|flag_1| set to \verb|e|, \verb|b|, \verb|p| or \verb|c,w| were removed. The stars that failed to meet the following quality criteria were also removed: (i) absolute value of \verb|correction_rv_sparv| $< 10$, \verb|correlation_coeff_sparv|~$> 10$, \verb|hrv_error_sparv|~$< 8$, \verb|algo_con_madera|~$\in [0, 2, 3, 4]$, \verb|snr_med_sparv| $\geq 20$ and \verb|fe_h_chisq_gauguin|~$<1.4$, and (ii) valid \verb|hrv_sparv|, \verb|teff_cal_madera|, \verb|logg_cal_madera| and \verb|fe_h_gauguin|. The RAVE~DR6 comparison sample contains 261~798 stars.

\section{Sample selection: atmospheric parameter trends\label{app:Acc}}
The dependence on the atmospheric parameters of the systematic differences between \gdrthree\ and the comparison samples (see Appendix~\ref{app:GB}) is studied in Sect.~\ref{sect:aptrends}. To avoid mixing cross-dependences, specific samples are used for the effective temperature, the surface gravity and the metallicity, respectively. They are described below.

\begin{figure*}[tbp]
\centering
\includegraphics[width=0.99 \hsize]{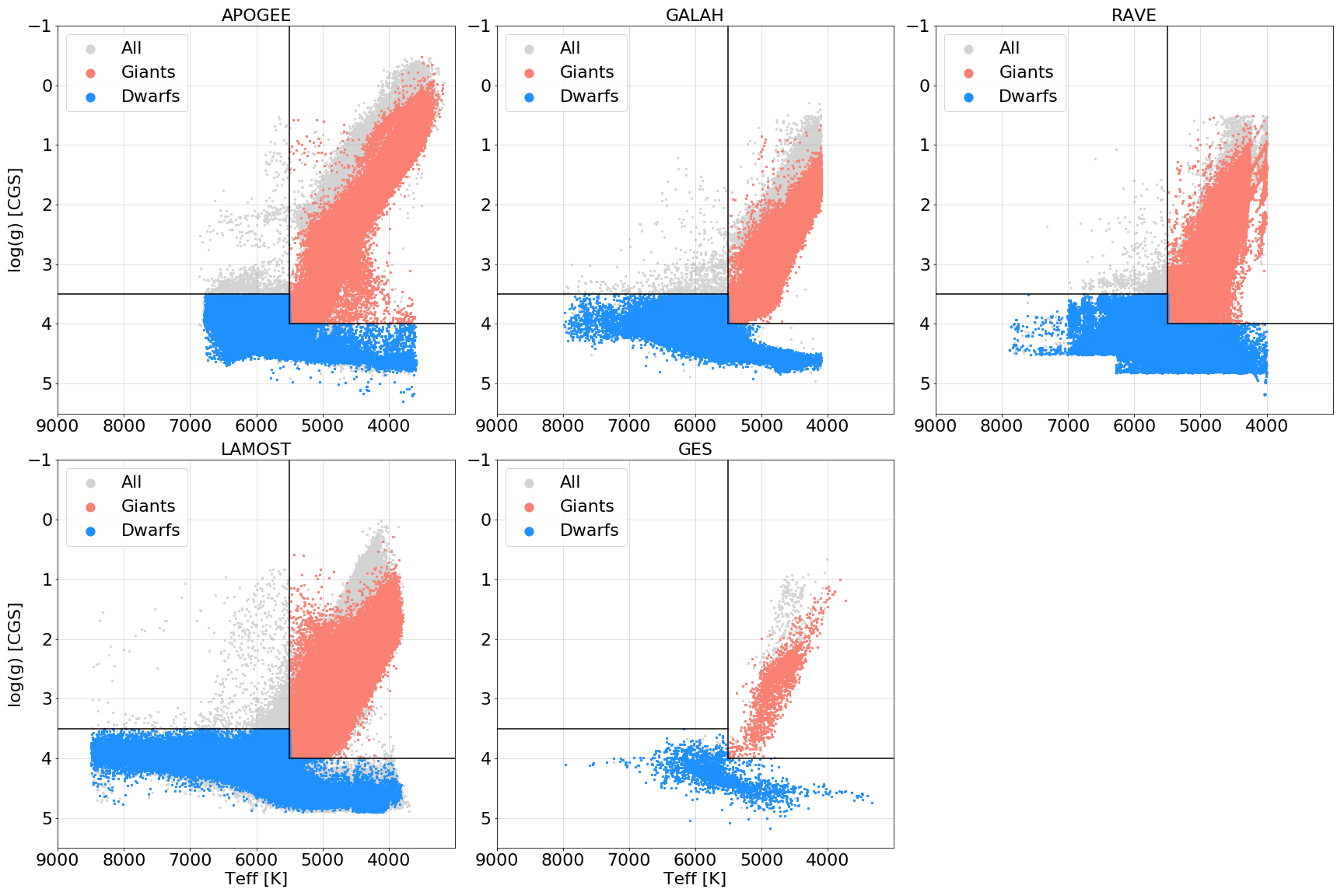}
\caption{Kiel diagrams of the comparison samples (see Appendix~\ref{app:GB}). The metal-rich giant and metal-rich dwarf samples, used to assess the radial velocity systematic differences as a function of surface gravity and effective temperature, are shown as salmon and blue dots, respectively. The stars selected in none of the samples are shown as grey dots.\label{fig:appC1}}
\end{figure*}

The surface gravity and the effective temperature trends are assessed with samples of metal-rich giants and dwarfs respectively. The selection is based on the atmospheric parameters provided by the different comparison catalogues. The metal-rich giants are selected as: $\Teff \leq 5500$~K, $\logg < 4.0$ and $\FeH \in[-0.5, 0.5]$~dex. The metal-rich dwarfs are selected as: $\FeH \in[-0.5, 0.5]$~dex and either $\Teff \leq 5500$~K and $\logg \geq 4.0$ or $\Teff > 5500$~K and $\logg \geq 3.5$. Some of the comparison catalogues contain few bright stars. To somehow homogenise the bright limit, the samples were trimmed at \verb|grvs_mag|~$> 8.5$~mag. The LAMOST sample was also restricted to \verb|grvs_mag|~$< 12$~mag, to avoid the radial velocity offset occurring at \verb|grvs_mag|~$> 12.5$~mag (see Sect.~\ref{sect:magtrend}). Figure~\ref{fig:appC1} presents the Kiel diagrams of the comparison samples. The metal-rich giant and metal-rich dwarf samples are shown as salmon dots and blue dots, respectively.

\begin{figure*}[tbp]
\centering
\includegraphics[width=0.99 \hsize]{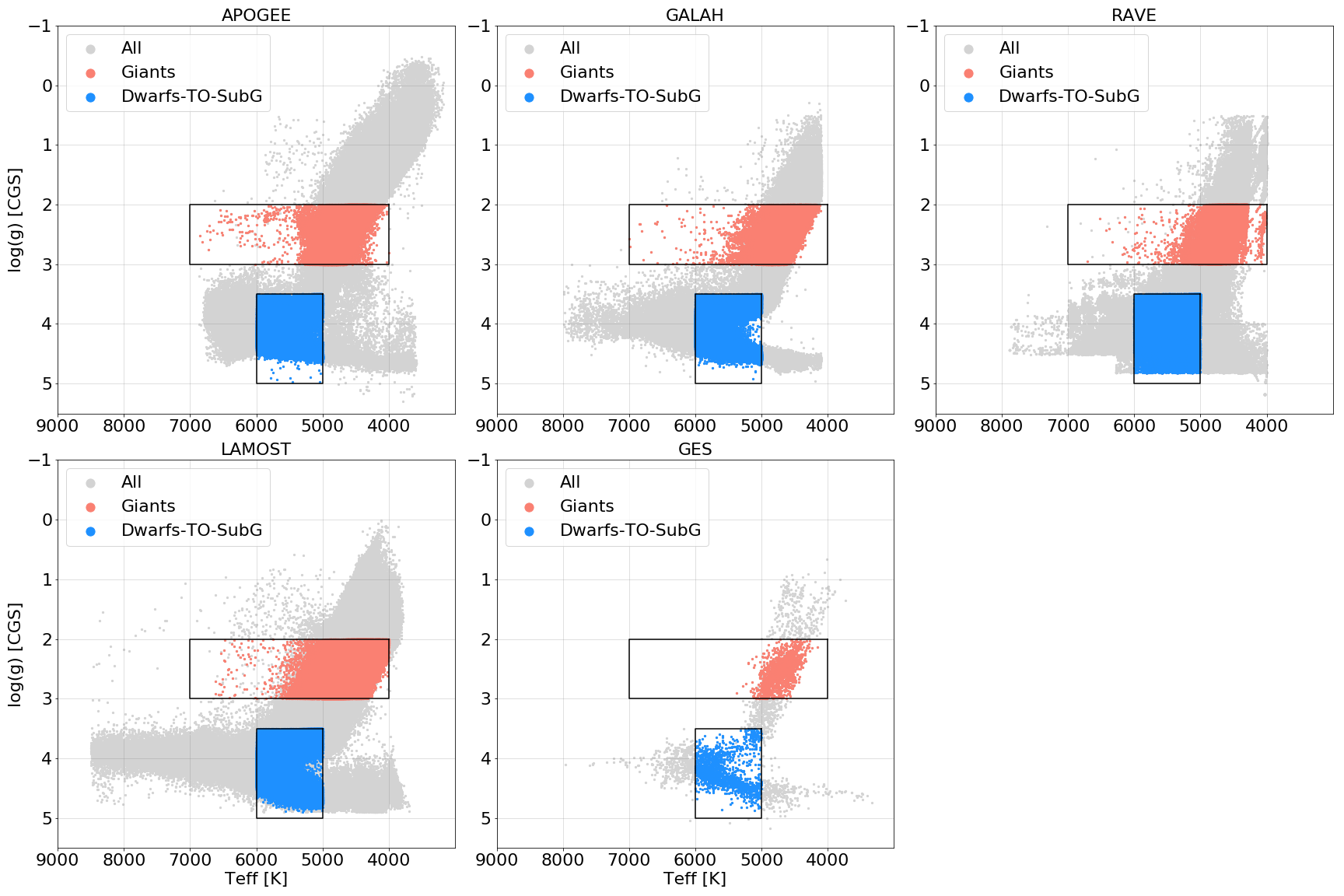}
\caption{Kiel diagrams of the comparison samples (see Appendix~\ref{app:GB}). The giant and dwarf samples, used to assess the radial velocity systematic differences as a function of metallicity, are shown as salmon and blue dots, respectively. The stars selected in none of the samples are shown as grey dots.\label{fig:appC2}}
\end{figure*}

In order to decouple the gravity/temperature trends on the one hand and the metallicity trends on the other hand, the latter was assessed with groups of giants and dwarfs selected in "narrow windows" in the Kiel diagram. They are shown in Fig.~\ref{fig:appC2} as salmon (giants) and blue (dwarfs) dots, respectively. The magnitude cuts used for the gravity/temperature samples also apply here, that is \verb|grvs_mag|~$> 8.5$~mag and for the LAMOST catalogue: \verb|grvs_mag|~$< 12$~mag.

\section{Sample selection: formal uncertainties\label{app:Unc}}
\begin{figure}[]
\centering
\includegraphics[width=0.99 \hsize]{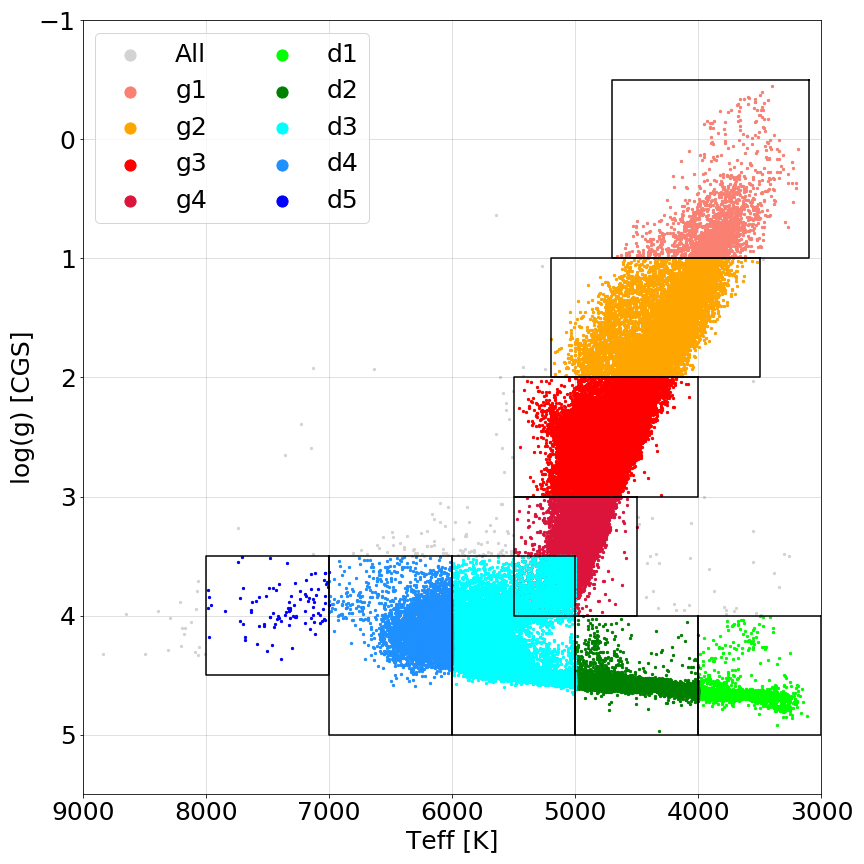}
\caption{Selection in the Kiel diagram of the different giant and dwarf star samples used to assess the reliability of the formal uncertainties of the \gdrthree\ radial velocities. The stars selected in none of the samples are shown as grey dots.\label{fig:appUnC}}
\end{figure}

In Sect.~\ref{sect:uncertainties}, the reliability of the formal uncertainties is assessed using the APOGEE~DR17 comparison sample (see Appendix~\ref{app:GB}). In order to minimise the number of radial velocity variable stars, two additional selection criteria are applied: at least four APOGEE measurements (\verb|NVISITS|~$\geq 4$) and a scatter of the individual APOGEE radial velocities \verb|VSCATTER| $\leq 0.5$~\kms. The 94~309 stars meeting these criteria are further split in several dwarf and giant star samples: from g1 at the top of the giant branch, $\logg \in [-0.5, 1.0[$, to g4 at the bottom, $\logg \in [3.0, 4.0[$, and from d1 at the cool end of the main sequence, $\Teff \in [3000, 4000[$~K, to d5 at the hot end, $\Teff \in [7000, 8000[$~K. We note that the g4 and d3 samples partly overlap and therefore share some of their stars. The selections are based on the APOGEE~DR17 effective temperatures and surface gravities. Fig.~\ref{fig:appUnC} shows the selection of the giant and dwarf star samples in the Kiel diagram.

\section{Sample selection: median formal precision\label{app:Prec}}
\begin{figure}[]
\centering
\includegraphics[width=0.99 \hsize]{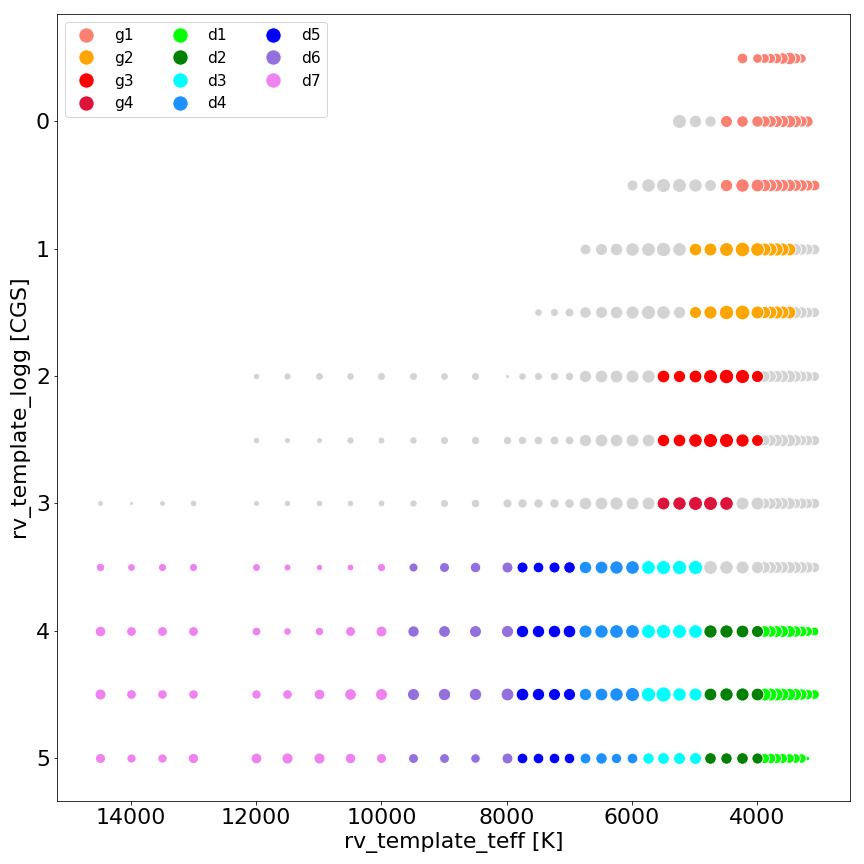}
\caption{Selection in the (rv\_template\_teff, rv\_template\_logg) plane of the different giant and dwarf star samples used to estimate the median formal precisions. The size of the dots is proportional to the number of stars with the combination of (rv\_template\_teff, rv\_template\_logg) parameters. The combinations of template parameters selected in none of the samples are shown as grey dots.\label{fig:appPrec}}
\end{figure}

In Sect.~\ref{sect:precision}, the median formal precision is estimated using the \starPublished\ stars with a radial velocity published in \gdrthree. The parameters of the templates are used to split the data in several dwarf and giant star samples: from g1 at the top of the giant branch, \verb|rv_template_logg| $\in [-0.5, 0.5]$ to g4 at the bottom, \verb|rv_template_logg| $= 3.0$ and from d1 at the cool end of the main sequence, \verb|rv_template_teff| $\leq 3750$~K, to d7 at the hot end, \verb|rv_template_teff| $\geq 10000$~K. Fig.~\ref{fig:appPrec} shows the selection of the giant and dwarf star samples in the (\verb|rv_template_teff|, \verb|rv_template_logg|) plane.

\end{appendix}

\end{document}